\documentclass[aps,prb,superscriptaddress,reprint,nofootinbib]{revtex4-2}

\usepackage{microtype}

\usepackage{xcolor}

\newcommand{\sgn}{\textup{sgn}}

\newenvironment{tentative-environment}{\color{blue}}{}

\usepackage{float}
\usepackage{graphicx}
\usepackage{amsmath}
\usepackage{amsthm}
\usepackage{array}
\usepackage{enumitem}
\usepackage{amssymb}

\usepackage[colorlinks=true,linkcolor=blue,citecolor=blue,urlcolor=blue,pdftitle={RC glass}]{hyperref}

\newcommand{\llangle}{\langle\!\langle}
\newcommand{\rrangle}{\rangle\!\rangle}

\newtheorem*{theorem*}{Theorem}
\newtheorem{theorem}{Theorem}
\newtheorem*{remark*}{Remark}

\def\EstarClean{-0.41} 
\def\EstarDis{-0.39}

\def\TstarClean{0.92}
\def\TstarDis{0.8} 

\def\Tcrit{0.89}

\def\EZeroDis{-0.70}

\def\VFTA{1.01} 
\def\VFTDelta{0.98}
\def\VFTT{0.78}

\usepackage[sectionbib]{bibunits}

\defaultbibliographystyle{apsrev4-2}
\defaultbibliography{references}

\begin{document}
\title{Saddles-to-minima topological crossover and glassiness in the Rubik's Cube}

\author{Alex Gower}
\affiliation{TCM Group, Cavendish Laboratory, University of Cambridge, Cambridge CB3 0HE, United Kingdom}
\affiliation{Nokia Bell Labs, Broers Building, 21 J.J. Thomson Avenue, Cambridge CB3 0FA, United Kingdom}
\author{Oliver Hart}
\affiliation{Department of Physics and Center for Theory of Quantum Matter, University of Colorado, Boulder, Colorado 80309, USA}
\author{Claudio Castelnovo}
\affiliation{TCM Group, Cavendish Laboratory, University of Cambridge, Cambridge CB3 0HE, United Kingdom}

\begin{abstract}
Slow relaxation and glassiness have been the focus of extensive research attention, along with popular and technological interest, for many decades. While much understanding has been attained through mean-field and mode-coupling models, energy landscape paradigms, and real-space descriptions of dynamical heterogeneities and facilitation, a complete framework about the origins and existence of a glass transition is yet to be achieved. In this work we propose a discrete model glass-former, inspired by the famous Rubik's Cube, where these questions can be answered with surprising depth. 
By introducing a swap-move Monte Carlo algorithm, we are able to access thermal equilibrium states above and below the temperature of dynamical arrest. Exploiting the discreteness of the model, we probe directly the energy-resolved connectivity structure of the model, and we uncover a saddles-to-minima topological crossover underpinning the slowing down and eventual arrest of the dynamics. We demonstrate that the smooth behaviour in finite-size systems tends asymptotically to a sharp step-change in the thermodynamic limit, identifying a well-defined energy threshold where the onset of stretched exponential behaviour and the dynamical arrest come to coincide. Our results allow us to shed light on the origin of the glassy behaviour, and how this is resolved by changing connectivity upon introducing swap-moves. 
\end{abstract}

\maketitle
%
%

\begin{bibunit}

\section{Introduction}
The emergence of long relaxation timescales in physical phenomena has puzzled and fascinated the scientific community -- as well as the general public -- for over a century~\cite{cavagnaSupercooledLiquidsPedestrians2009}. 
The difference in timescales between systems that appear at first sight deceivingly similar can be gargantuan~\cite{angellPerspectiveGlassTransition1988,angellFormationGlassesLiquids1995}. The origin of slow dynamics and glassiness has been the focus of much research over the years, with numerous models and mechanisms proposed and substantial understanding developed. Yet, a complete theoretical framework to diagnose, explain, and predict these systems remains in the making (for a non-comprehensive selection of recent reviews, see Refs.~\onlinecite{biroliPerspectiveGlassTransition2013,garrahanAspectsNonequilibriumClassical2018,speckDynamicFacilitationTheory2019,arceriStatisticalMechanicsPerspective2020,guiselinGlassStateMatter2022,berthierModernComputationalStudies2023}). 

Real-space studies of glassy systems provide in-depth understanding based on modelling the behaviour of microscopic defects and excitations. They have led to the identification of dynamical facilitation processes and heterostructures, which are believed to underpin the exceptionally slow relaxation observed in these systems~\cite{berthierRealSpaceOrigin2003,garrahanAspectsNonequilibriumClassical2018}. 

Complementary insight has been gained by relating the slow relaxation to complex dynamics on rugged energy landscapes~\cite{walesExploringEnergyLandscapes2018,razaComputerSimulationsGlasses2015}. 
Energy landscape analyses offer the advantage of an agnostic approach that does not require model-specific understanding of the defect structure or possible quasiparticle description of the system. They are also relevant across a diverse range of fields, including deep neural networks~\cite{chaudhariEnergyLandscapeDeep2017} and protein folding~\cite{walesEnergyLandscapesApplications2004}. 

These studies have highlighted several characteristic temperatures in the phase diagram of glass-forming systems. Two are of particular importance in the discussion below: the `onset temperature', below which heterogeneous dynamical processes and stretched-exponential relaxation are first observed~\cite{berthierRealSpaceOrigin2003}; and the (lower) `glass transition' temperature, below which relaxation processes become so slow as to be effectively inaccessible~\cite{angellPerspectiveGlassTransition1988,angellFormationGlassesLiquids1995}. 

In recent years it has been proposed that the appearance of slow relaxation phenomena may be related to a topological crossover in the instability index of stationary points (a proxy for the connectivity of configuration space, when organised as a function of the energy of the system)~\cite{broderixEnergyLandscapeLennardJones2000,angelaniSaddlesEnergyLandscape2000,cavagnaFragileVsStrong2001,cavagnaRoleSaddlesMeanfield2001,grigeraGeometricApproachDynamic2002,coslovichLocalizationTransitionUnderlies2019,rosDynamicalInstantonsActivated2021}. 
This topological crossover takes place between two distinct regions along the energy axis: one at high energy where most configurations are saddles (i.e., they have energy-decreasing as well as increasing connections), and one at low energy where they are mostly minima. The crossover energy 
between these two regimes relates directly to the typical energy $E^*$ at which the system freezes when thermally annealed, due to the sudden appearance of exceptionally large relaxation timescales. 
The rapidity of this growth, namely the difference between strong and fragile glasses~\cite{debenedettiSupercooledLiquidsGlass2001,angellFormationGlassesLiquids1995}, has been proposed to relate to the magnitude of the typical energy barriers in the system at the energy in which it undergoes the topological crossover~\cite{cavagnaFragileVsStrong2001}. 

This idea suggests a relationship between the slow dynamics of a system and the connectivity of its configuration network. However, in order to investigate the topological crossover in detail, one ideally ought to access the structure of such network not only above but also below $E^*$ -- which is generally difficult precisely due to the exceedingly long timescales required to thermally explore energies $E < E^*$. Indeed, the study in Ref.~\onlinecite{grigeraGeometricApproachDynamic2002} was enabled by the discovery of swap-moves that allow efficient equilibrium sampling of glassy polydisperse soft sphere systems~\cite{grigeraFastMonteCarlo2001} below $E^*$ (see also Refs.~\onlinecite{gutierrezStaticLengthscaleCharacterizing2015,berthierEfficientSwapAlgorithms2019,coslovichLocalizationTransitionUnderlies2019} for further developments on swap-move studies of slow dynamics). 
Finding glassy systems where alternative dynamics are known that allow efficient thermalisation below a glass or jamming transition is generally a tall order~\cite{berthierModernComputationalStudies2023}. Moreover, in order to investigate the proposed topological crossover scenario, it would be ideal to attain this thermalisation capability in a discrete system, for two practical reasons: (i) the connectivity of the configuration network is immediately defined, without the need to go through stationary states~\cite{broderixEnergyLandscapeLennardJones2000} (this allows, for instance, the exploration of how configurations are connected to each other, not just their local stability); and (ii) the alternative (swap-move) dynamics can be studied on an equal footing, allowing one to see how changes in the connectivity result in the suppression of the topological crossover and in faster relaxation. To date, swap-move studies have mainly been limited to models with continuous dynamics (see however Refs.~\onlinecite{gutierrezAcceleratedRelaxationSuppressed2019,alfaro-mirandaSWAPAlgorithmLattice2024} for recent discrete examples). 

Here we uncover a discrete microscopic model -- a variant of the ferromagnetic 6-state Potts model inspired by the famous Rubik's Cube -- that exhibits a dramatic increase in its relaxation timescales at low temperature $T \lesssim T^*$ (seemingly of the Vogel-Fulcher-Tammann form~\cite{hTemperaturabhangigkeitsgesetzViskositatFlussigkeiten1921,tammannAbhangigkeitViscositatTemperatur1926,fulcherAnalysisRecentMeasurements1925}), leading to a dynamical arrest at a finite fraction ($E^*$) of the known ground state energy, even in systems as small as the $3 \times 3 \times 3$ cube with as few as $N=54$ spins. We further discover suitable swap-moves (combinations of the well-known slice-rotations which act as elementary dynamical moves in the original system) that, when added to the dynamics, allow the system to thermalise efficiently at any temperature. In this setting, we are able to study in detail the energy-resolved connectivity of the configuration network across the glass transition, and demonstrate that it exhibits a topological crossover around an energy scale 
$\simeq E^*$, 
similar to the one observed in Ref.~\onlinecite{grigeraGeometricApproachDynamic2002}. We find that the topological crossover is indeed removed if the connectivity is modified to include swap-moves, and we quantitatively compare the connectivity properties of the two dynamics. 

Our study enables us to shed light on the origin of the slow dynamics in the model. The typical size of the energy barriers incurred in the dynamics of the system grows as temperature is lowered; however, for energies larger than $E^*$ (above the topological crossover), there remain viable relaxation routes that avoid such barriers and the system can easily thermalise. When $T < T^*$, a topological crossover occurs and energy barriers become unavoidable; by this point the typical (equilibrium) barrier has grown to be substantially larger than temperature, leading to an abrupt dynamical arrest (characteristic of fragile glass behaviour~\cite{cavagnaFragileVsStrong2001}). We propose a simple toy mechanism for the system's arrest at energy $E^*$ based on a recursive argument resulting in a dynamical attractor towards $(T^*, \, E^*)$. 
The introduction of swap-move dynamics in the model introduces a finite density of pathways for the system to avoid the energy barriers, thus relieving the slow dynamics and allowing it to efficiently thermalise at all temperatures. 

In our study, we are able to directly relate the onset and glass transition temperatures to specific features in the energy dependence of the proportion of saddles and minima configurations. While these proportions appear to depend smoothly on the energy of finite size systems, and are therefore devoid of any threshold feature, we find that their behaviour converges to a step function when extrapolated to infinite system size, occurring at a well-defined finite energy. In this limit, the onset and glass transition temperatures come to coincide -- resolving in an elegant way the conundrum about which of these two temperatures most significantly captures the appearance of glassy behaviour in the system. 

It is remarkable and interesting to have uncovered such glassy phenomenology and its understanding in terms of a topological crossover of the discrete configuration network connectivity, in a system as well-known and widespread as the Rubik's cube. 
Slice-rotations are closely reminiscent of loop updates characteristic of dimer, colouring, vertex and other constrained models~\cite{ritortGlassyDynamicsKinetically2003,franzRelaxationProcessesEntropic1997, williamsParamagneticGlassTransitions2012,leeRubikCubeProblem2008}; it will be interesting to see in the future whether a similar investigation and understanding may be developed for such models, as well as models with more local dynamics. 
%
%

\begin{figure}[t]
    \centering
    \includegraphics[width=0.5\textwidth]{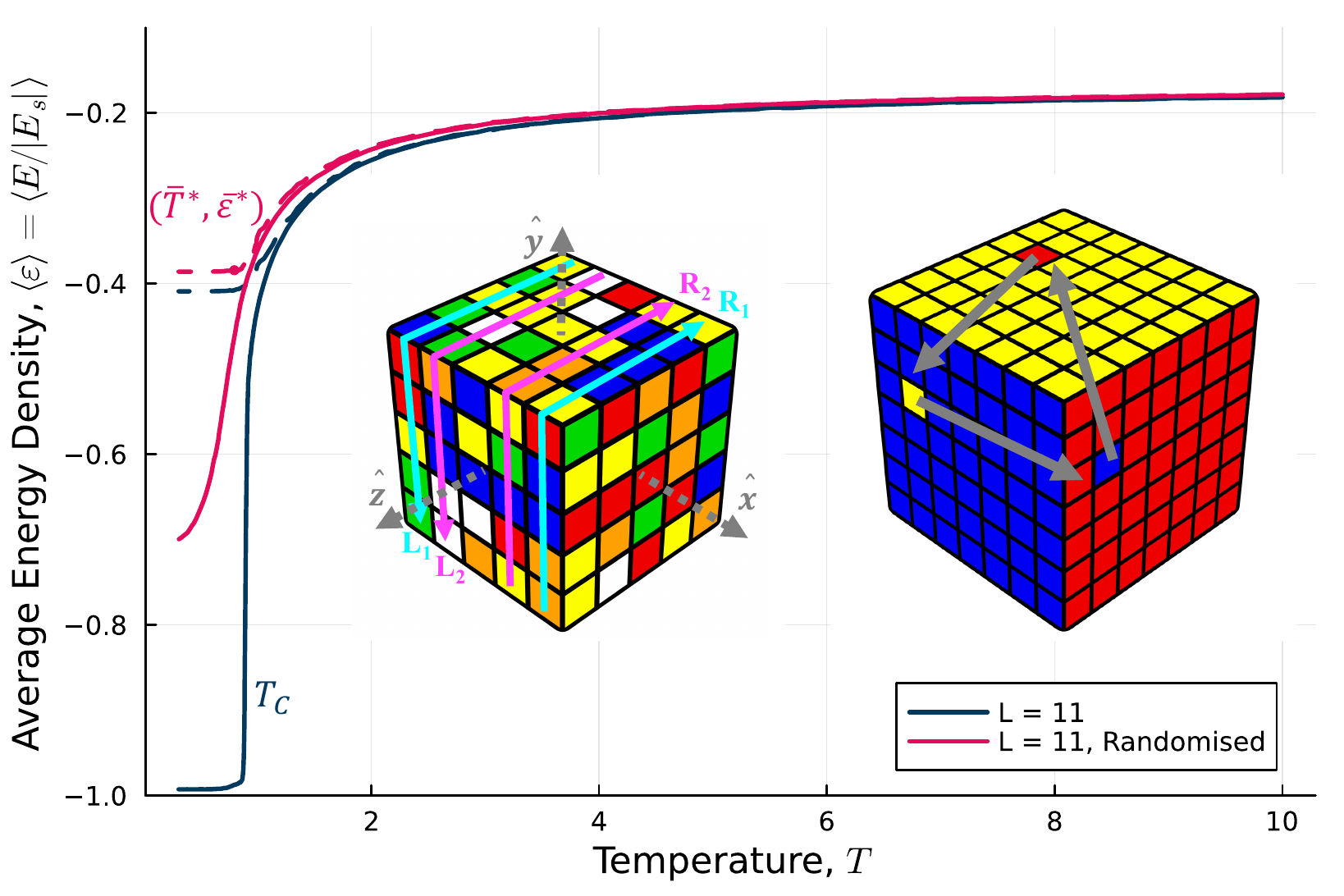}
    \caption{Energy vs temperature curves obtained from thermal anneals of an $L=11$ RC, using slice-rotations (dashed lines) and swap-moves (solid lines), for both the original cube (blue) and the randomised cube (red), as defined in the main text. The data are averaged over 50 histories and, for the randomised cube, each history corresponds to a different realisation of disorder. Left inset: Illustration of slice-rotations around the $x$-axis on a $5 \times 5 \times 5$ Rubik's Cube. Right inset: Example of a swap-move (see Supp.~Mat.~Note~\ref{sm-sec:swap-moves} for a complete list). 
    }
    \label{fig:relaxed-anneals}
\end{figure}

\section{\label{sec:model}Model}
In this work, we frame the Rubik's Cube (RC) as a $6$-state Potts model (i.e., with $N=6L^2$ Potts spins $\sigma_i \in \{1, \dots, 6 \}$ over the six faces of an $L \times L \times L$ cube). We label the system configurations as $\mathbf \sigma = \{ \sigma_i \}\vert_{i=1,\dots,6L^2}$, and generate the configuration space by applying kinetically constrained dynamics corresponding to the well-known RC moves (which we refer to as `slice-rotations') to the solved cube. These dynamics impose several non-trivial but well-defined constraints on the allowed Potts configurations, collectively referred to as the so-called `First Law of Cubology'~\cite{gower-constraints-of-lxlxl-rubiks-cube} (see Supp.~Mat.~Note~\ref{sm-sec:swap-moves}).
%
Some of these constraints are straightforward (for instance having a conserved equal number of spins of each Potts value), others less so (for instance Potts values are conserved within groups of 24 Potts spins). Note that permuting Potts spins with identical values is equivalent to the identity operation, and so in this sense the configuration space corresponds to the set of visually distinct configurations of the cube that are reachable from the solved configuration via a sequence of slice-rotations.

We introduce an energy of the standard nearest-neighbour Potts ferromagnetic form (with $J=1$ as the reference energy scale), neglecting for convenience the interactions between spins on different faces of the cube: 
\begin{equation}\label{eqn:potts-energy}
    E(\mathbf \sigma) = - \sum_{f=1}^6 \sum_{\langle i, j \rangle \in f} \delta_{\sigma_i \sigma_j}
    \, , 
\end{equation}
where $f$ labels the six faces, $i,j \in \{1,\dots,N \}$, and $\delta_{\sigma_i \sigma_j}$ is a Kronecker delta between spin values. The solved configuration of the RC has energy $E_{s} = -12L(L-1)$. Neglecting the constraints imposed by the slice-rotation moves, the (infinite-temperature) energy density $\epsilon = E/\vert E_s \vert$ of a randomly sampled cube configuration is $\epsilon_\infty = -1/6$; including the constraints does not seem to alter this value and indeed we find $\langle \epsilon_{\infty} \rangle \simeq -0.1666665(3)$, numerically averaged over $10^9$ MC steps for the $L=11$ cube. We choose to study odd-$L$ cubes and to fix the value of the central spin on each face (which merely precludes trivial rotations of the whole system). Accounting for both clockwise and anticlockwise $90$° slice-rotations, there are $6(L-1)$ configurations reachable from any given one. This defines the connectivity of configuration space and thereby the dynamics of the RC under slice-rotations. 

Attempts to solve the RC (i.e., to reach its lowest-energy configuration) by thermal annealing using a Markovian Metropolis Monte Carlo algorithm result in a surprising and remarkably strong dynamical arrest, reminiscent of glassy behaviour (Fig.~\ref{fig:relaxed-anneals}). Even exploiting parallel tempering and in systems as small as $L=3$, containing just $N=54$ spins, we observe the appearance of a plateau at finite energy density 
(see Supp.~Mat.~Notes~\ref{sm-sec:parallel-tempering},\ref{sm-sec:system-size-dependence}). For the $L=11$ cube, this plateau occurs at an energy density $\epsilon^* \simeq {\EstarClean}$, which is notably far from the ground-state energy ($\epsilon_0 = -1$). 
This plateau sets rapidly at temperatures lower than $T^* \simeq {\TstarClean}$,
and it appears to be robust to increasing the annealing time. A study of the system-size dependence of $\epsilon^*$ and $T^*$ can be found in Supp.~Mat.~Note~\ref{sm-sec:system-size-dependence}.

Taking advantage of the properties of the group generated by combinations of slice-rotations, one can show that: (i) the group contains swap-moves that involve as few as $3$ spins (see, e.g., the right inset in Fig.~\ref{fig:relaxed-anneals}); and (ii) the group has an alternative presentation where combinations of an appropriate set of such swap-moves generate the entire group (see Supp.~Mat.~Note~\ref{sm-sec:swap-moves}; further explicit examples of sequences of slice-rotations equivalent to swap-moves can be found in Ref.~\onlinecite{gower-constraints-of-lxlxl-rubiks-cube}). Therefore, we can implement a modified Monte Carlo algorithm in terms of these swap-moves which does not alter the configuration space nor the energy of the system.
However, the change in connectivity greatly enhances the ability of the system to reach thermal equilibrium throughout the entire temperature range of interest. This unveils a sharp first-order transition at a finite critical temperature $T_c \simeq {\Tcrit} < T^*$, across which the system quickly reaches the solved configuration (see Fig.~\ref{fig:relaxed-anneals}). 
The focus of this work is to investigate the origin of this stark change in dynamical properties of the system, and we relegate a brief characterisation of the thermodynamic properties and phase transition to Supp.~Mat.~Note~\ref{sm-sec:phase-transition}. 

For our purpose, it is convenient to thermodynamically suppress the transition in order to access a broader range of temperatures below $T^*$ where we can equilibrate the system using swap-moves without ordering, and then use slice-rotations to characterise the nature and origin of the slow dynamics. To do this, we start from a configuration with randomly re-arranged spin values (equivalent to a random re-stickering of the cube) and then generate the ensemble of allowed configurations by applying slice-rotations to it. Note that this ensemble of configurations must then not include the solved configuration of the original cube in order to suppress the transition. We refer to this alternative system as the `randomised RC'. The final results are then averaged over the initial choice of disorder. As illustrated in Fig.~\ref{fig:relaxed-anneals} (and Supp.~Mat.~Note~\ref{sm-sec:phase-transition}), the outcome of the randomisation is to suppress the phase transition (i.e., the singularity in thermodynamic quantities for the original RC disappears), while maintaining the starkly different equilibration properties between swap-moves and slice-rotations: the former easily reach thermodynamic equilibrium (down to some potentially degenerate ground state with $\overline E_0 > E_s$), whereas the latter exhibit a sudden dynamical arrest at an energy density $\overline \epsilon^* \simeq {\EstarDis}$ which is finitely larger than the ground state energy density $\overline \epsilon_0 \simeq {\EZeroDis}$, for temperatures below $\overline T^* \simeq {\TstarDis}$ (here the overline indicates the disorder average). Hereafter, we focus on the randomised RC. 

We note that the values of $T^*$ and $\epsilon^*$ are not strictly speaking sharply defined. At the very least in finite-size systems one generally expects that, if it were possible to access much slower thermal anneals using slice-rotations, the behaviour of the system ought to eventually converge to the equilibrium one. However, this is not seen in accessible simulations, even for relatively small systems, and it appears to be possible to assign reasonably well-defined values to $T^*$ and $\epsilon^*$. We shall subsequently see how one can relate these quantities to crisply defined features in the configuration connectivity of the system, when studied as a function of energy, and how the latter allow us to define precise threshold values in the thermodynamic limit. 
%
%

\section{Slow dynamics}\label{sec:slow-dynamics}
\begin{figure}[t]
    \centering
     \includegraphics[width=0.5\textwidth]{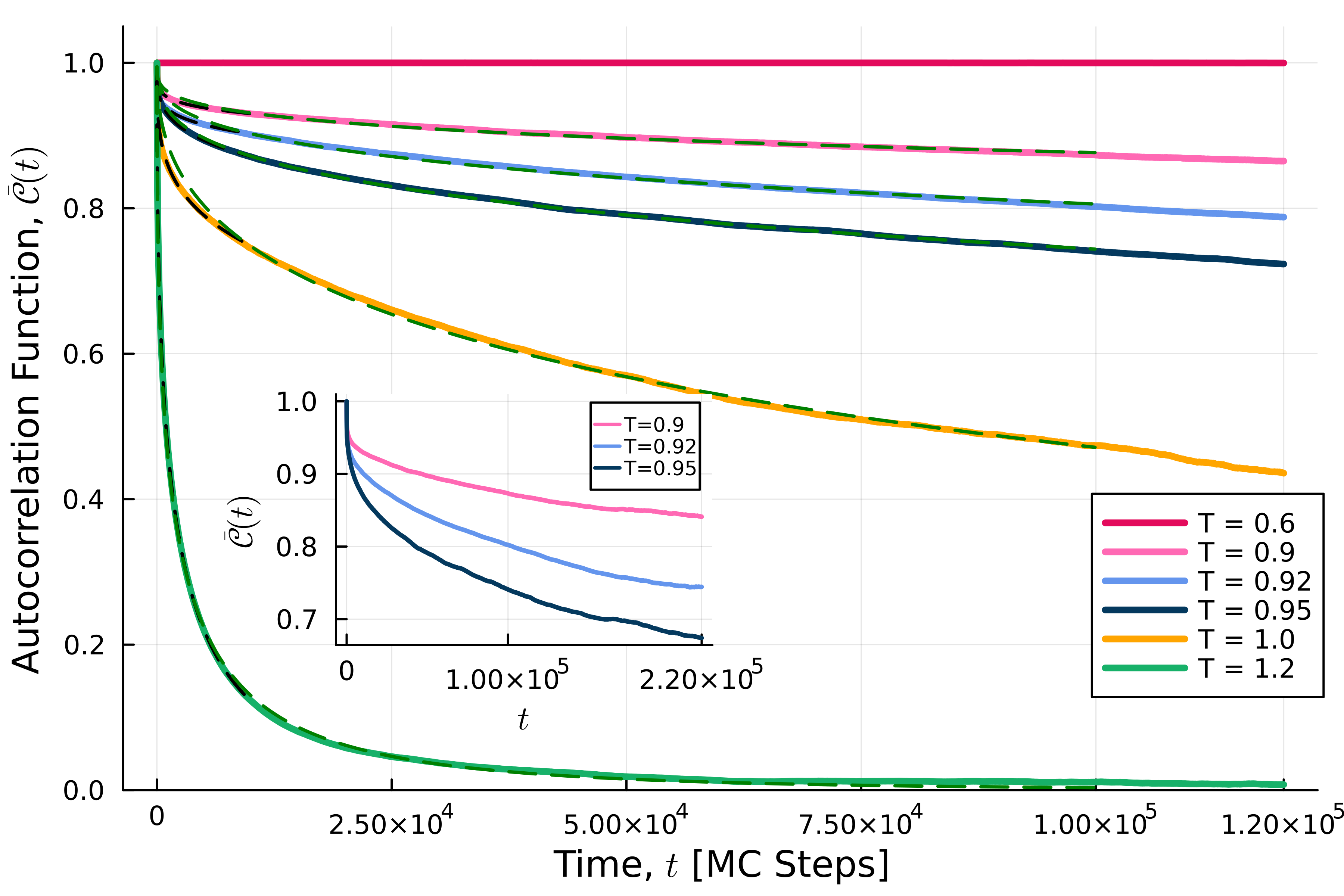}
    \caption{Rescaled autocorrelation function $\overline{\mathcal{C}}(t)$ for the randomised $L=11$ RC (averaged over 50 instances of disorder) at different temperatures, where we first equilibrate the cube using swap-moves and then evolve it for time $t$ using slice-rotations. Dashed lines illustrate examples of stretched exponential fits used to extract the exponent $\beta$ and relaxation time $\tau$ shown in Fig.~\ref{fig:relaxation-time-stretching-exponent} (black corresponds to fits up to $t=10^4$, green corresponds to fits up to $t=10^5$). Inset: The autocorrelation function for temperatures close to $\overline T^*$ (averaged over 100 instances of disorder) for longer evolution times shows no clear evidence of distinct $\alpha$ and $\beta$ relaxation~\cite{goldsteinViscousLiquidsGlass1969}.}
    \label{fig:autocorrelation-function}
\end{figure}

Let us first characterise the slow dynamics by studying the behaviour of the autocorrelation function: 
\begin{equation}
C(t) = \frac{1}{N-6} \sideset{}{'}\sum\limits_{i=1}^{N-6} \delta_{\sigma_i(0) \sigma_i(t)}
\, ,
\end{equation}
where the primed summation excludes the six central spins, whose values are fixed by construction. 
Note that $C(\infty)$ does not vanish. Indeed, for the original cube one finds $\langle C(\infty) \rangle \simeq \frac{1}{6}$, as expected for randomly sampled Potts 6-spins with equal probability. The analogous result for the randomised cube in the large $L$ limit is $\overline{\langle C(\infty) \rangle} \sim 0.201$, as demonstrated in Supp.~Mat.~Note~\ref{sm-sec:autocorrelation-function-limits}. 
For convenience, we therefore plot $\overline{\mathcal{C}}(t) := [\overline C(t) - \overline{\langle C(\infty)\rangle}]/[1 - \overline{\langle C(\infty)\rangle}]$ in Fig.~\ref{fig:autocorrelation-function}, which decays from $1$ to $0$ as a function of time. 
We equilibrate the randomised RC at a given temperature using swap-moves and then evolve it for time $t$ using slice-rotations (at the same temperature). 

\begin{figure}[t]
    \centering
    \includegraphics[width=\linewidth]{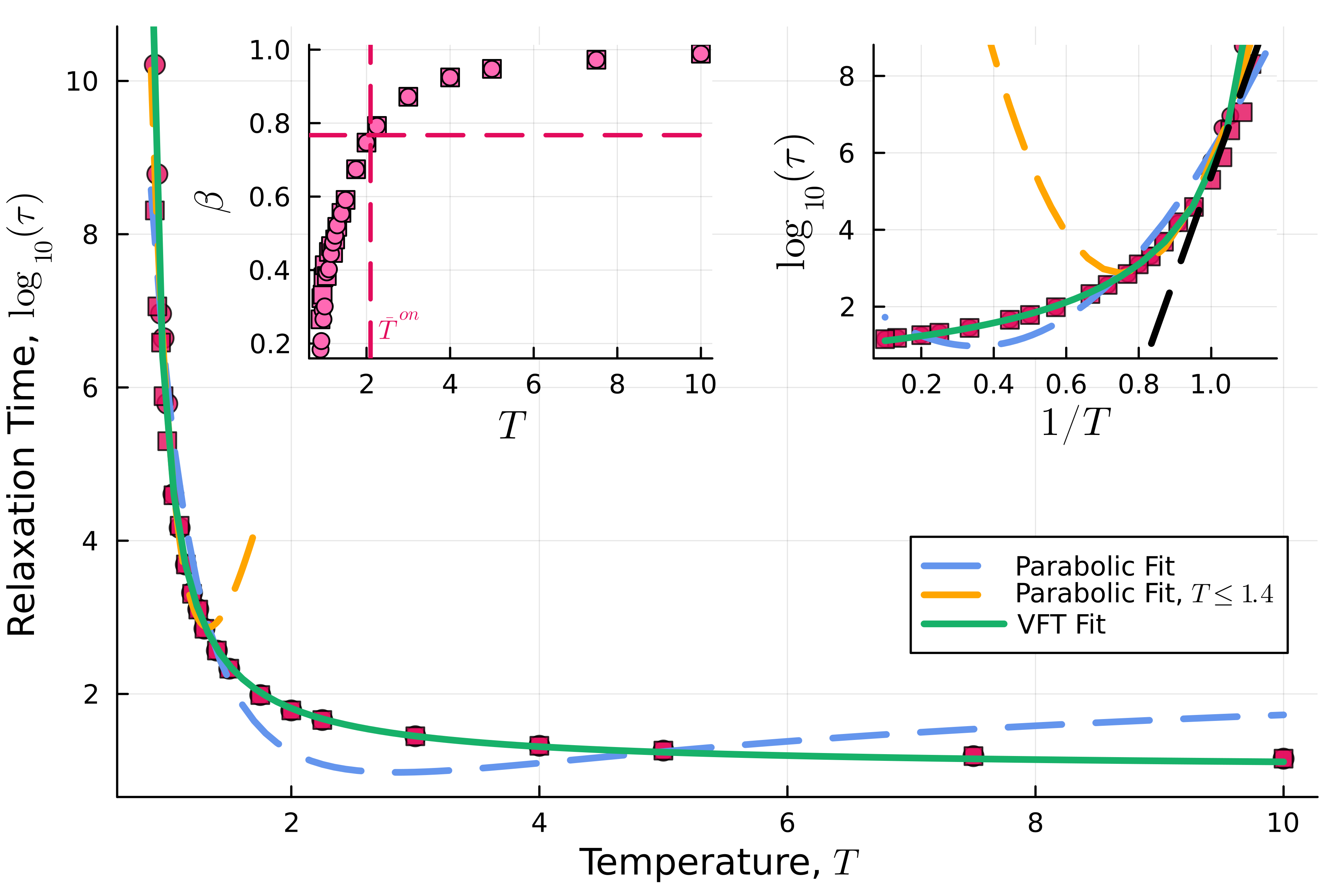}
    \caption{Temperature dependence of the relaxation time $\tau(T)$ obtained from stretched exponential fits up to $t=10^4$ (circles) and $t=10^5$ (squares) of the $L=11$ randomised RC autocorrelation function data illustrated in Fig.~\ref{fig:autocorrelation-function}. Note the logarithmic vertical axis in base $10$. Fits (for the combined $t=10^4$ and $t=10^5$ data) are presented for the VFT (green solid line) and parabolic (blue and orange dashed lines) functional forms, as discussed in the main text. Left inset: Temperature dependence of the stretching exponent $\beta(T)$; we also plot the onset temperature $\overline T^{\rm on}$ (red dashed line), defined as the temperature $\beta(T=\overline T^{\rm on}) = 0.75\times\beta(T=\infty)$ where the autocorrelation function decay exhibits significantly stretched exponential behaviour (Supp.~Mat.~Note~\ref{sm-sec:system-size-dependence}). Right inset: $\tau(T)$ data plotted against $1/T$, to highlight the departure from Arrhenius behaviour; we also plot the activated behaviour expected for typical energy barriers $\overline{\Delta E^*} \simeq 26$ at the topological crossover, $\tau \propto \exp(\overline{\Delta E^*}/T)$ (black dashed line), discussed in Sec.~\ref{sec:discussion}.}
    \label{fig:relaxation-time-stretching-exponent}
\end{figure}

We carefully look for signs of distinct $\alpha$ and $\beta$ relaxation processes~\cite{goldsteinViscousLiquidsGlass1969} by running longer evolution times for $T \sim \overline T^*$, but found no clear evidence of it (see inset in Fig.~\ref{fig:autocorrelation-function}). Note that the energy barriers near the glass transition are much larger than temperature (Supp.~Mat.~Note~\ref{sm-sec:modal-energy-differences}), and it is likely that the $\alpha$ relaxation is pushed beyond our accessible simulation time scales.
Nonetheless, we note that the decay of the autocorrelation function is no longer fit well by a single stretched exponential~\cite{kohlrauschTheorieElektrischenRuckstandes1854} for $T \lesssim 1$ (see also Supp.~Mat.~Note~\ref{sm-sec:autocorrelation-function-fits}). Pragmatically, we compare single stretched exponential fits both up to $t=10^4$ and $t=10^5$. The extracted fitting parameters are displayed as circles and squares, respectively, in Fig.~\ref{fig:relaxation-time-stretching-exponent}. We clearly observe a dramatic increase in relaxation time $\tau$ and a decrease in stretching exponent $\beta$ as temperature is lowered to $\sim \overline T^*$, consistent with the generic behaviour of a fragile glass approaching a glass transition, and we note that this behaviour is relatively robust between the two fitting timescales. 

Due to the rapid slowing down of the dynamics, we were only able to reliably fit the autocorrelation decay for temperatures $T \geq 0.9$. Notwithstanding this limitation, our results uncover an impressive increase in relaxation time $\tau(T)$ by more than 8 orders of magnitude. 
The temperature dependence of $\tau(T)$ (combining the $t=10^4$ and $t=10^5$ data) is best fit by a function of the Vogel-Fulcher-Tammann~\cite{hTemperaturabhangigkeitsgesetzViskositatFlussigkeiten1921,tammannAbhangigkeitViscositatTemperatur1926,fulcherAnalysisRecentMeasurements1925} (VFT) form $A \, e^{\Delta/(T-T_0)}$, with $A \simeq {\VFTA}$, $\Delta \simeq {\VFTDelta}$, and $T_0 \simeq {\VFTT}$. The curvature of the data in the inset of Fig.~\ref{fig:relaxation-time-stretching-exponent} clearly precludes fits of the Arrhenius form~\cite{arrheniusUeberDissociationswaermeUnd1889,arrheniusUeberReaktionsgeschwindigkeitBei1889} $A \, e^{\Delta/T}$, and a satisfactory fit of the parabolic form~\cite{elmatadCorrespondingStatesStructural2009} $A \, e^{\Delta/T + \Delta'/T^2}$ could not be found for the full range of temperatures. For comparison, we also repeat the parabolic fit after including only $T \leq 1.4$; the agreement at low temperatures is improved, at the expense of a more pronounced departure at higher temperatures. All fitting parameters (including for the $t=10^4$ and $t=10^5$ data separately) are available in Supp.~Mat.~Note~\ref{sm-sec:tau-fit-parameters}.
%
%

\section{Saddles-to-minima topological crossover}
Using swap-moves to equilibrate the randomised RC at a given temperature (including $T<\overline T^*$), we are able to stochastically sample configurations of different energies. For each configuration, we then use swap-moves or slice-rotations to gain insight into the respective connectivities of configuration space, resolved along the energy axis. 

We define the saddle index, $K$, of a configuration as its number of energy-decreasing connections, ranging between $0$ and the maximal degree of the configuration network, $z$: $K=0$ corresponds to a minimum; $K=1$ corresponds to a configuration with only one move that decreases its energy; and $K\geq2$ corresponds to a generic saddle. This is the discrete analogue of the saddle index introduced for systems with continuous degrees of freedom in Refs.~\onlinecite{broderixEnergyLandscapeLennardJones2000,angelaniSaddlesEnergyLandscape2000,cavagnaFragileVsStrong2001,grigeraGeometricApproachDynamic2002,coslovichLocalizationTransitionUnderlies2019}. We note that $K\geq 2$ and not $K=1$ configurations are identified as generic saddles, since only the former can facilitate activated processes between local minima and thereby promote dynamical exploration of the energy landscape. (Further discussion about $K=1$ configurations can be found in Supp.~Mat.~Note~\ref{sm-sec:K=1_configurations}.) 

\begin{figure}[t]
    \centering
    \includegraphics[width=0.5\textwidth]{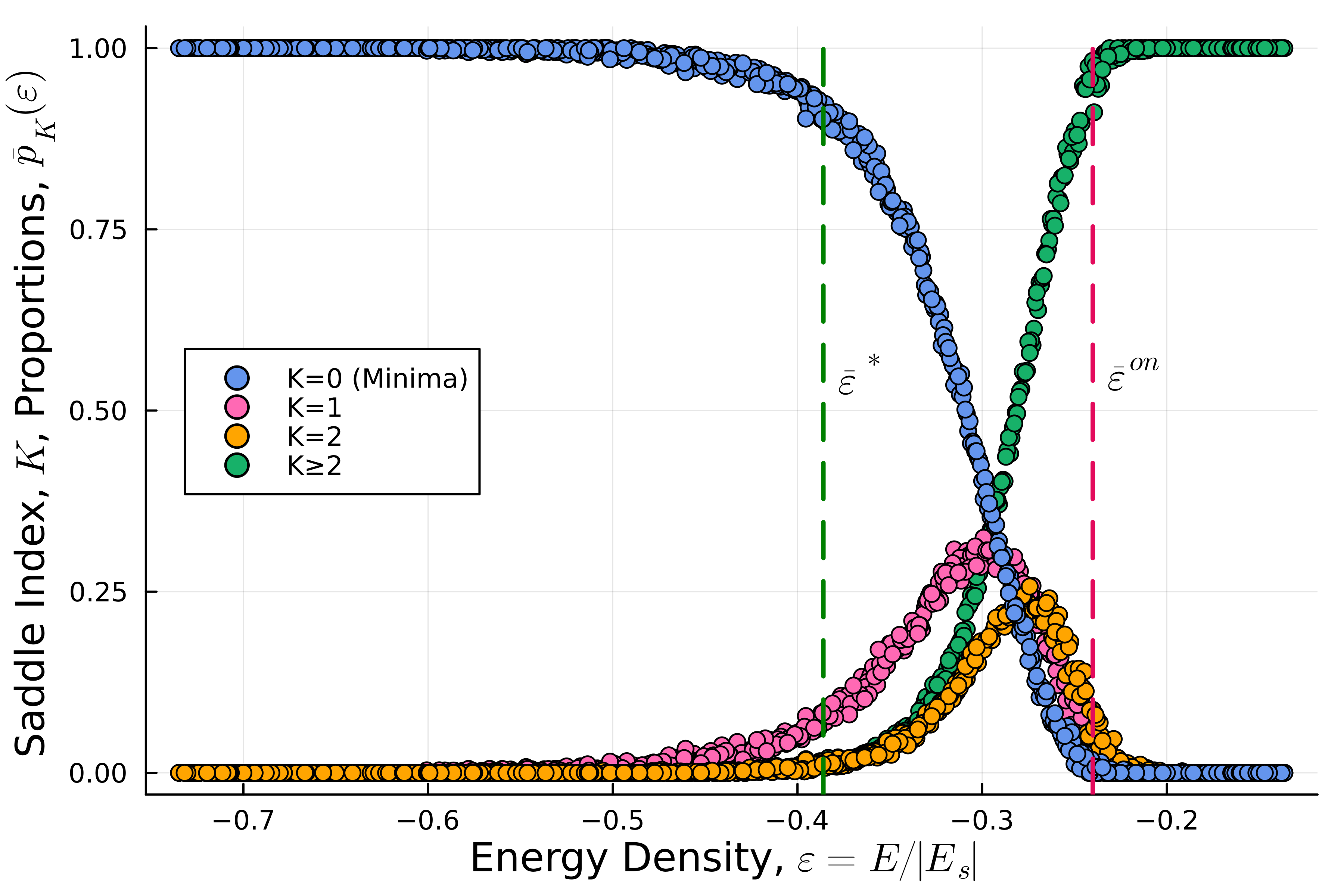}
    \caption{Saddles-to-minima topological crossover in the $L=11$ slice-rotation cube, demonstrated by the behaviour of the proportion of stochastically sampled configurations with saddle index $K = 0, 1, 2$ and $K\geq 2$, as a function of energy density $\epsilon = E/\vert E_s \vert$.}
    \label{fig:saddle-index-proportions}
\end{figure}

The central result of our work is illustrated in Fig.~\ref{fig:saddle-index-proportions} (all energy and configuration connectivity data were extracted from the randomised RC, averaged over 80 instances of disorder). 
When we use slice-rotations, the proportion of saddles is strongly suppressed in favour of minima as energy is decreased. 
This suppression appears to occur rapidly, and the energy 
below which we fail to observe, within our statistics, any $K \geq 2$ configurations correlates well with the value $\overline \epsilon^*$ that we found in the dynamically arrested regime. Correspondingly, the equilibrium temperature when this energy threshold is reached correlates well with $\overline T^*$. Similarly, we find that the energy 
below which we first observe a non-negligible proportion of $K=0$ minima configurations aligns well with the energy density $\overline\epsilon^{\rm on}$ (and corresponding equilibrium temperature $\overline T^{\rm on}$) below which the dynamics exhibit significantly stretched exponential behaviour. (Definitions and comparisons between these energy scales for different system sizes can be found in Supp.~Mat.~Note~\ref{sm-sec:system-size-dependence}.) 

\begin{figure}[t]
    \centering
    \includegraphics[width=0.5\textwidth]{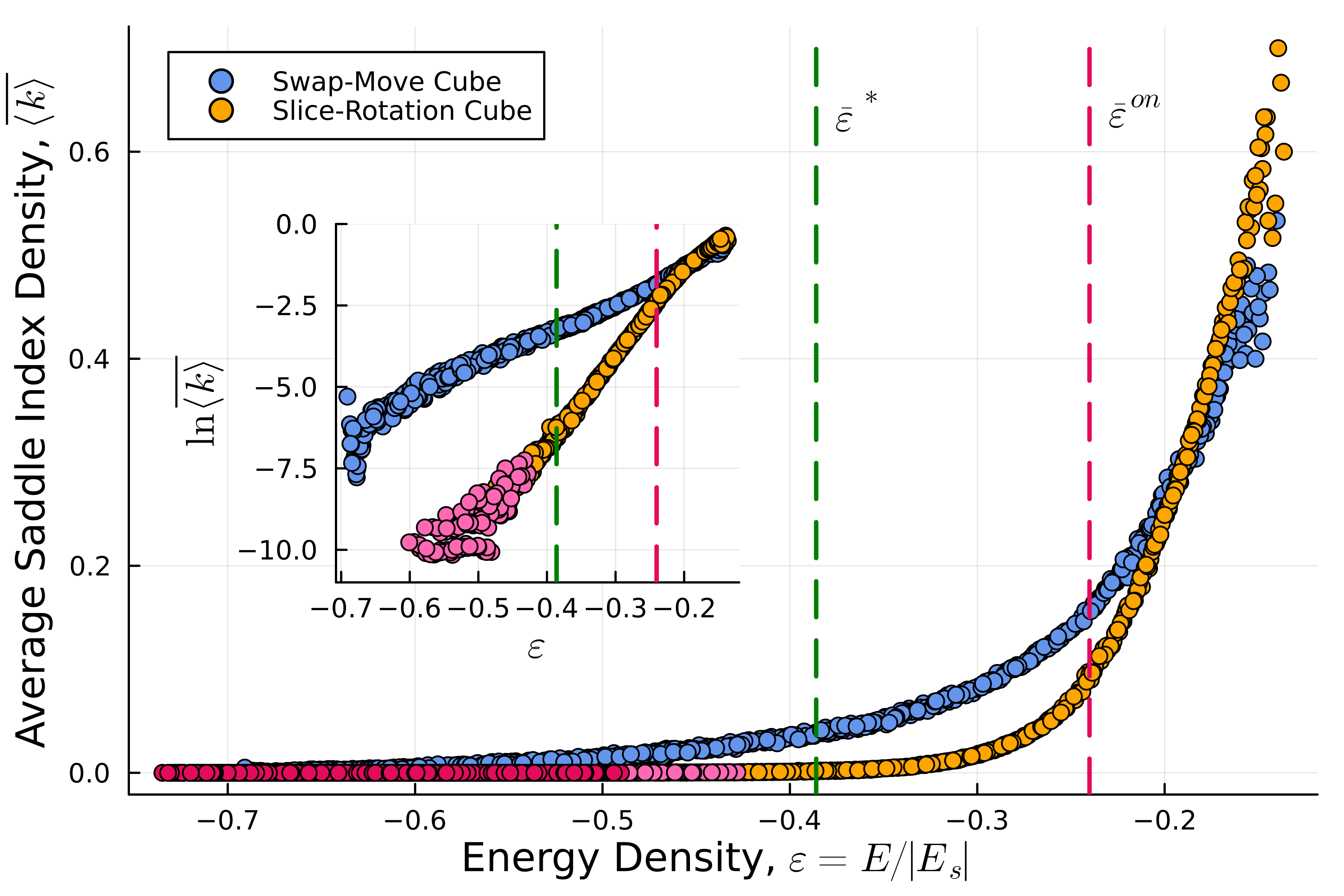}
    \caption{Average saddle index density $\overline{\langle k \rangle} = \overline{\langle K/z \rangle}$ against energy density $\epsilon$ for the $L=11$ slice-rotation cube (red, pink, and orange) and swap-move cube (blue). For visualisation purposes, we colour-coded the slice-rotation cube data as: 
    red, when we do not observe any values of $K>0$ within our sampling statistics; 
    pink, when we do not observe any values of $K\geq2$ within our sampling statistics; 
    and orange, when at least an instance of $K \geq 2$ was observed. 
    Inset: Same data on a log-linear scale to demonstrate that the topological crossover appears to be exponential in both models.}
    \label{fig:saddle-index-density-against-energy}
\end{figure}

This is in stark contrast with the behaviour we observe when we use swap-moves. In the latter, the large connectivity of the configuration network prevents us from being able to compute the swap-move saddle indices $K$ exactly (Supp.~Mat.~Note~\ref{sm-sec:swap-move-saddle-index-proportions}), and we are only able to sample $K$ statistically. To compare the two models, we consider the average saddle index density $\overline{\langle k \rangle} = \overline{\langle K/z \rangle}$, where $\langle\, \cdot\, \rangle$ denotes an average over sampled configurations at a given energy and $z$ is the maximal degree of the configuration network ($z=6(L-1)=60$ for slice-rotations with $L=11$, and is significantly larger for swap-moves, see Supp.~Mat.~Note~\ref{sm-sec:growth-of-swap-moves-with-L}). 
This is illustrated in Fig.~\ref{fig:saddle-index-density-against-energy}. 
We find that $\overline{\langle k \rangle}$ remains appreciable at all energy densities for swap-moves, contrary to the strong suppression we observe when $\epsilon \lesssim \overline\epsilon^*$ for slice-rotations. This is consistent with the absence of a saddles-to-minima topological crossover for swap-moves. In our simulations, both behaviours appear to be smooth and exponential in energy (see inset in Fig.~\ref{fig:saddle-index-density-against-energy}), akin to Ref.~\onlinecite{fabriciusDistanceInherentStructures2002}, and in contrast to the sharp crossovers argued in Refs.~\onlinecite{broderixEnergyLandscapeLennardJones2000,angelaniSaddlesEnergyLandscape2000,grigeraGeometricApproachDynamic2002}. 

\begin{figure}
    \centering
    \includegraphics[width=0.5\textwidth]{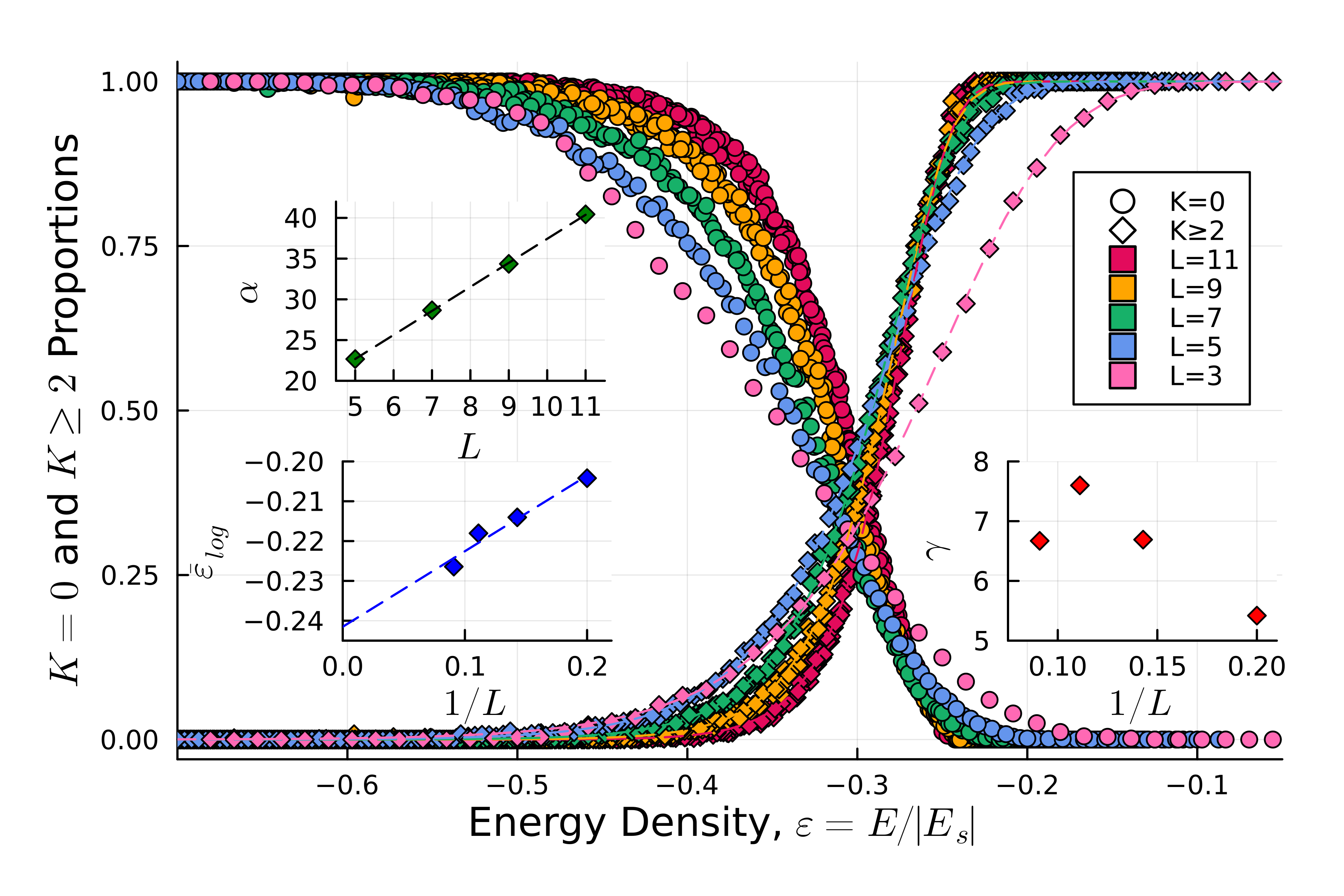}
    \caption{Proportion of stochastically sampled configurations with saddle index $K=0$ and $K \geq 2$,  as a function of energy density, for varying system size $L$. The $K\geq 2$ curves are well fit by generalised logistic functions of the form $1 - (1 + e^{\alpha(\epsilon - \overline\epsilon_{\rm log})})^{-\gamma}$ (shown as dashed lines). The fitting parameters are displayed in the insets; a linear fit for $\overline\epsilon_{\rm log}$ as a function of $1/L$ is presented as a guide to the eye. $\alpha$ scales linearly with $L$ as $\alpha \simeq 7.93 + 2.95L$ and these functions converge to Heaviside step functions as $L \rightarrow \infty$, denoting a change from crossover to threshold behaviour in the thermodynamic limit.}
    \label{fig:combined-saddle-index-proportions-finite-size-scaling}
\end{figure}

In Fig.~\ref{fig:combined-saddle-index-proportions-finite-size-scaling}, we plot the proportion of $K=0$ (minima) and $K \geq 2$ (saddles) when using slice-rotations for varying system size $L$. The $K\geq 2$ curves are well fit by generalised logistic functions of the form $1- (1 + e^{\alpha(\epsilon - \overline\epsilon_{\rm log})})^{-\gamma}$, with $\alpha$ linear in $L$, 
and $\overline\epsilon_{\rm log}, \, \gamma$ not scaling significantly with $L$ (see insets). These functions converge to Heaviside step functions as $L \rightarrow \infty$, with the step occurring at $\epsilon = \lim_{L\to\infty} \overline\epsilon_{\rm log}$, suggesting that the smooth exponential crossover behaviour is replaced by a sharp transition in the thermodynamic limit. Here, $\overline\epsilon^*$ and $\overline\epsilon^{\rm on}$ converge to a common value (Supp.~Mat.~Note~\ref{sm-sec:system-size-dependence}) which represents the energy threshold below which configurations with $K\geq 2$ form a set of measure zero (and correspondingly $K=0$ minima form a set of measure one). 
%

\begin{figure}[!ht]
    \centering
    \includegraphics[width=0.5\textwidth]{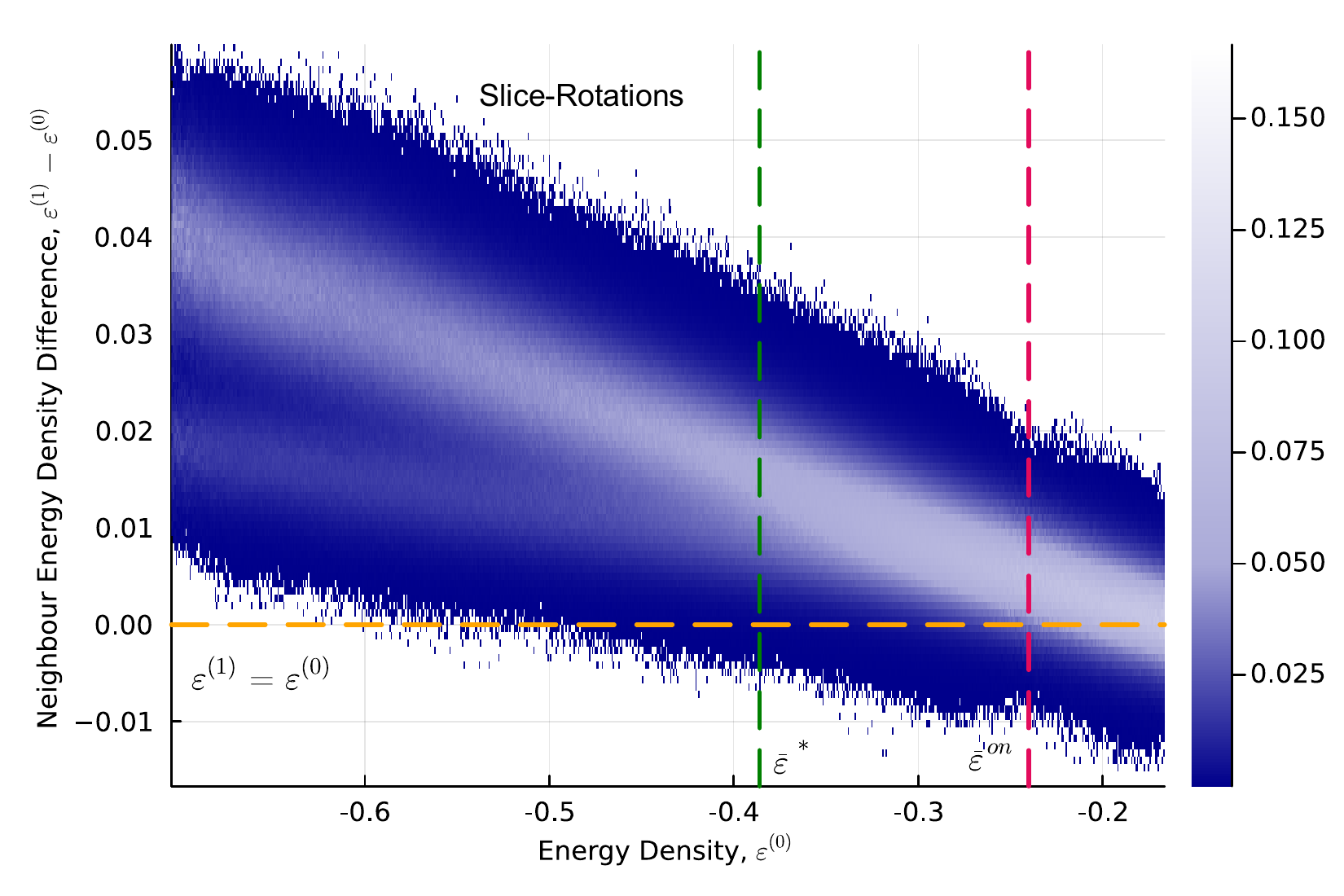}
    \includegraphics[width=0.5\textwidth]{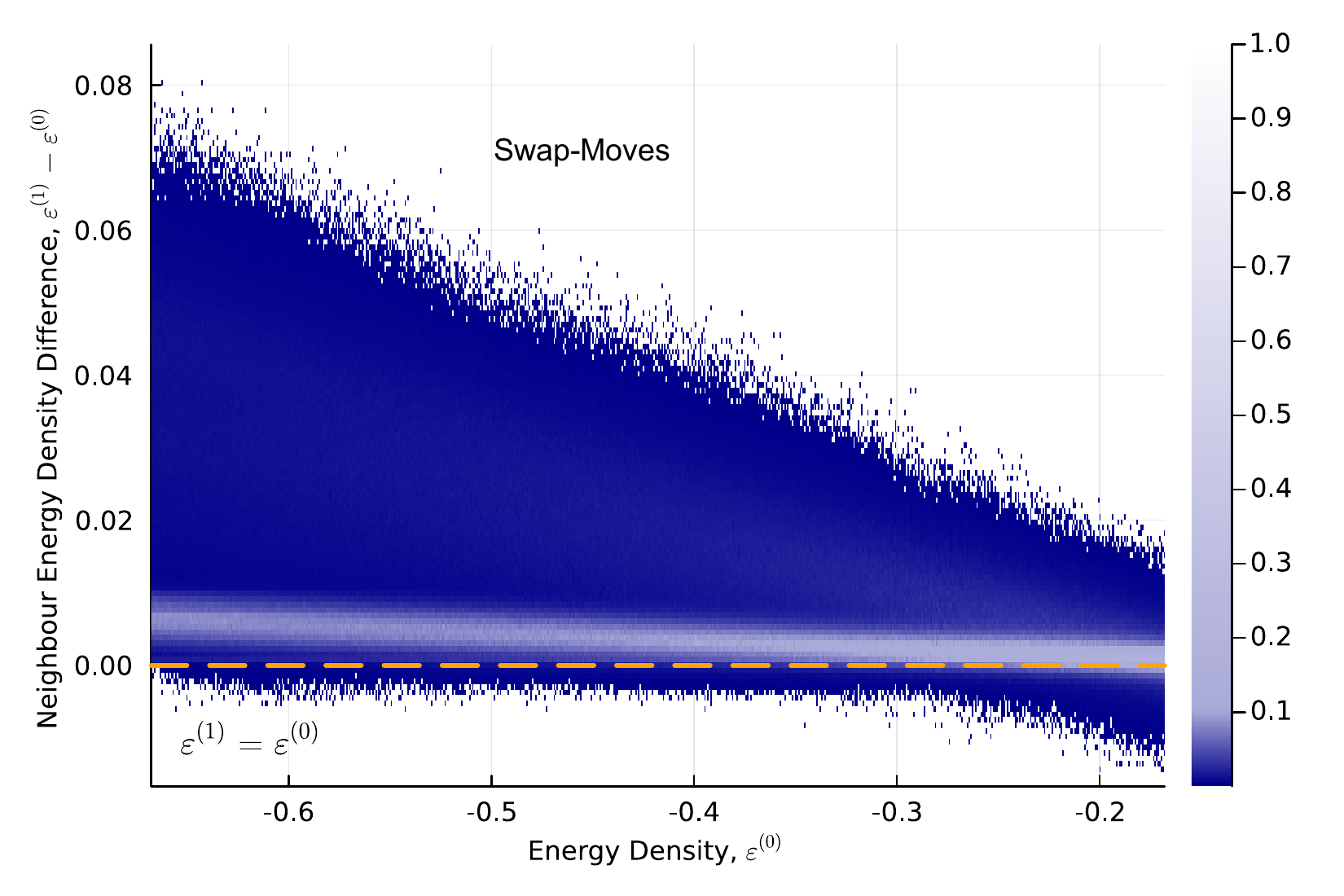}
    \caption{Histograms of the difference $\epsilon^{(1)} - \epsilon^{(0)}$ between energy densities, where 
    $\epsilon^{(0)}$ corresponds to stochastically sampled $L=11$ RC configurations in equilibrium (using swap-move MC) and $\epsilon^{(1)}$ corresponds to the energy density obtained after applying a single slice-rotation (top panel) or swap-move (bottom panel). 
    The histograms are normalised independently for each value of $\epsilon^{(0)}$. The slice-rotation cube histograms clearly depart above the $\epsilon^{(1)} = \epsilon^{(0)}$ line for energy densities $\epsilon^{(0)} \lesssim \overline\epsilon^*$ (green dashed vertical line), indicating a suppression of energy-decreasing connections. For comparison, we also indicate the value of $\overline\epsilon^{\rm on}$ (red dashed vertical line). The bimodal structure of the slice-rotation histograms for low energy densities corresponds to the typical energy costs of internal compared to external layer slice-rotations (see, e.g., $R_1$ vs.~$R_2$ in Fig.~\ref{fig:relaxed-anneals}, Supp.~Mat.~Note~\ref{sm-sec:predicted-energy-density-costs} and Ref.~\onlinecite{gower-constraints-of-lxlxl-rubiks-cube}) for the RC sufficiently close to its lowest energy state. 
    }
    \label{fig:E0-E1-connectivities}
\end{figure}

In Fig.~\ref{fig:E0-E1-connectivities} we take a closer look at the energy-resolved connectivity of configuration space, and we investigate the magnitude of the associated energy changes. We stochastically sample configurations of energy density $\epsilon^{(0)}$ in equilibrium using swap-moves, and then build histograms of the energy densities $\epsilon^{(1)}$ of the configurations that are connected to them by single slice-rotations (top panel) or swap-moves (bottom panel). For the slice-rotation case we systematically sample $\epsilon^{(1)}$ of all neighbours ($z=60$ for the $L=11$ RC) for each $\epsilon^{(0)}$ configuration. For the swap-move case, since $z$ is prohibitively large, we sample a random subset of 1000 configurations of the neighbouring $\epsilon^{(1)}$ values. 

We observe that, for slice-rotations, the histograms depart above the $\epsilon^{(1)}=\epsilon^{(0)}$ line for energy densities $\epsilon^{(0)} \lesssim \overline\epsilon^*$ (consistent with the extreme sparsity of energy-decreasing connections observed in simulations in this range of $\epsilon^{(0)}$ values, see Supp.~Mat.~Note~\ref{sm-sec:boltzmann-factor-reweighting}). We find analogous behaviour for energy densities $\epsilon^{(2)}$ of configurations connected to equilibrium $\epsilon^{(0)}$ configurations by two slice-rotations (Supp.~Mat.~Note~\ref{sm-sec:E2-E0-connectivities}). However, the departure is suppressed if we study the energy density differences $\epsilon^{(2)} - \epsilon^{(1)}$ (Supp.~Mat.~Note~\ref{sm-sec:E2-E1-connectivities}), indicating that sampled configurations with energy density $\epsilon^{(1)}$ are atypical compared to the thermodynamic equilibrium ensemble. For the swap-moves, the histograms show a finite probability of $\epsilon^{(1)} < \epsilon^{(0)}$ for all $\epsilon^{(0)}$ values. 

Notice that the maximum slice-rotation energy density cost $\Delta \epsilon_\text{max}$ is finite ($\sim 0.07$ for the $L=11$ cube) and small relative to the depth of the below-crossover region of the energy landscape $\overline\epsilon^* - \overline\epsilon_0$ ($\sim 0.61$ for the $L=11$ cube). Therefore, configurations sufficiently below the crossover connect to configurations above the crossover via a (long) sequence of (exponentially rare) below-crossover saddle configurations. That is, the saddles-to-minima topological crossover does not mean that below-crossover minima are typically connected to above-crossover saddles. 
%
%

\section{\label{sec:discussion}Discussion}
\begin{figure}[ht]
    \centering
    \includegraphics[width=0.5\textwidth]{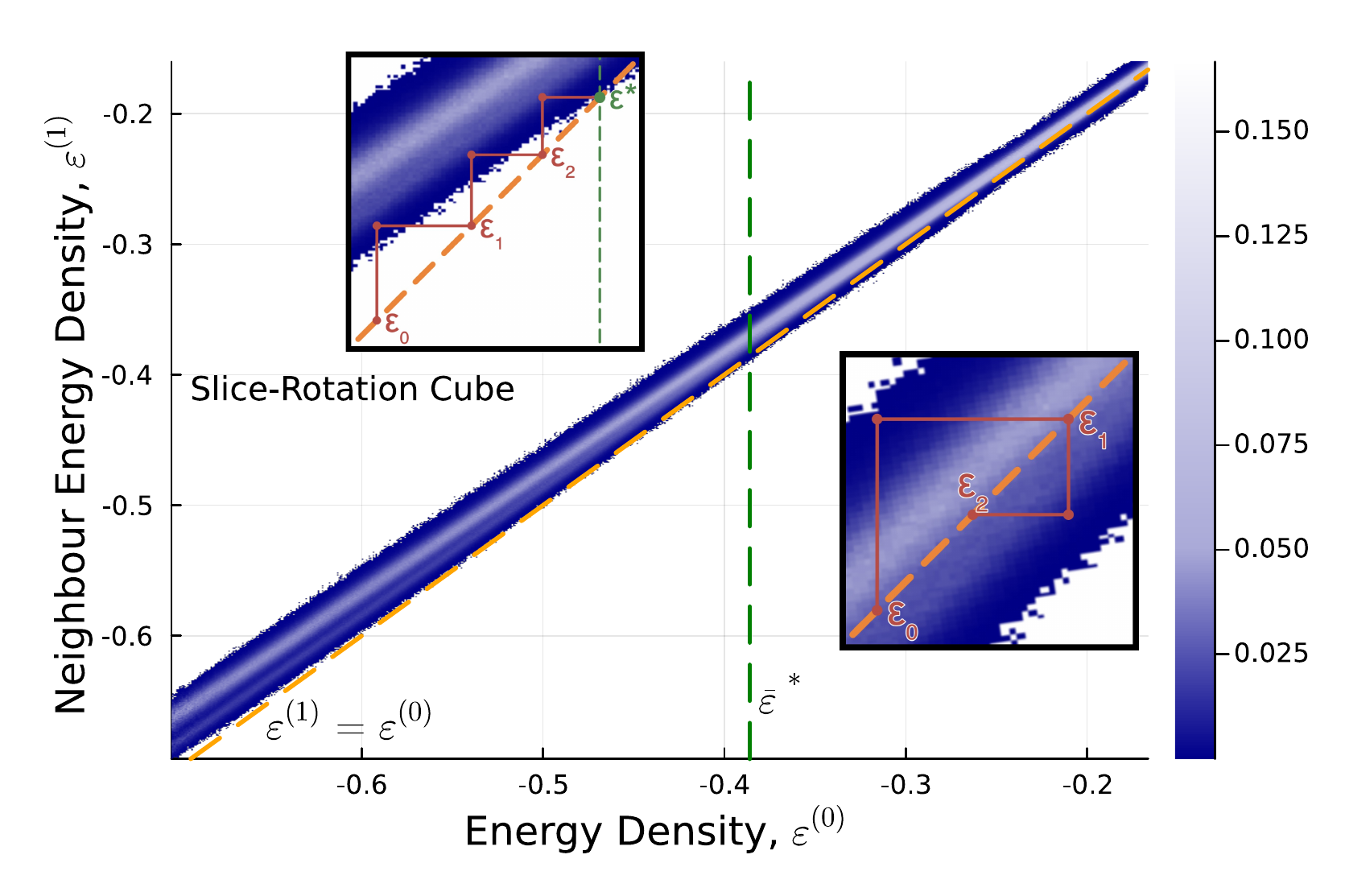}
    \caption{Histograms of the same energy densities ($\epsilon^{(0)}$ and $\epsilon^{(1)}$) as in Fig.~\ref{fig:E0-E1-connectivities} (top panel). Top-left inset: Cartoon depicting the dynamical attraction of the system at low energies towards $\epsilon^*$, for slice-rotation connectivity. Bottom-right inset: Same $\epsilon^{(1)}$ vs. $\epsilon^{(0)}$ histograms window as in the top-left inset, but for swap-move connectivity, showing how a similar formulation of an attractor does not work in this case (other than towards the ground state of the system).}
    \label{fig:spiral-attractors}
\end{figure}

Looking at the behaviour observed in Fig.~\ref{fig:E0-E1-connectivities} (top panel), it is tempting to propose the following effective argument for the slowing down of the dynamics at low temperature, and for the significance of $T^*$ and $\epsilon^*$. One can imagine the system with slice-rotation connectivity starting in equilibrium at some energy density $\epsilon^{(0)} < \epsilon^*$. From this, it will reach a mean energy density $\epsilon^{(1)} > \epsilon^{(0)}$ since energy-decreasing connections are statistically suppressed. Initially, the configurations that the system explores at this new energy are atypical (see Supp.~Mat.~Note~\ref{sm-sec:E2-E1-connectivities}). However, the system is able to move between configurations close to this new energy -- exploiting the surplus of energy-decreasing connections in these atypical configurations (see Fig.~\ref{fig:E1-E2-connectivities} in Supp.~Mat.~Note~\ref{sm-sec:E2-E1-connectivities}) -- and one may assume that it eventually thermalises at $\epsilon^{(1)}$. The process then starts over upon resetting $\epsilon^{(0)} = \epsilon^{(1)}$, etc. If we take this picture to the letter, as pictorially illustrated in Fig.~\ref{fig:spiral-attractors}, we see that the slice-rotation connectivity causes the process to have an attractor at energy density $\epsilon^*$ -- thus explaining the freezing of the energy at this specific value -- whereas the swap-move connectivity only has an attractor in the ground state of the system. 

More quantitatively, we can follow a similar argument to the one proposed in Ref.~\onlinecite{cavagnaFragileVsStrong2001} and consider the modal energy barrier of $\overline{\Delta E^*} = (\epsilon^{(1)} - \epsilon^{(0)}) |E_s| \simeq 0.02 |E_s| \simeq 26$ at $\epsilon^{(0)}=\overline\epsilon^*$. This places the slice-rotation cube in the $\overline{\Delta E^*} \gg \overline T^* \simeq \TstarDis$ regime associated with fragile glass behaviour~\cite{cavagnaFragileVsStrong2001}. As depicted in Fig.~\ref{fig:E0-E1-connectivities}, the typical energy barriers encountered by the slice-rotation cube increase as the temperature and equilibrium energy decrease (see also Supp.~Mat.~Note~\ref{sm-sec:modal-energy-differences}). However, for $T \gtrsim \overline T^*$ there remains a finite density of routes for the system to lower its energy without having to overcome such barriers. As $T \to \overline T^{*+}$, these routes become rarer than the probability to overcome typical equilibrium energy barriers (which are by now large compared to temperature). This suggests a sudden onset of activated relaxation around $\overline T^*$, causing the dynamics to freeze out sharply as $T$ is lowered. To confirm this scenario, we plot in Fig.~\ref{fig:relaxation-time-stretching-exponent} (right inset, black dashed line) the activated behaviour $\propto \exp(\overline{\Delta E^*}/T)$ expected at the topological crossover based on the argument above, 
which leads to
reasonable quantitative agreement. 

It has been argued in the literature~\cite{berthierRealSpaceOrigin2003,fraggedakisInherentstateMeltingOnset2023} that a qualitative change in the dynamical behaviour happens in glassy systems at an onset temperature $T^\textrm{on}$, below which real-space dynamical heterogeneities appear. This temperature relates to the emergence of stretched exponential behaviour, $\beta \lesssim 1$. In our configuration connectivity analysis, we find that such onset temperature (left inset of Fig.~\ref{fig:relaxation-time-stretching-exponent}) correlates well with the appearance of an appreciable density of minima, see Fig.~\ref{fig:saddle-index-proportions}, and a finite typical equilibrium energy barrier greatly exceeding the thermal energy, see Fig.~\ref{fig:E0-E1-connectivities}. 

In continuous models, swap-moves act as non-local discrete connections on an otherwise continuous energy landscape. Therefore, they cannot be analysed with the same Hessian eigenvalue formalism~\cite{grigeraGeometricApproachDynamic2002, angelaniSaddlesEnergyLandscape2000, broderixEnergyLandscapeLennardJones2000} used to examine the landscape itself. On the contrary, in our discrete model the energy landscape generated by swap-moves is on equal footing to that generated by slice-rotations (both are discrete connections between configurations), and we have a natural way to compare the (easily equilibrating) swap-move connectivity with the glass-forming slice-rotation connectivity. 
In this vein, we observe in Fig.~\ref{fig:saddle-index-density-against-energy} that the swap-move cube's average saddle index density has a non-zero measure for all energies, and in Fig.~\ref{fig:E0-E1-connectivities} that a finite density of energy-decreasing connections is sampled at all energies. 
The results present a picture of swap-moves facilitating equilibration by systematically increasing the proportion of configurations which are saddles. This suppresses the saddles-to-minima crossover energy density $\epsilon^*$ to the ground state energy, and curtails the activated slowing down discussed above for slice-rotations by providing a finite density of energy decreasing connections at all temperatures. 

It is perhaps notable that we are able to learn so much about the long-time dynamical behaviour of the Rubik's Cube from an investigation of the one-step (nearest-neighbour) connectivity of the configuration network. A priori, one might have expected that an understanding of the long range behaviour of the connectivity is needed (see also Supp.~Mat.~Note~\ref{sm-sec:E2-E0-connectivities}). 
A possible reason for this result may be the size of the typical updates, which involve order $L$ spins in a system of order $L^2$ spins. 
It will be interesting to see if a similar insight may be obtained in systems with more local dynamics. 
%
%

\section{Conclusion}
In this work, we investigated the dynamics of the Rubik's Cube under slice-rotations, and we demonstrated behaviour typical of fragile glass-formers, with a sharp super-exponential dynamical arrest at some characteristic temperature $T^*$. 
Fitting the decay of the equilibrium autocorrelation function using a stretched exponential form, we found a characteristic relaxation timescale that increases by over eight orders of magnitude in the temperature range accessed in this study. The overall behaviour is captured quantitatively well by a Vogel-Fulcher-Tammann form. 

We uncovered suitable swap-moves that allow us to reach equilibrium efficiently at all temperatures, and afford us the rare opportunity to explore the behaviour of the system both above and below the glass transition. 
With this, we investigated the connectivity of configuration space and discovered a topological crossover from a saddle-dominated energy regime to a minima-dominated energy regime, akin to those observed in earlier work~\cite{broderixEnergyLandscapeLennardJones2000,angelaniSaddlesEnergyLandscape2000,cavagnaFragileVsStrong2001,grigeraGeometricApproachDynamic2002,coslovichLocalizationTransitionUnderlies2019}. The crossover follows a smooth exponential behaviour rather than a sharp threshold~\cite{fabriciusDistanceInherentStructures2002} in systems of finite size. However, in the thermodynamic limit we find that the saddles proportion curves converge to a step-function at a finite energy density (and therefore at a well-defined temperature, in equilibrium) that identifies a sharp glass transition in the dynamical behaviour of the system. Correspondingly, we observe that in the thermodynamic limit, the onset temperature $T^{\rm on}$ and the temperature $T^*$ where the dynamical arrest occurs come to coincide with the glass transition. 

By exploiting the discreteness of the model, we were able to directly contrast the original slice-rotation connectivity with the swap-moves connectivity, shedding light into the origin of the slow relaxation dynamics in the former and the suppression of the topological crossover in the latter. Unlike in continuous models, we were also able to clarify that the relevant energy density scale $\epsilon^*$ of the dynamical arrest coincides with when the proportion of $K \geq 2$ and not $K=1$ saddles becomes strongly suppressed. We found that our system behaves as a fragile glass, in the sense proposed in Ref.~\onlinecite{cavagnaFragileVsStrong2001}, where the system suddenly incurs energy barriers that are large compared to temperature as it moves through the topological crossover. We also proposed a toy attractor argument to explain the freezing at the threshold energy for temperatures below the crossover. 

Our study of the Rubik's Cube as a glass former opens avenues for further research, for instance exploring the possible presence and relevance of our observed topological crossover in quantum settings~\cite{garrahanAspectsNonequilibriumClassical2018}. It would also be desirable to see whether a real-space description of the RC behaviour in terms of defects and dynamical heterostructures and facilitation could be found to provide a complementary understanding of the glassy behaviour observed in our study of the energy landscape dynamics~\cite{berthierRealSpaceOrigin2003}. 
%
%

\section{Acknowledgements}
We are grateful to Giulio Biroli, Andrea Cavagna, Claudio Chamon, Juanpe Garrahan, and Roderich Moessner for insightful discussions. 
This work is supported in part by the Engineering and Physical Sciences Research Council (EPSRC) grant No.~EP/V062654/1. 
%
%


\putbib
\end{bibunit}
%
%

\clearpage
\newpage
%
%
\appendix

\setcounter{figure}{0}
\renewcommand{\thefigure}{S\arabic{figure}}
\renewcommand{\appendixname}{SM Note}

\begin{bibunit}

\onecolumngrid
\vspace{\columnsep}
\begin{center}
{\Large\bf Supplementary Material for:} 

\vspace{0.2 cm}
{\large\bf ``Saddles-to-minima topological crossover and glassiness in the Rubik’s Cube''}

\vspace{0.5 cm}
Alex Gower,$^{1,2}$
Oliver Hart,$^{3}$ 
and 
Claudio Castelnovo$^{1}$

\vspace{0.3 cm}
$^{1}$T.C.M. Group, Cavendish Laboratory, JJ Thomson Avenue, Cambridge CB3 0HE, United Kingdom

$^{2}${Nokia Bell Labs, Broers Building, 21 J.J. Thomson Avenue, Cambridge CB3 0FA, United Kingdom}

$^{3}$Department of Physics and Center for Theory of Quantum Matter, University of Colorado, Boulder, Colorado 80309, USA
\end{center}
\vspace{\columnsep}
\twocolumngrid
%
%

\section{Swap-Moves}\label{sm-sec:swap-moves}
We design the swap-moves such that the typical energy change incurred is small compared to the energy difference between the crossover energy and the ground state energy. This enhances the connectivity between below-crossover configurations (with each added connection in this region necessarily introducing an energy-decreasing connection to one of them). It therefore increases the proportion of saddles in this regime and suppresses the threshold energy density $\epsilon^*$.

The ensemble of Rubik's Cube configurations generated by swap-moves must be identical to that generated by slice-rotations. Note that this ensemble is not simply equivalent to arbitrary permutations of facelets; for instance, a facelet on a `corner piece' can never become a `centre piece', as we explain below. The constraints imposed on the facelet positions by the slice-rotations of the cube are known in the Rubik's Cube literature as the `First Law of Cubology'. As a prerequisite for this project, we proved this law (correcting previous attempts in the literature) for the general $L \times L \times L$ cube~\cite{gower-constraints-of-lxlxl-rubiks-cube}. Note that each swap-move is equivalent to a finite sequence of slice-rotations whose length is upper bounded by the `God's Number' of the RC, which grows with system size as $L^2/\log(L)$~\cite{demaineAlgorithmsSolvingRubik2011}. Explicit examples of sequences of slice-rotations equivalent to swap-moves can be found in Ref.~\onlinecite{gower-constraints-of-lxlxl-rubiks-cube}.

We first state the law and relevant background knowledge in Supp.~Mat.~Note~\ref{sec:first-law-of-cubology}; we then describe the class of swap-moves used in this paper in Supp.~Mat.~Note~\ref{sec:swap-move-types}; and finally we demonstrate that these generate the same configuration space as slice-rotations in Supp.~Mat.~Note~\ref{sec:swap-moves-span}. 
%
%

\subsection{First Law of Cubology: Sufficient constraints of an \texorpdfstring{$L \times L \times L$}{L x L x L} cube}\label{sec:first-law-of-cubology}
The fundamental constraints of the $L \times L \times L$ cube are most easily phrased in terms of the (permutations of the) cubelet positions and their orientations. By `cubelets' we mean each of the $L^3 - (L-2)^3$ individual sub-cubes of the overall Rubik's Cube. Note that while the \textit{edge} and \textit{corner} cubelets have several facelets each (2 and 3 respectively) and therefore require an additional \textit{orientation} parameter to fully describe their configuration, all the remaining \textit{centre} cubelets have just a single facelet each and thus no orientation parameter is needed.
%
%

\subsubsection{Independent cubelet subsystems}
The cubelets of an $L \times L \times L$ Rubik's cube decompose into several `independent cubelet subsystems'. By this we mean that the positions of cubelets can only be permuted within their individual subsystem and not between distinct subsystems.

The independent subsystems are classified below~\footnote{The derivation of this decomposition is provided in our supplementary work in Ref.~\onlinecite{gower-constraints-of-lxlxl-rubiks-cube}.}.
\begin{figure}[ht]
    \label{fig:facelet_classification_labelled}
    \centering
    \includegraphics[width=0.6\linewidth]{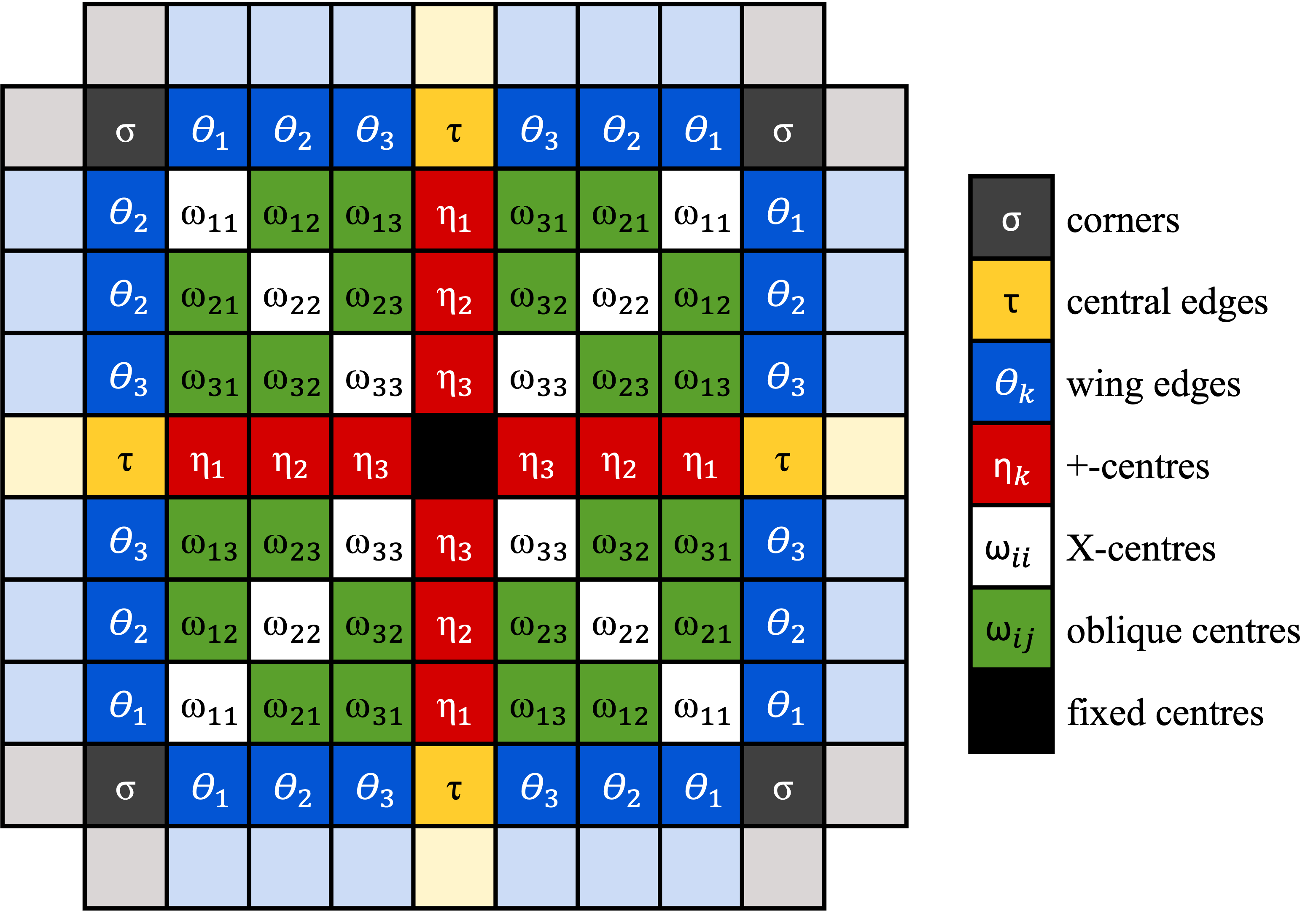}
    \caption{Classification of the various types of facelets on a given face of an $L=9$ cube. Edges and corners have more than one facelet per cubelet, while centres have exactly one facelet per cubelet.}
\end{figure}
The constraints imposed by the rotations lead to the following subclasses of cubelets and facelets on a cube with odd $L$: 
\begin{enumerate}[label=(\roman*)]
    \item fixed centres (6 cubelets, 6 facelets),
    \item corner pieces (8 cubelets, $24$ facelets),
    \item central edges (12 cubelets, $24$ facelets),
    \item wing edges ($12(L-3)$ cubelets, $24(L-3)$ facelets),
    \item centre pieces ($6[(L-2)^2-1]$ cubelets, $6[(L-2)^2-1]$ facelets).
\end{enumerate}
The wing edges and centre pieces also further decompose into a number of subsystems that cannot mutually exchange facelets: 
\begin{itemize}
    \item wing edges form $(L-3)/2$ independent subsystems containing 24 cubelets, 48 facelets each,
    \item centre piece facelets decompose into
    \begin{itemize}
        \item +-centres, $(L-3)/2$ independent subsystems containing 24 cubelets, 24 facelets
        \item $X$-centres, $(L-3)/2$ independent subsystems containing 24 cubelets, 24 facelets
        \item oblique centres, $\frac{(L-3)}{2}[\frac{(L-3)}{2} - 1]$ independent subsystems containing 24 cubelets, 24 facelets
    \end{itemize}
\end{itemize}
Therefore, excluding the immovable fixed centres and the trivial case for $L=1$, we have that the number of independent cubelet subsystems for an odd-$L$ cube is given by: 
\begin{equation}
    N_{\textup{L, odd}} = \frac{L^2 -2L +5}{4} 
    \, . 
\end{equation}
%
%

\subsubsection{Cubelet permutations}
We can now describe permutations of the cubelets within their respective independent cubelet subsystems using the following notation:
\begin{itemize}
    \item corner pieces, permutations $\sigma \in S_8$,
    \item central edge pieces, permutations $\tau \in S_{12}$,
    \item wing edge pieces, permutations $\theta_k \in S_{24}$,  \\ $k \in \{ 1, 2, \ldots,  (L-3)/2  \}$,
    \item $+$-centre pieces, permutations $\eta_k \in S_{24}$, \\ $k \in \{ 1, 2, \ldots, (L-3)/2 \}$,
    \item $X$-centre pieces, permutations $\omega_{ii} \in S_{24}$, \\ $i \in \{ 1, 2, \ldots,  (L-3)/2  \}$,
    \item oblique centre pieces, $\omega_{ij} \in S_{24}$, \\ $i,j \in \{ 1, 2, \ldots,  (L-3)/2  \}$, and $i \neq j$,
\end{itemize}
%
%

\subsubsection{Cubelet orientations}
In addition to the constraints on the allowed permutations of the cubelets, those cubelets that contain multiple facelets per cubelet also have a rotational degree of freedom which we call \textit{orientation}. Namely, the corners can be rotated three times before returning to their original orientation, parameterised by an element of $\mathbb{Z}/3\mathbb{Z}$. Analogously, each of the edges contains two facelets per cubelet and may be rotated twice before returning to their original position, and the orientation of each is therefore described by an element of $\mathbb{Z}/2\mathbb{Z}$.

We therefore describe orientations of the cubelets using~\footnote{The precise conventions used to uniquely map configurations onto the cubelet permutation and orientation description are detailed in Ref.~\onlinecite{gower-constraints-of-lxlxl-rubiks-cube}.}:
\begin{itemize}
    \item corner pieces, orientations $x_i \in \{ 0, 2\pi/3, 4\pi/3 \} \cong \mathbb{Z}/3\mathbb{Z}$, $1 \leq i \leq 8$,
    \item central edge pieces, orientations $z_i \in \{0, \pi \} \cong \mathbb{Z}/2\mathbb{Z}$, $1 \leq i \leq 12$,
    \item wing edge pieces, orientations $y_i(k) \in \{0, \pi \} \cong \mathbb{Z}/2\mathbb{Z}$, $1 \leq i \leq 24$, $k \in \{ 1, 2, \ldots,  (L-3)/2  \}$. 
\end{itemize}
%
%

\subsubsection{First Law of Cubology}
Using the above description of configurations in terms of cubelet permutations and orientations, we are finally in a position to state the First Law of Cubology. We proved the sufficiency of these constraints in our supplementary work in Ref.~\onlinecite{gower-constraints-of-lxlxl-rubiks-cube}. 

\begin{theorem}[First Law of Cubology]\label{thm:first-law-of-cubology}

    A valid configuration~\footnote{A configuration is \textit{valid} if it is in the orbit of the solved configuration of the cube under the action of $\mathbf{G}_L$.} of a cube with odd linear dimension $L \geq 3$, parameterised using permutations $\{ \sigma, \tau, \theta_k, \eta_k, \omega_{ij} \}$ and orientations $\{ x_i, z_i, y_i(k) \}$ must satisfy:

    \begin{subequations}
        \begin{equation}
            \sgn (\sigma) = \sgn (\tau) = \sgn(\omega_{ii})
            \label{eqn:perm-sufficient-constraint-1}
        \end{equation}
        
        \begin{equation}
              \sgn (\eta_k) = \sgn (\sigma) \sgn (\theta_k)
              \label{eqn:perm-sufficient-constraint-2}
        \end{equation}

        \begin{equation}
             \sgn(\omega_{ij}) = \sgn(\sigma) \sgn(\theta_i) \sgn(\theta_j), \  (i \neq j)
            \label{eqn:perm-sufficient-constraint-3}       
        \end{equation}

        \begin{equation}
             \sum_i x_i = 0 \mod 2\pi
            \label{eqn:orientation-sufficient-corners}
        \end{equation}

        \begin{equation}
             \sum_i z_i = 0 \mod 2\pi
            \label{eqn:orientation-sufficient-central-edges}
        \end{equation}

        \begin{equation}
            y_i(k) = 0 \, . 
            \label{eqn:orientation-sufficient-wing-edges}
        \end{equation}
    \end{subequations}

\end{theorem}
%
%

\subsection{Classification of Swap-Moves}\label{sec:swap-move-types}
A corollary of Theorem~\ref{thm:first-law-of-cubology} (demonstrated in Ref.~\onlinecite{gower-constraints-of-lxlxl-rubiks-cube}) is that all subgroups $\mathbf C_X$ of $\mathbf G_L$ (the Rubik's Cube group of slice-rotations) that only permute cubelets within their independent subsystem $X$ (whilst leaving the rest of the cube unchanged) are isomorphic to alternating groups i.e., $\mathbf C_X \cong \mathbf A_{|X|}$ (where $|X|$ is the number of cubelets in the subsystem $X$). This means that there does not exist a sequence of slice-rotations whose net effect is to swap 2 cubelets, but there do exist sequences that 3-cycle 3 cubelets within an independent subsystem. We use these as our first class of swap-move.

\begin{remark*}
\textbf{Swap-Move Class 1: 3-Cycles}
$$P_3(X, \{i,j,k\}), \ i,j,k \in \{ 1,...,|X|\} \ \textup{for any} \ X$$
(where $|X|$ is the number of cubelets in subsystem $X$) is a class of swap-moves which 3-cycles the positions of 3 cubelets indexed by $\{i,j,k\}$ within a single independent cubelet subsystem $X$, whilst preserving each cubelet's orientation.
\end{remark*}

Another corollary of Theorem~\ref{thm:first-law-of-cubology} is that there does not exist a sequence of slice-rotations whose net effect is to rotate a single cubelet, but there do exist composite rotations that rotate (in opposite directions) the orientations of 2 cubelets within an independent subsystem. We use these as our second class of swap-move.

Note that $y_i(k)=0$ in Theorem~\ref{thm:first-law-of-cubology} means that `wing edge cubelets' have no orientation freedom (i.e., their orientation is fixed by their position), hence we do not (and must not) rotate these in order to generate $\mathbf G_L$.

\begin{remark*}
\textbf{Swap-Move Class 2: Opposite Orientation Rotations }
$$O_2(X, \{i,j\}), \ i,j \in \{ 1,...,|X| \} \ \textup{for} \ X \in X_{rot}$$
(where $|X|$ is the number of cubelets in subsystem $X$) is a class of swap-moves which rotates cubelets indexed by $\{i,j\}$ in subsystem $X$ of rotatable cubelets by one unit in an anti-clockwise and clockwise direction respectively. $X_{rot} = \{ \sigma, \tau \}$ for odd $L$ (i.e., only the corners and central edges are freely rotatable).
\end{remark*}

Finally, we note that, while the 3-cycles provide the most delicate cubelet permutation transformation possible (in the sense of affecting the fewest facelets), they preserve the parities $\sgn(X)$ of all subsystems. Therefore they do not allow the cube to transition between regions of configuration space with differing $\{ \sgn(X) \}$ and thus cannot generate all of $\mathbf G_L$. To remedy this, we introduce a final class of swap-moves that performs 2-cycles on multiple independent cubelet subsystems simultaneously, in such a way that all constraints in Theorem~\ref{thm:first-law-of-cubology} are still satisfied.

\begin{remark*}
\textbf{Swap-Move Class 3: Coupled Subsystem 2-Cycles}
\begin{align*}
    &C(\{X_i \}, \{ \{j,k \}_i \}) = \prod_{X_i \in \{X_i \}} P_2(X_i, \{j,k\}_i), \\
    &\quad j,k \in\{1,...,|X_i|\} \\
    &\quad \textup{for } \{ X_i \} \subseteq \{ X \} \textup{ s.t. Theorem~\ref{thm:first-law-of-cubology} is satisfied}
\end{align*}
(where $|X_i|$ is the number of cubelets in subsystem $X_i$) is a class of swap-moves which swaps the positions of cubelets indexed by $\{j,k\}_i$ within independent cubelet subsystem $X_i$, but chooses a coupled set of independent subsystems $\{X_i \}$ such that the constraints in Theorem~\ref{thm:first-law-of-cubology} are still satisfied.
\end{remark*}

By observing that the independent variables of the `First Law of Cubology' constraints in Theorem~\ref{thm:first-law-of-cubology} are $\{\textup{sgn}(\sigma), \{\textup{sgn}(\theta_k)|_{k \leq \frac{L-3}{2}} \}\}$, we can derive all required coupled subsystem 2-cycles by first fixing all possible combinations of the these subsystems to 2-cycle, and then using Theorem~\ref{thm:first-law-of-cubology} again to derive the other subsystems which must be 2-cycled in each case in order to respect the fundamental constraints of the cube.

Note that while the `3-cycles' and `opposite orientation rotations' swap-moves affect $O(1)$ facelets and can therefore only produce an $O(1)$ energy cost, the `coupled subsystem 2-cycles' affect $O(L^2)$ facelets~\footnote{This $L^2$ scaling can be deduced from Eq.~\eqref{eqn:perm-sufficient-constraint-3} in Theorem~\ref{thm:first-law-of-cubology}, or, more simply, from the fact that an extensive number of the $O(L^2)$ independent sub-systems may require 2-cycles.} in the worst case and may produce an $O(L^2)$ energy cost. 

In this work, we chose to include all swap-moves defined above in our swap-move dynamics; as demonstrated in the main text, we found that it results in the suppression of the saddles-to-minima topological crossover, thereby allowing us to thermalise the RC at all temperatures. 
In principle, we cannot exclude the possibility that an appropriate subset of the swap-moves may be sufficient to efficiently thermalise the system within reasonable computational efforts. An investigation of this possibility is left for future work. 
%
%

\subsection{Swap-move and slice-rotation reachable configurations are equivalent}\label{sec:swap-moves-span}
%
%

\subsubsection{All swap-moves can be represented as slice-rotations}
\begin{enumerate}
    \item All swap-moves produce configurations which respect the constraints of Theorem~\ref{thm:first-law-of-cubology}.
    \item Since the constraints of Theorem~\ref{thm:first-law-of-cubology} are sufficient, this means all satisfying configurations can be generated by a sequence of slice-rotations.
\end{enumerate}
%
%

\subsubsection{All slice-rotations can be represented as swap-moves}
\begin{enumerate}
    \item Showing that all slice-rotations can be represented as swap-moves is equivalent to showing that all configurations satisfying Theorem~\ref{thm:first-law-of-cubology} can be generated by swap-moves.
    \item This can be done with the following procedure:
    \begin{enumerate}
        \item Use the required coupled subsystem 2-cycle swap-move $C(\{X_i \}, \{ \{j,k \}_i \})$ to generate a configuration with the same permutation signatures $\{ \sgn(X) \}$ as the desired satisfying configuration. This is possible because $C(\{X_i \}, \{ \{j,k \}_i \})$ can generate all combinations of permutation signatures $\{ \sgn(X) \}$.
        \item Use the required 3-cycle swap-moves $P_3(X, \{i,j,k\})$ for each subsystem $X$ to generate a configuration with the same cubelet positions as the desired satisfying configuration. This is possible because 3-cycles $P_3(X, \{i,j,k\})$ can generate all permutations of each subsystem $X$ which preserve the signature $\sgn(X)$.
        \item Use the required opposite orientation rotation swap-moves $O_2(X, \{i,j\})$ to generate a configuration with the same cubelet orientations as the desired satisfying configuration (which is therefore the satisfying configuration itself). This is possible because $O_2(X, \{i,j\})$ can generate all combinations of cubelet rotations which satisfy the modular constraints in Theorem \ref{thm:first-law-of-cubology}.
    \end{enumerate}
\end{enumerate}
%
%

\subsection{Proposal ratio}
The most natural way to implement the proposal of swap-moves in the Metropolis Monte Carlo algorithm would be to propose each swap-move with equal probability. 
However, there are many more types of $C(\{X_i \}, \{ \{j,k \}_i \})$ swap-move than $P_3(X, \{i,j,k\})$ or $O_2(X, \{i,j\})$~\footnote{One can show that the number of $C(\{X_i \}, \{ \{j,k \}_i \})$ swap-moves scales exponentially with $L$, the number of $P_3(X, \{i,j,k\})$ swap-moves scales polynomially, and the number of $O_2(X, \{i,j\})$ swap-moves is constant.}. Also the $C(\{X_i \}, \{ \{j,k \}_i \})$ swap-moves have at worst $O(L^2)$ energy costs and so have far lower $e^{-\Delta E/T}$ acceptance rates than the others with $O(1)$ energy-costs. From the arguments presented in the main text, we also anticipate the moves with lower energy costs to be more significant in increasing the proportion of below-crossover saddle configurations, thereby suppressing the saddles-to-minima crossover.

Therefore, in order to reduce the rejection ratio and to increase the efficiency of our MC algorithm, we add a fixed bias against $C(\{X_i \}, \{ \{j,k \}_i \})$ moves for our proposed swap-moves. Since coupled subsystem 2-cycles are their own inverse, this bias does not violate the detailed balance property required for the MC algorithm to converge to the Boltzmann distribution.

Specifically, our swap-move proposal scheme (using a bias of $-0.6$) can be described as follows (where $z$ is the total number of swap-moves, and $|C|$, $|P_3|$ and $|O_2|$ are the total number of each class of swap-move):
\begin{itemize}[itemsep=5pt, parsep=0pt, topsep=-1pt]
    \item Propose a random $C(\{X_i \}, \{ \{j,k \}_i \})$ swap-move with probability, $p = \frac{|C|}{z} - 0.6$.
    \item If the above swap-move is not proposed, propose a random $P_3(X, \{i,j,k\})$ swap-move with relative probability $p = \frac{|P_3|}{|P_3| + |O_2|}$ or a random $O_2(X, \{i,j\})$ swap-move with relative probability $p = \frac{|O_2|}{|P_3| + |O_2|}$.
\end{itemize}
This scheme should not impact any fundamental results of our work, and the decreased rejection-ratio will just mean fewer swap-move MC steps are required to equilibrate the cube.
%
%

\subsection{Swap-moves increase saddle proportion due to small \texorpdfstring{$\Delta \epsilon$}{Delta epsilon} not large \texorpdfstring{$z$}{z}}
Naively one might assume that it is simply the increase in connections in the swap-move configuration network (from a maximal degree of $z = 60$ for slice-rotations to $z \sim 10^{37}$ for swap-moves~\footnote{It should be noted that the increase to $z \sim 10^{37}$ for swap-moves is still negligible compared to the total number of configurations $N_\text{tot} \sim 10^{425}$ for the $L=11$ cube, and that the ratio $z/N_\text{tot}$ decays rapidly with $L$ (see Supp.~Mat.~Note~\ref{sm-sec:growth-of-swap-moves-with-L})} for the $L=11$ cube) which is responsible for the increased proportion of saddles at low energies and associated suppression of the dynamical arrest. In fact, the more important difference is the reduced typical~\footnote{Note also that, for slice-rotations, $\Delta \epsilon_\text{max} \simeq 0.07$ since $\Delta E_\text{max} = +8L$ ($=\!+88$ for the $L=11$ cube) because an internal layer slice-rotation (e.g., $R_2$ in Fig.~\ref{fig:relaxed-anneals}) disrupts $8L$ bonds and so a maximum of $8L \times \delta_{\sigma_i \sigma_j}$ bond energy contributions can be broken by a single slice-rotation.} energy density changes of swap-moves (from slice-rotation modal $\Delta \epsilon \lesssim 0.04$ to swap-move modal $\Delta \epsilon \lesssim 0.005$, as displayed in Fig.~\ref{fig:E0-E1-connectivities}) resulting from their careful design as moves which disrupt a minimal number of spins. 

Since there are exponentially more configurations above the crossover ($\epsilon > \epsilon^*$) than below, this means that connections added at random to the configuration network by increasing $z$ will almost never connect two below-crossover configurations. The latter is necessary in order to increase the proportion of below-crossover saddles~\footnote{Only by connecting two below-crossover configurations do we introduce an energy-decreasing connection to one of them, and thereby produce a saddle.}. Only by constraining the added connections to involve small energy density changes $\Delta \epsilon$, can we bias the added connections to lie below the crossover. Also, since the number of configurations grows exponentially with energy density $\mathcal{N}(\epsilon) \sim e^{\lambda \epsilon}$ (see Supp.~Mat.~Note~\ref{sm-sec:exponential-growth-of-configurations-with-energy}), adding connections with smaller $\Delta \epsilon$ will be more efficient at producing saddles, as it allows the $\mathcal{N}(\epsilon)$ configurations at $\epsilon$ to connect to a much larger pool of configurations at $\epsilon - \Delta \epsilon$. The relative size of this pool shrinks exponentially with larger $\Delta \epsilon$ as $\mathcal{N}(\epsilon - \Delta \epsilon) / \mathcal{N}(\epsilon) \sim e^{-\lambda \Delta \epsilon}$. 
%
%

\section{Parallel tempering anneal of the original Rubik's Cube}\label{sm-sec:parallel-tempering}
A parallel tempering algorithm (using slice-rotation local moves) applied to the original (non-randomised) RC also results in a dynamical arrest akin to that in the original Metropolis Monte Carlo algorithm.

\begin{figure}[ht!]
    \centering
    \includegraphics[width=\linewidth]{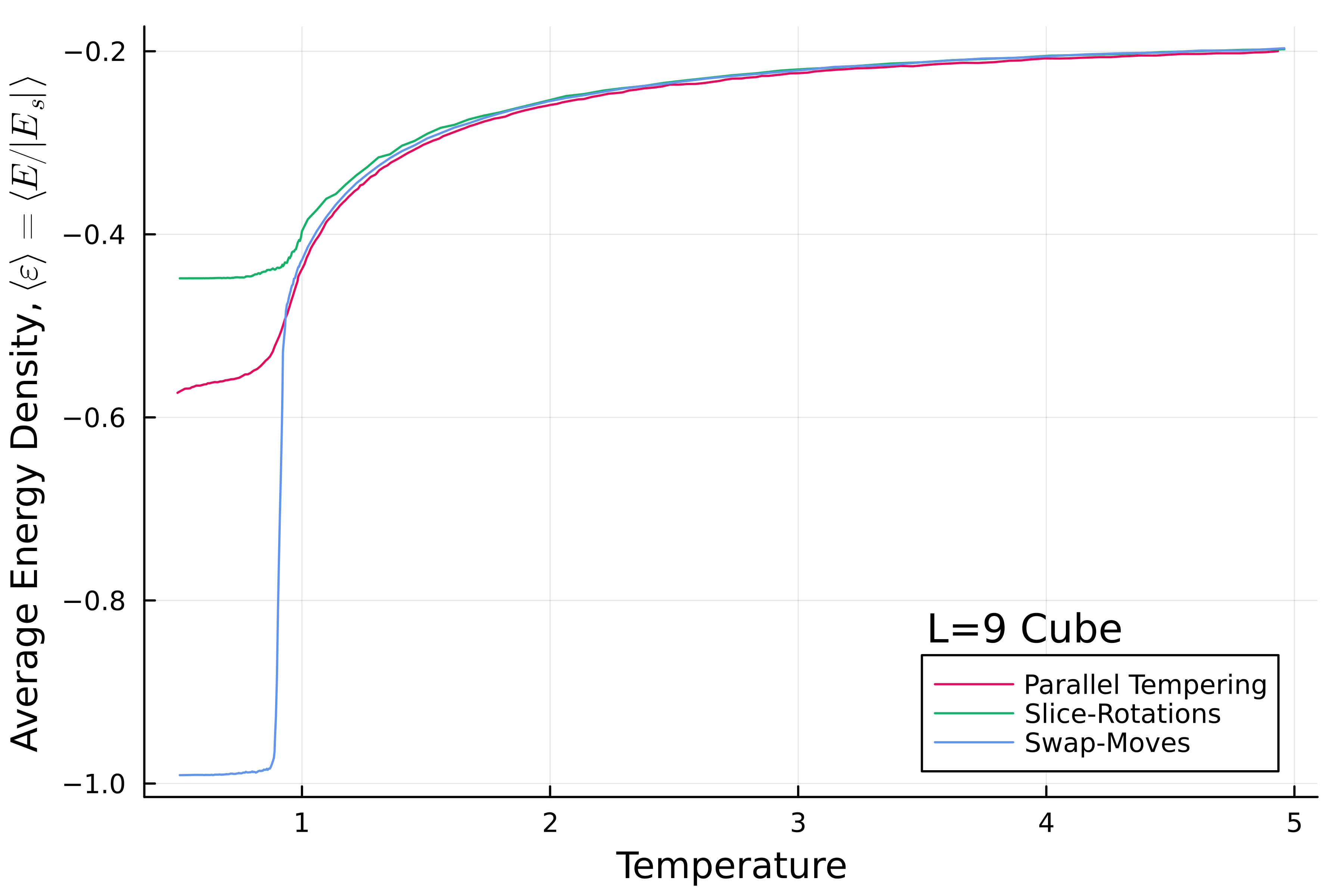}%
    \caption{Equilibrium value of the energy $\langle E(T) \rangle$ as a function of temperature for the $L=9$ cube, displaying a sharp transition to the solved configuration at $T = T_c$ for the swap-move MC algorithm, but not for the parallel tempering algorithm nor for the slice-rotation algorithm. (Data for swap-move MC and slice-rotation MC are the same as the $L=9$ data in Fig.~\ref{fig:relaxed-anneal-system-size-dependence}.)}
    \label{fig:parallel-tempering}
\end{figure}

As seen in Fig.~\ref{fig:parallel-tempering}, at intermediate temperatures, the swap-move average energies agree with the parallel tempering results. The parallel tempering simulations however (like the original slice-rotation MC algorithm) fall out of equilibrium before the ordering transition that one observes in the swap-move simulations. Note that the parallel tempering simulations fall out of equilibrium after the slice-rotation MC algorithm does, i.e., parallel tempering equilibrates the cube down to lower temperatures than the slice-rotation MC algorithm.

For computational convenience, this study was carried out for an $L=9$ cube, while the rest of the paper focuses on an $L=11$ cube. Similar results are expected for any system size, demonstrating that parallel tempering is unable to efficiently equilibrate the Rubik's Cube at low temperatures across the phase transition. 
%
%

\section{System-size dependence of slice-rotation dynamics}\label{sm-sec:system-size-dependence}
In this section we first present the system-size dependence of the dynamics of the RC, and extract the parameters $\overline T^{\rm on}$, $\overline\epsilon^{\rm on}$ (associated with the onset of stretched exponential behaviour) and $\overline T^*$, $\overline \epsilon^*$ (associated with the sharp dynamical arrest) for each system size. We then show that these energy densities can be related to the energy densities $\overline \epsilon_\otimes$ and $\overline \epsilon_\times$, respectively, which correspond to crisply defined features in the saddle index proportions resolved as a function of energy. Finally, we show that these parameters all converge to the same value in the $L \rightarrow \infty$ thermodynamic limit, supporting our claims in the main text that the exponential crossover behaviour tends asymptotically to a sharp step-function threshold in this limit.

\begin{figure}[ht]
    \centering
    \includegraphics[width=\linewidth]{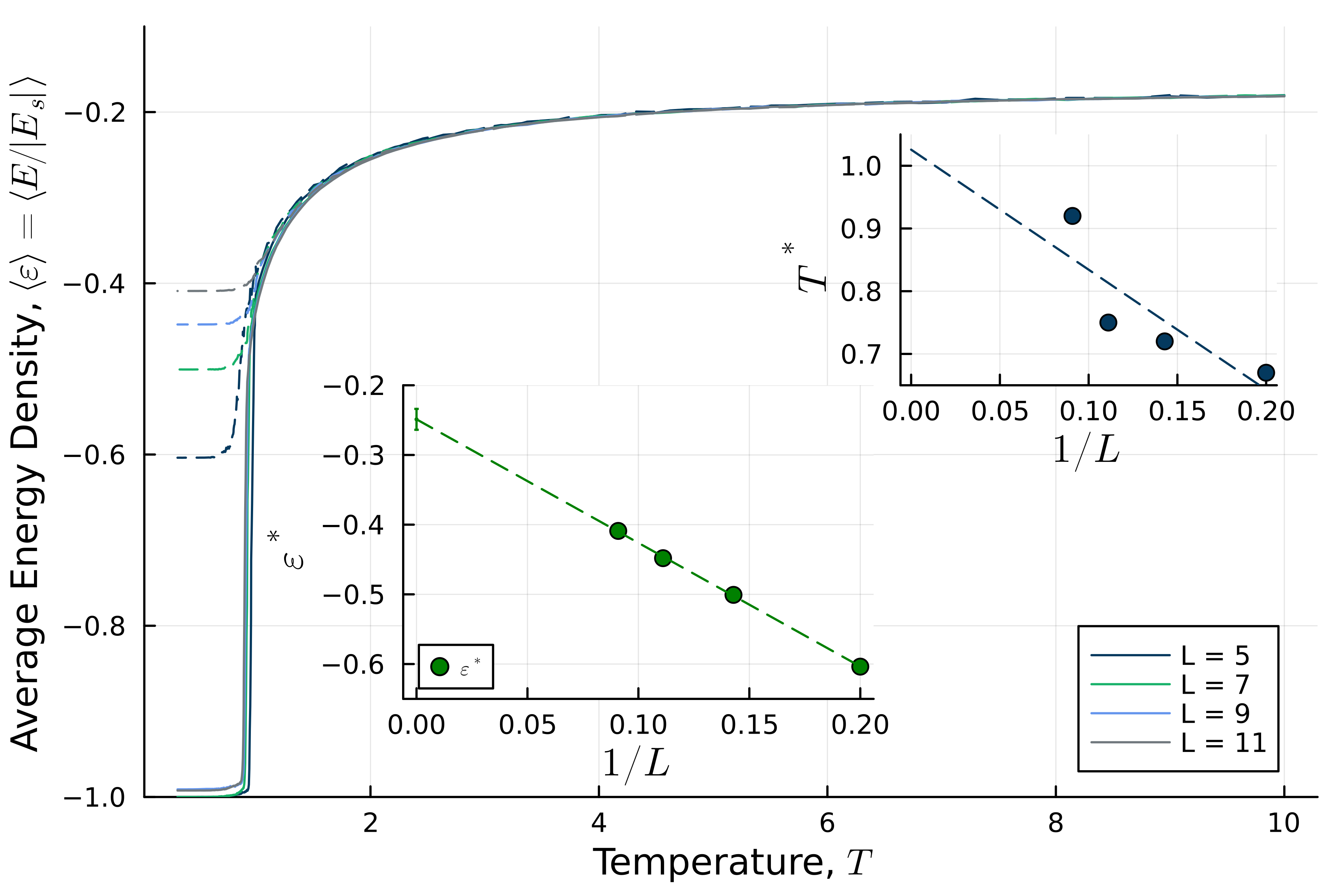}
    \includegraphics[width=\linewidth]{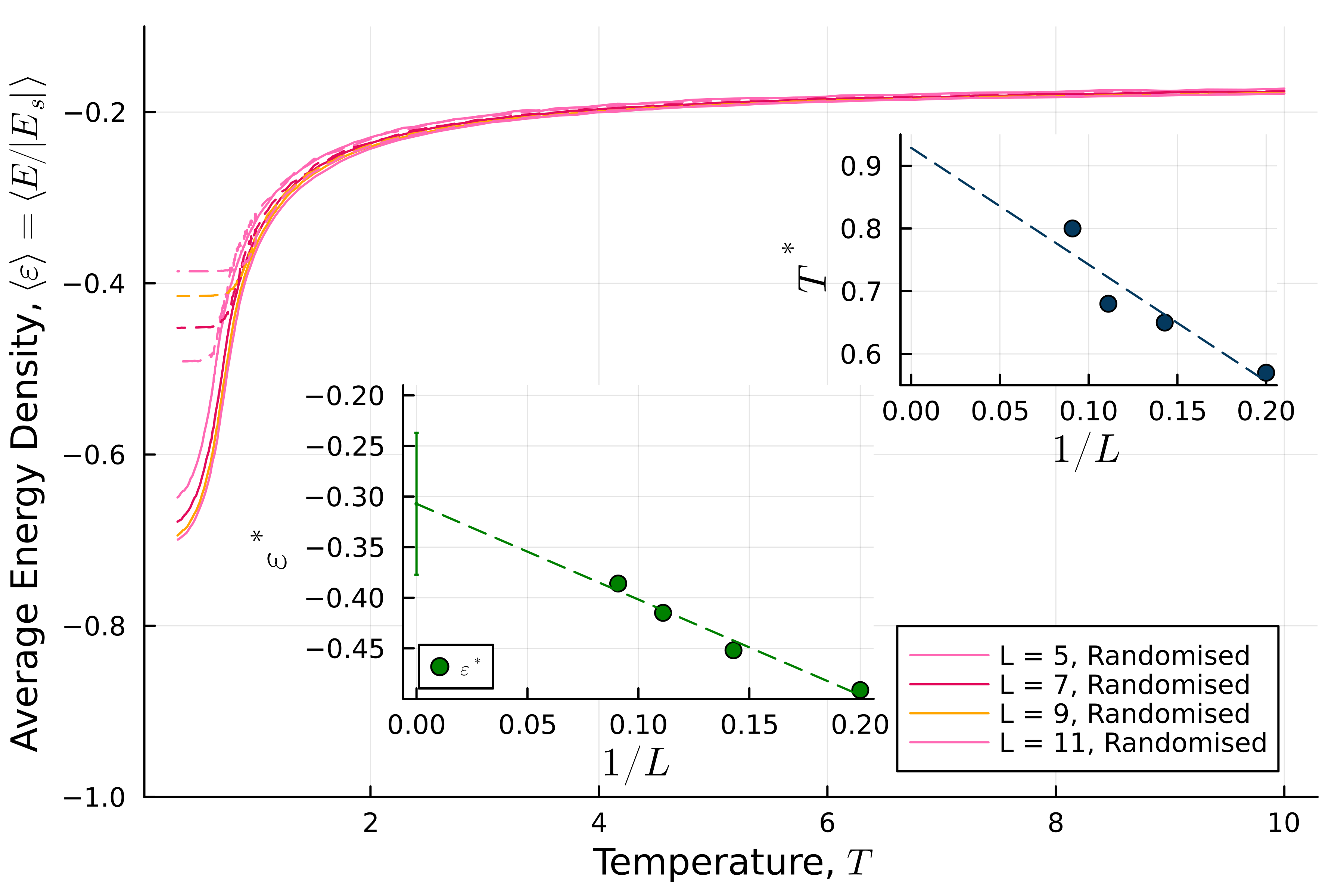}
    \caption{Energy vs. temperature curves obtained from thermal anneals of Rubik's Cubes with varying $L$, using slice-rotations (dashed lines) and swap-moves (solid lines), for both the original RC (top panel) and the randomised RC (bottom panel), as defined in the main text. The data are averaged over 50 histories and, for the randomised cube, each history has a different realisation of disorder. (Left inset) $\epsilon^* = \min \langle \epsilon \rangle$ for the slice-rotation anneals as a function of $L$, with a linear fit (dashed) and the $95\%$ confidence interval at $\frac{1}{L} = 0$. (Right inset) $T^*$ (defined as $\langle \epsilon(T^*) \rangle = \epsilon^*$) for the slice-rotation anneals as a function of $L$ with a linear fit (dashed) as a guide to the eye.}
    \label{fig:relaxed-anneal-system-size-dependence}
\end{figure}

In Fig.~\ref{fig:relaxed-anneal-system-size-dependence} we present thermal anneals for RC models of varying system size $L$. For the slice-rotation dynamics, we extract $\epsilon^*$ (defined as the minimum average energy density reached during the anneal, $\epsilon^* = \min \langle \epsilon \rangle$) and $T^*$ (defined as the temperature at which the average energy sharply plateaus at $\epsilon^*$, $\langle \epsilon(T^*) \rangle = \epsilon^*$).

\begin{figure}[ht]
    \centering    \includegraphics[width=\linewidth]{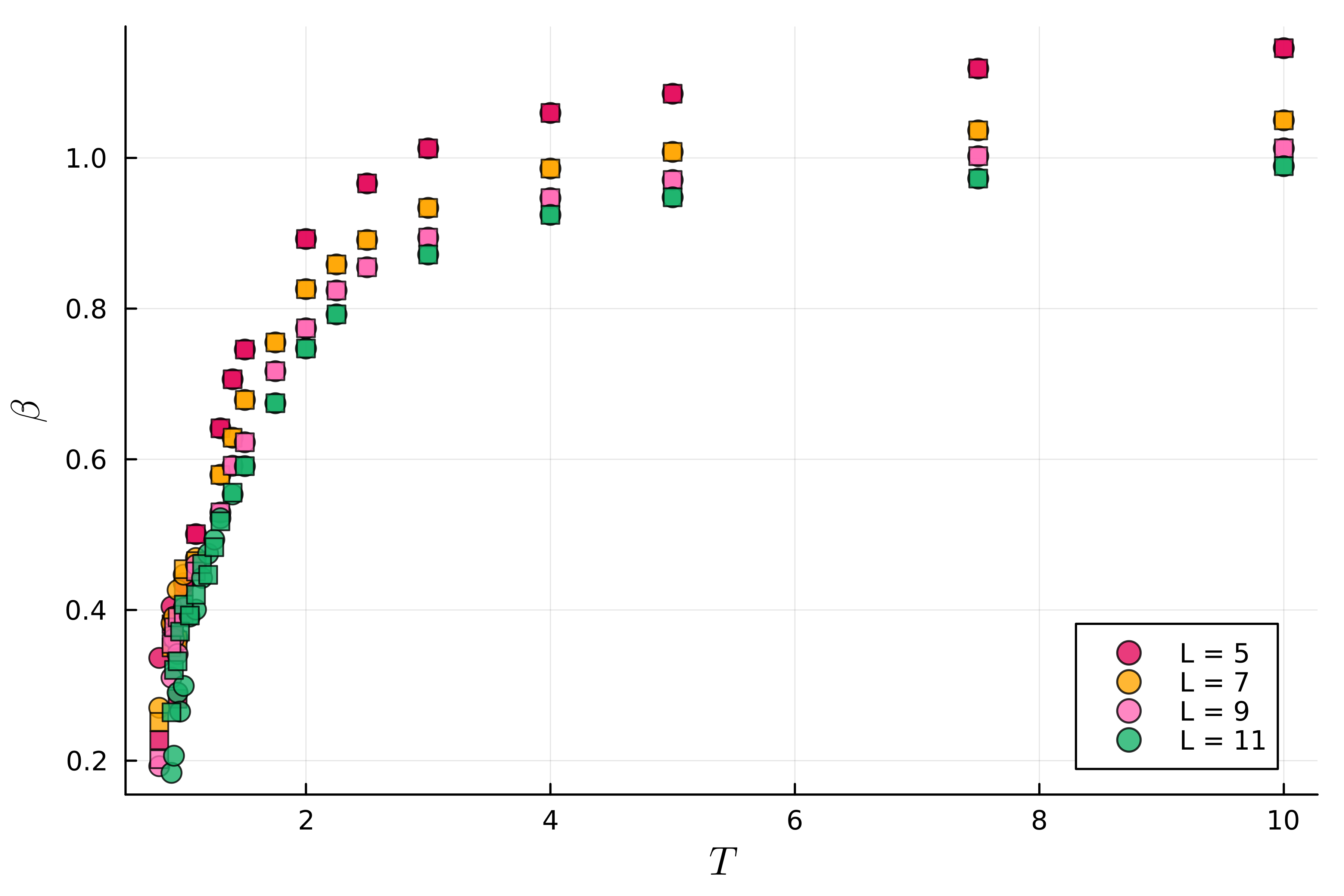}

    \includegraphics[width=\linewidth]{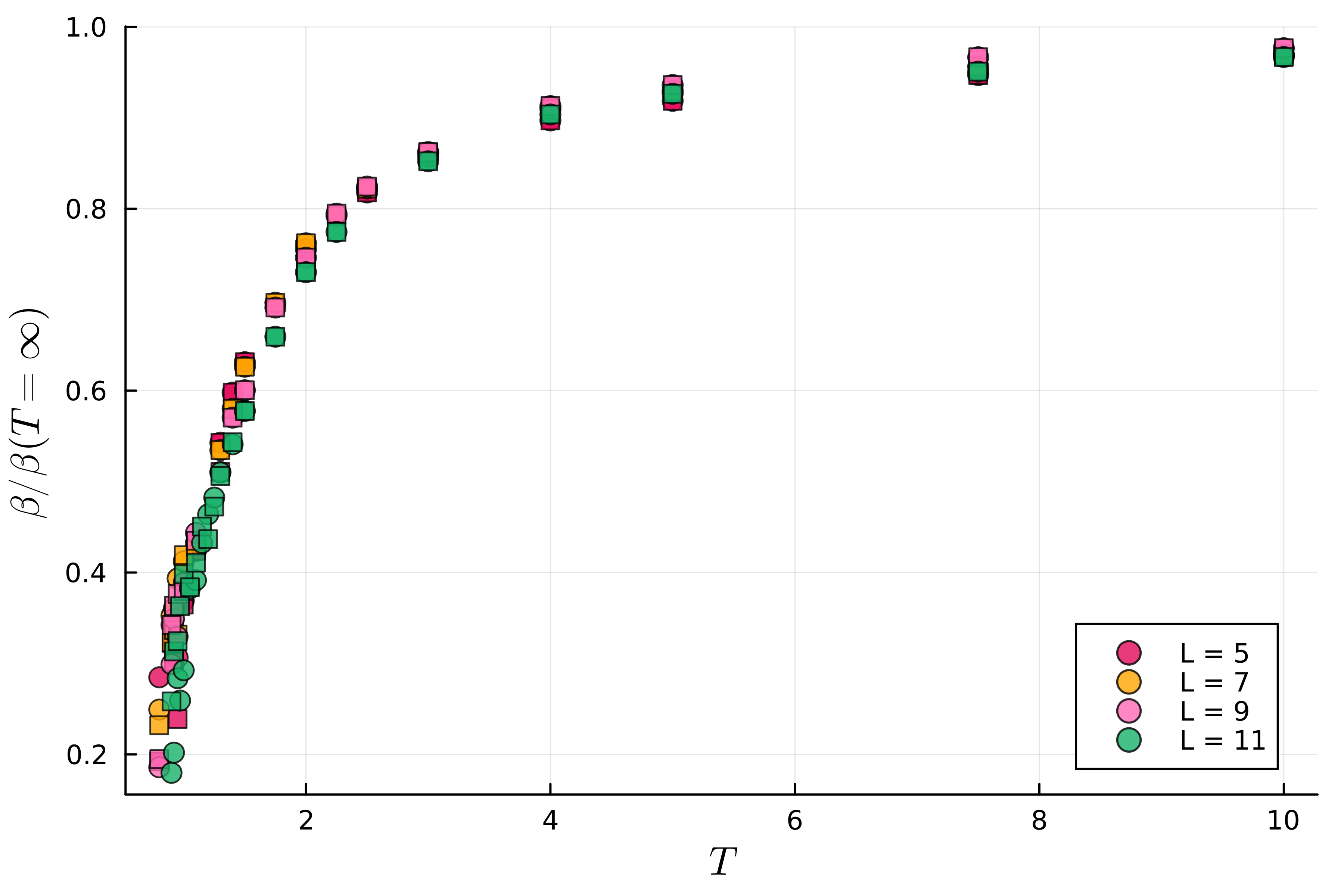}
    \caption{(Top panel) Temperature dependence of the stretching exponent $\beta(T)$ associated with stretched exponential fits of the autocorrelation function decay for the randomised RC with varying system size $L$. $\beta(T\rightarrow \infty)$ asymptotes to values larger than $1$ at small system size. (Bottom panel) Rescaling the $\beta(T)$ curves by $\beta(T=\infty)$ results in a collapse onto a universal curve.}
    \label{fig:beta-system-size-dependence}
\end{figure}

In the top panel of Fig.~\ref{fig:beta-system-size-dependence}, we plot  the temperature dependence of the stretching factors $\beta(T)$ associated with stretched exponential fits of the autocorrelation function decay for varying $L$ (as detailed in Sec.~\ref{sec:slow-dynamics} of the main text). Interestingly, we find that $\beta(T \rightarrow \infty)$ asymptotes to values larger than $1$ at small system size. We therefore adopt the convention of $\beta(T=T^{\rm on}) := 0.75 \times \beta(T=\infty)$ to define the onset of stretched exponential behaviour in the dynamics (we checked that the behaviour is robust to an arbitrary threshold choice around this value of $0.75$). In the bottom panel of Fig.~\ref{fig:beta-system-size-dependence} we see that this $\beta(T=\infty)$ normalization collapses $\beta(T)$ for all $L$ onto the same universal curve. We extract $\epsilon^{\rm on}$ as the equilibrium energy density associated with $T^{\rm on}$.

\begin{figure}[ht]
    \centering    \includegraphics[width=\linewidth]{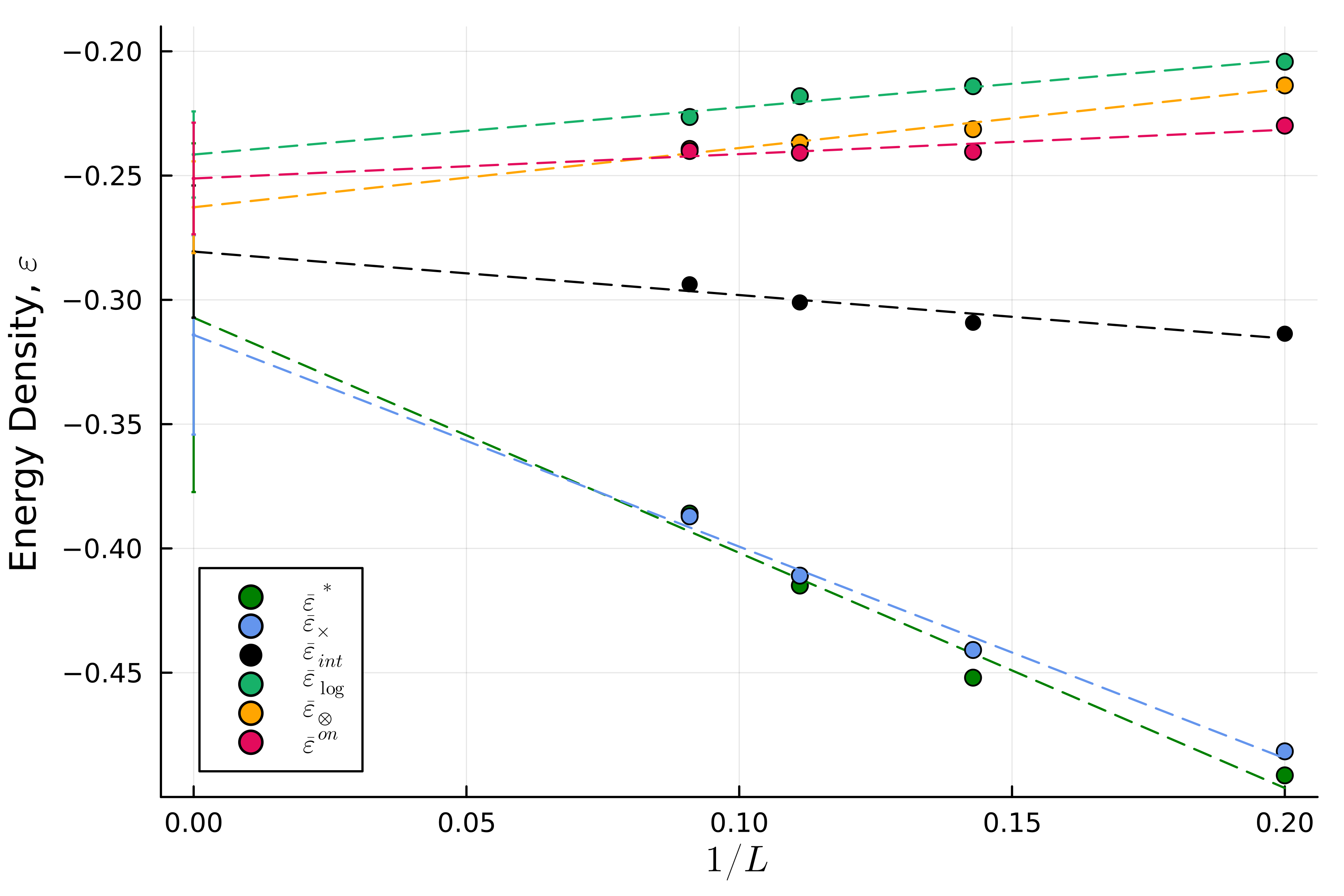}
    \caption{System-size dependence of: $\overline\epsilon^*, \overline\epsilon^{\rm on}$ (defined in the main text in relation to the slice-rotation dynamics of the randomised RCs), and $\overline\epsilon_\otimes, \overline\epsilon_\times, \overline\epsilon_{\rm log}, \overline\epsilon_{\rm int}$ (defined in the main text in relation to the energy dependence of the saddle index proportions). Linear fits are presented as dashed lines, along with their $95\%$ confidence intervals at $\frac{1}{L}=0$. We observe $\overline\epsilon^{\rm on} \simeq \overline\epsilon_\otimes$ and $\overline\epsilon^* \simeq \overline\epsilon_\times$ for all $L$, and that all parameters converge within the confidence intervals as $L \rightarrow \infty$.}
    \label{fig:epsilon-parameters-system-size-dependence}
\end{figure}

Taking inspiration from Figs.~\ref{fig:saddle-index-proportions} and~\ref{fig:combined-saddle-index-proportions-finite-size-scaling} in the main text, we pragmatically define the energy density $\epsilon_\otimes$ below which the proportion of configurations that are minima ($K=0$) becomes appreciable as $p_{K=0}(\epsilon_\otimes) := 0.01$. Analogously, we define the energy density $\epsilon_\times$ below which the proportion of saddles ($K \geq 2$) becomes non-appreciable as $p_{K\geq 2}(\epsilon_\times) := 0.01$. In Fig.~\ref{fig:epsilon-parameters-system-size-dependence} we show that $\overline\epsilon^{\rm on} \simeq \overline\epsilon_\otimes$ and $\overline\epsilon^* \simeq \overline\epsilon_\times$ for all $L$, suggesting that the behaviours observed in the dynamics can be related to features derived from the energy-resolved connectivity structure of the configuration networks. In addition, we observe that these energy density scales converge (within reasonable confidence intervals of their linear fits) as $L \rightarrow \infty$, consistent with the replacement of a crossover with a sharp threshold in the thermodynamic limit. For completeness, we also include the $\overline \epsilon_{\rm log}$ parameter associated with the generalised logistic function fits of $\overline p_{K \geq 2}(\epsilon)$ (see Fig.~\ref{fig:combined-saddle-index-proportions-finite-size-scaling} of the main text), and $\overline \epsilon_{\rm int}$ defined as the energy density where $\overline p_{K=0} = \overline p_{K \geq 2}$, and observe that they too converge within the confidence intervals at $\frac{1}{L} = 0$. 
%
%
%

\section{First order phase transition 
in the Rubik's Cube}
\label{sm-sec:phase-transition}
In this section we demonstrate that the original RC undergoes a first-order phase transition (to the solved configuration) below a critical temperature $T_c$, and we present a brief characterisation of the thermodynamic behaviour.

\begin{figure}[ht]
    \centering
    \includegraphics[width=\linewidth]{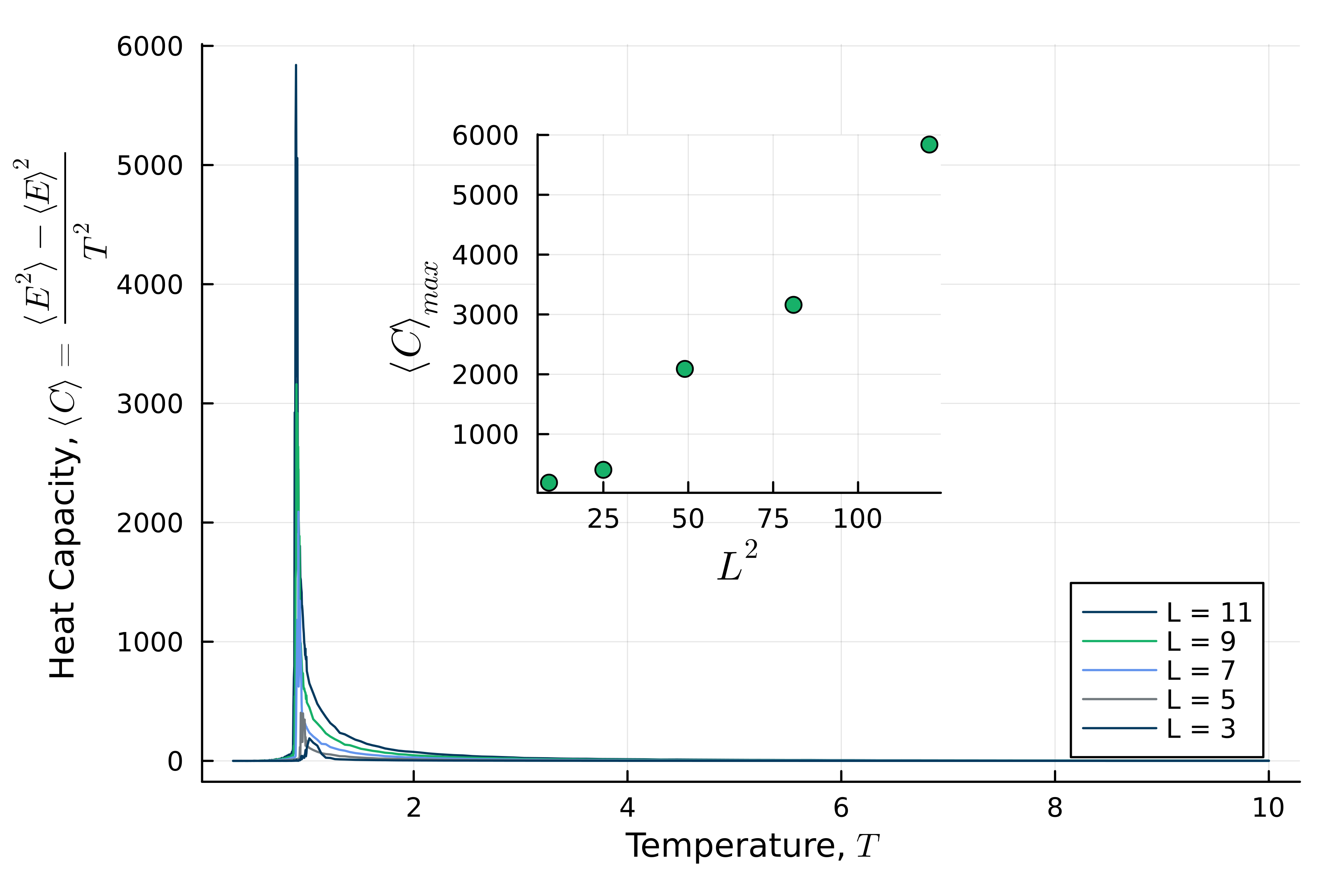}
\includegraphics[width=\linewidth]{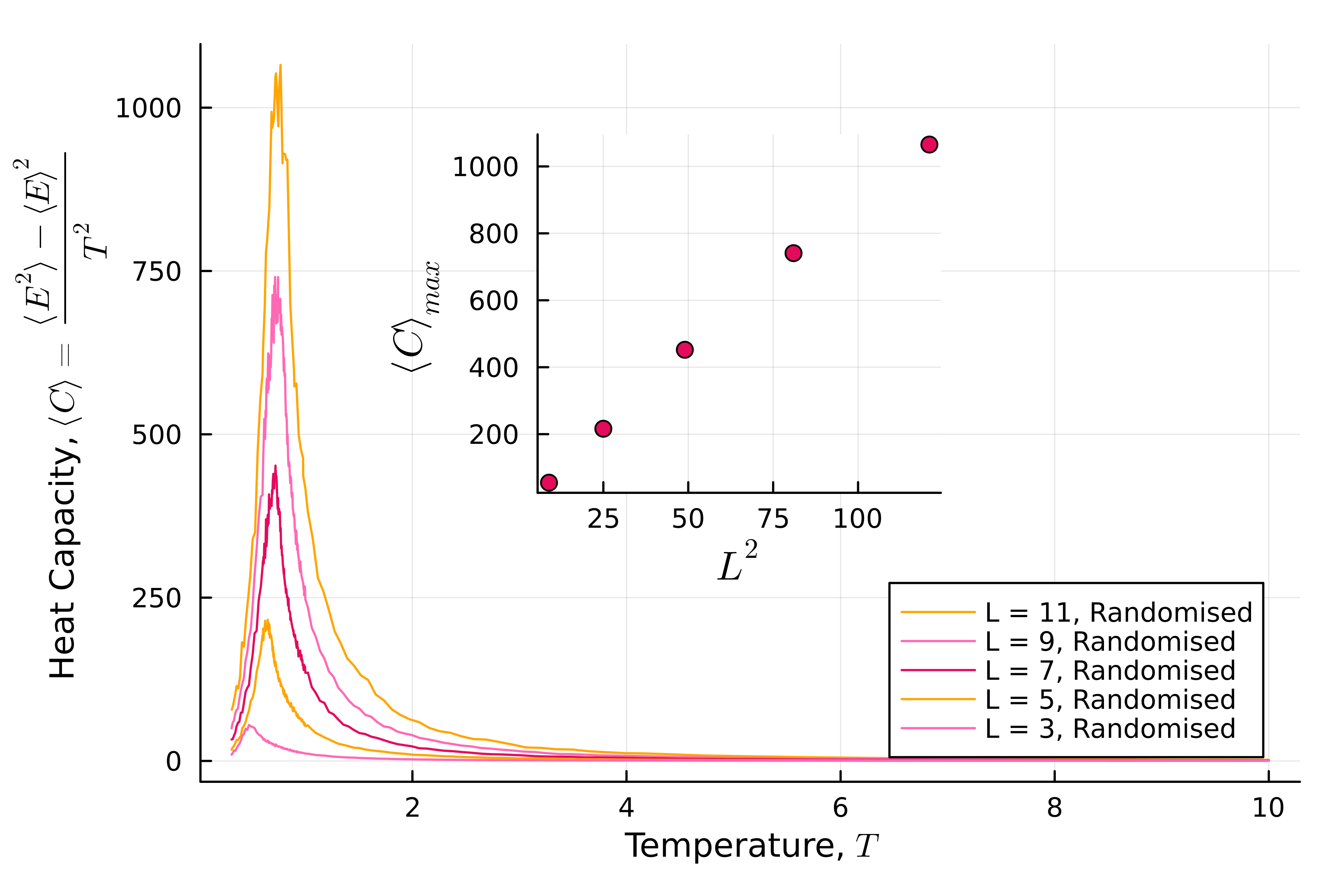}
    \caption{Heat capacity as a function of temperature for the original RC model (top panel) and for the randomised RC model (bottom panel), averaged over 50 histories (where, for the randomised cube, each history has a different realisation of disorder). The original RC model shows a singularity which becomes sharper with increasing $L$, indicative of a first-order phase transition, while the randomised RC model does not. Insets: Peak height $\langle C \rangle_{\rm{max}}$ of the heat capacity as a function of volume, showing approximate linear behaviour.}
    \label{fig:heat-capacities}
\end{figure}
In Fig.~\ref{fig:heat-capacities} (top panel) we show that the original RC exhibits a peak in the heat capacity $\langle C \rangle$ at some temperature $T_c$, which becomes sharper for larger system sizes $L$, and with a peak height $\langle C \rangle_{\rm max}$ (obtained from a Gaussian fit near the peak) that grows linearly with the volume $L^2$ of the system. This is indicative of a first-order phase transition.

\begin{figure}[ht]
    \centering
    \includegraphics[width=\linewidth]{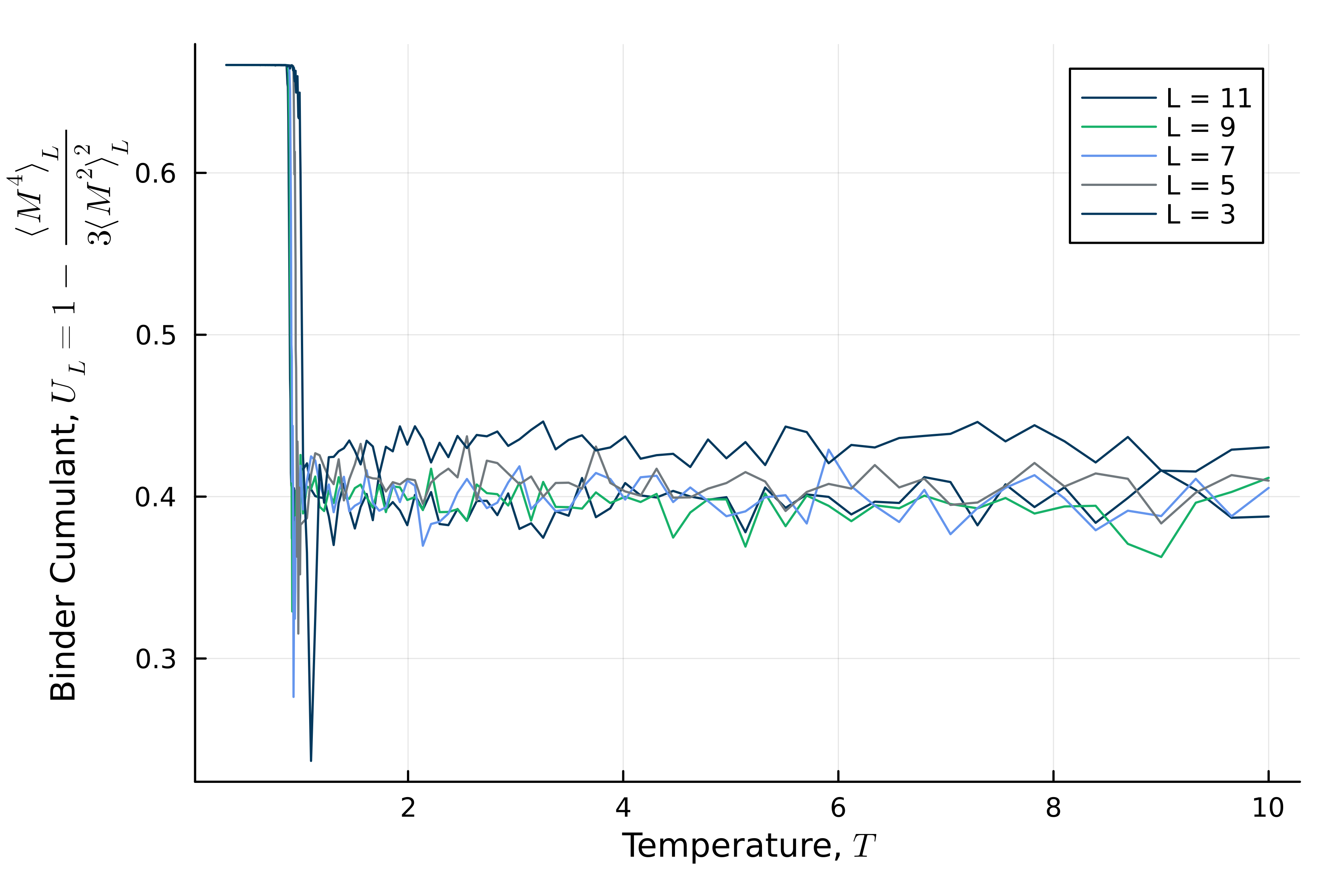}
\includegraphics[width=\linewidth]{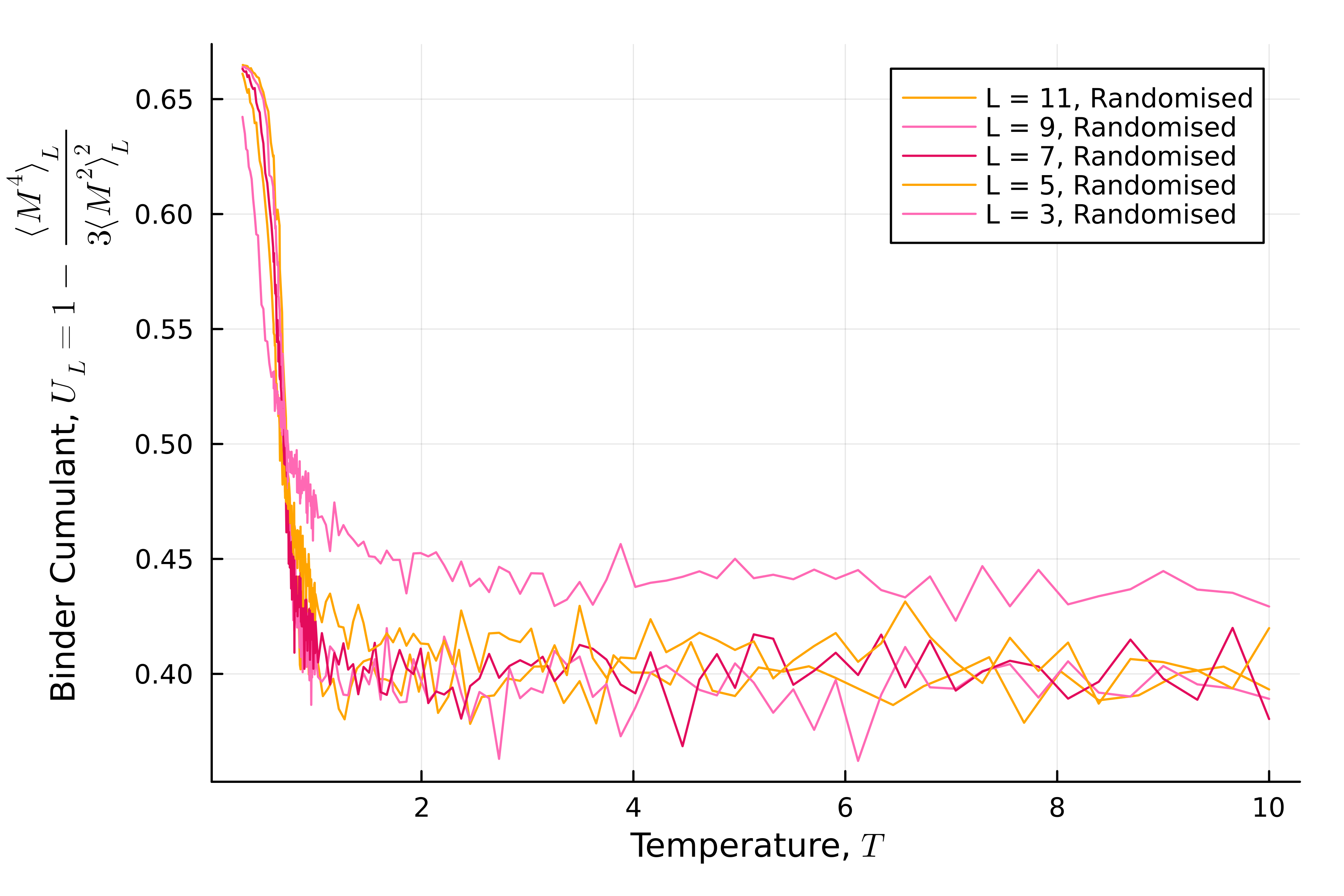}
    \caption{Binder cumulant as a function of temperature for the original RC model (top panel) and randomised RC model (bottom panel), averaged over 50 histories (where, for the randomised cube, each history has a different realisation of disorder). The original RC model shows a sharp decrease in $U_L$ at $T_c$ which becomes sharper with increased $L$, indicative of a first-order phase transition, while the randomised RC model does not.}
    \label{fig:binder-cumulants}
\end{figure}

We define an order parameter for the Rubik's Cube as:
\begin{equation}
    M^2 = \frac16 \sum_f |m_f|^2
    \, ,
\end{equation}
where the individual complex order parameters for each of the six faces, $m_f$, $f=1,\ldots,6$, are defined as
\begin{equation}
    m_f = \frac{1}{L^2} \sum_{i \in f} \exp{\left( \frac{2\pi i}{6} \sigma_{i} \right)}
    \, .
\end{equation}
Such an order parameter is insensitive to which colour orders on which face $f$. In the solved configuration, $\langle |m_f|^2 \rangle = 1 \, , \forall f$, while at infinite temperature $\langle |m_f|^2 \rangle = 0 \, , \forall f$. Hence, $0 \leq M^2 \leq 1$ and $M^2 = 1$ in the solved configuration.
Utilising this order parameter, we calculate the Binder cumulant:
\begin{equation}
    U_L = 1 - \frac{\langle M^4\rangle_L}{3\langle M^2 \rangle^2_L}
    \, , 
\end{equation}
and study its behaviour across the phase transition. 
Fig.~\ref{fig:binder-cumulants} (top panel) shows a sharp decrease in $U_L$ at $T_c$ (associated with a sharp change in order parameter $M$) which becomes sharper with increased $L$. This is consistent with a first-order phase transition, and we observe no indication of the crossing point characteristic of continuous transitions.

If we now contrast the behaviour of the original RC with the behaviour of the randomised RC, we observe that the peak in the heat capacity is significantly broader, as illustrated in Fig.~\ref{fig:heat-capacities} (bottom panel); it does not appear to sharpen for larger system sizes, and its position exhibits a notable drift with system size. Moreover, the sharp decrease in the Binder cumulant $U_L$ at the peak position becomes less pronounced in Fig.~\ref{fig:binder-cumulants} (bottom panel), and importantly no longer exhibits a dip characteristic of first order behaviour (nor does it exhibit a crossing characteristic of continuous transitions). 
For these reasons, we believe that the first-order phase transition is thermodynamically suppressed in the randomised RC model, and the model remains in a paramagnetic phase down to zero temperature. This enables us to probe a broader range of temperatures below $T^*$ where we can equilibrate the system using swap-moves (and use slice-rotations to study the slow dynamics) without undergoing any ordering transition in the system. 
%
%

\section{Uncorrelated limits of the autocorrelation function \texorpdfstring{$\overline C(t)$}{C(t)} in the large \texorpdfstring{$L$}{L} Limit}\label{sm-sec:autocorrelation-function-limits}
We define the autocorrelation function as in the main text: $$
C(t) = \frac{1}{N-6} \sideset{}{'}\sum\limits_{i=1}^{N-6} \delta_{\sigma_i(0) \sigma_i(t)} 
\, ,
$$
where 
the primed summation 
excludes the six central spins, whose values are trivially fixed. 

In the large $L$ limit, we need only consider the bulk of the cube~\footnote{In fact, up to subtle constraints on the orientation of corner pieces and wing edge pieces, the approximations used in this calculation work reasonably well for the cubelets on the boundary too.} (i.e., `centre pieces' as defined in Supp.~Mat.~Note~\ref{sm-sec:swap-moves}), as the number of facelets scales quadratically in the bulk and linearly on the boundary. (Obviously, we exclude the `fixed centres' from the calculation.) 

As we have shown in Ref.~\onlinecite{gower-constraints-of-lxlxl-rubiks-cube}, the centre pieces decompose into subsystems of $24$ facelets which are independent (in the sense that facelets cannot be exchanged between subsystems). The configurations of each such independent subsystem therefore correspond to elements of the $\mathbf{S}_{24}$ permutation group~\footnote{Note that the `First Law of Cubology' (see Supp.~Mat.Note.~\ref{sm-sec:swap-moves}) constrains which combinations of configurations different independent subsystems are able to exist in simultaneously, but in this calculation we can treat each subsystem independently since the autocorrelation function is only a function of a single facelet in a single subsystem.}. Therefore the probability that a given facelet is the same colour in two randomly sampled configurations is equivalent to the probability that a facelet remains the same colour after an arbitrary permutation from $\mathbf{S}_{24}$ is applied to its subsystem of $24$ facelets. 

If there are $X_i$ facelets of a given colour in a subsystem then the probability of a facelet having this colour after an arbitrary permutation is $p^{(i)}_{\rm same} = X_i/24$. Computing the expectation of this value over all $6$ possible colours in the subsystem (which each have $p_{i} = X_i/24$) we obtain:
$$
\langle p_{\rm same} \rangle 
= 
\sum_{i=1}^6 p_{i} p^{(i)}_{\rm same} 
= 6 \left( \frac{X_i}{24} \right) 
    \left( \frac{X_i}{24} \right) 
= 6 \left( \frac{X_i}{24} \right)^2 
\, .
$$
For the original RC we have by definition that $X_i = 4$ (i.e., 4 of the 24 facelets in each subsystem are of each colour) therefore 
$$
C_{\rm uncorrelated} \rightarrow \langle p_{\rm same} \rangle = \frac{1}{6} \sim 0.1667
\, . 
$$
For the randomised RC in the large $L$ limit we can consider the facelets of colour $i$ distributed binomially as $X_i \sim \textup{Bin}(n=24, p=\frac{1}{6})$ (or strictly as independent multinomial variables). Computing the average $\langle p_{same} \rangle$ over this disorder we obtain 
$$
C_{\rm uncorrelated} \rightarrow \overline{\langle p_{\rm same} \rangle} = \left( \frac{6}{24^2} \right) \overline{X_i^2} \sim 0.2014
\, .
$$ 
This is consistent with numerical simulations for large $L$ cubes; indeed we find $C_{\rm uncorrelated} = \overline{\langle C (T=\infty) \rangle} \simeq 0.2013$, numerically averaged over $10^8$ MC steps for the $L=100$ cube. 
%
%

\section{Autocorrelation function fits}\label{sm-sec:autocorrelation-function-fits}
\begin{figure}[ht]
    \centering
     \includegraphics[width=0.5\textwidth]{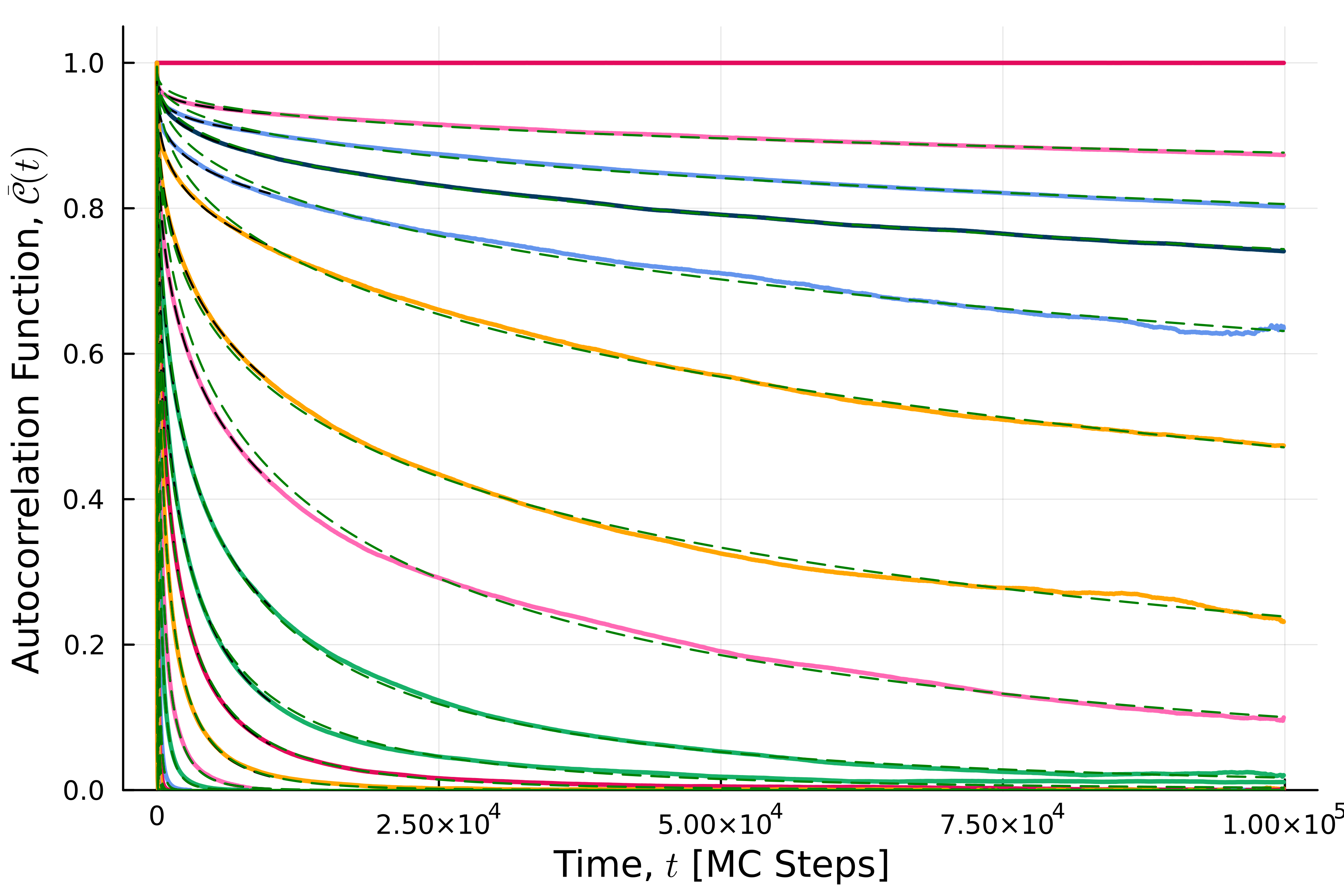}
     \includegraphics[width=0.5\textwidth]{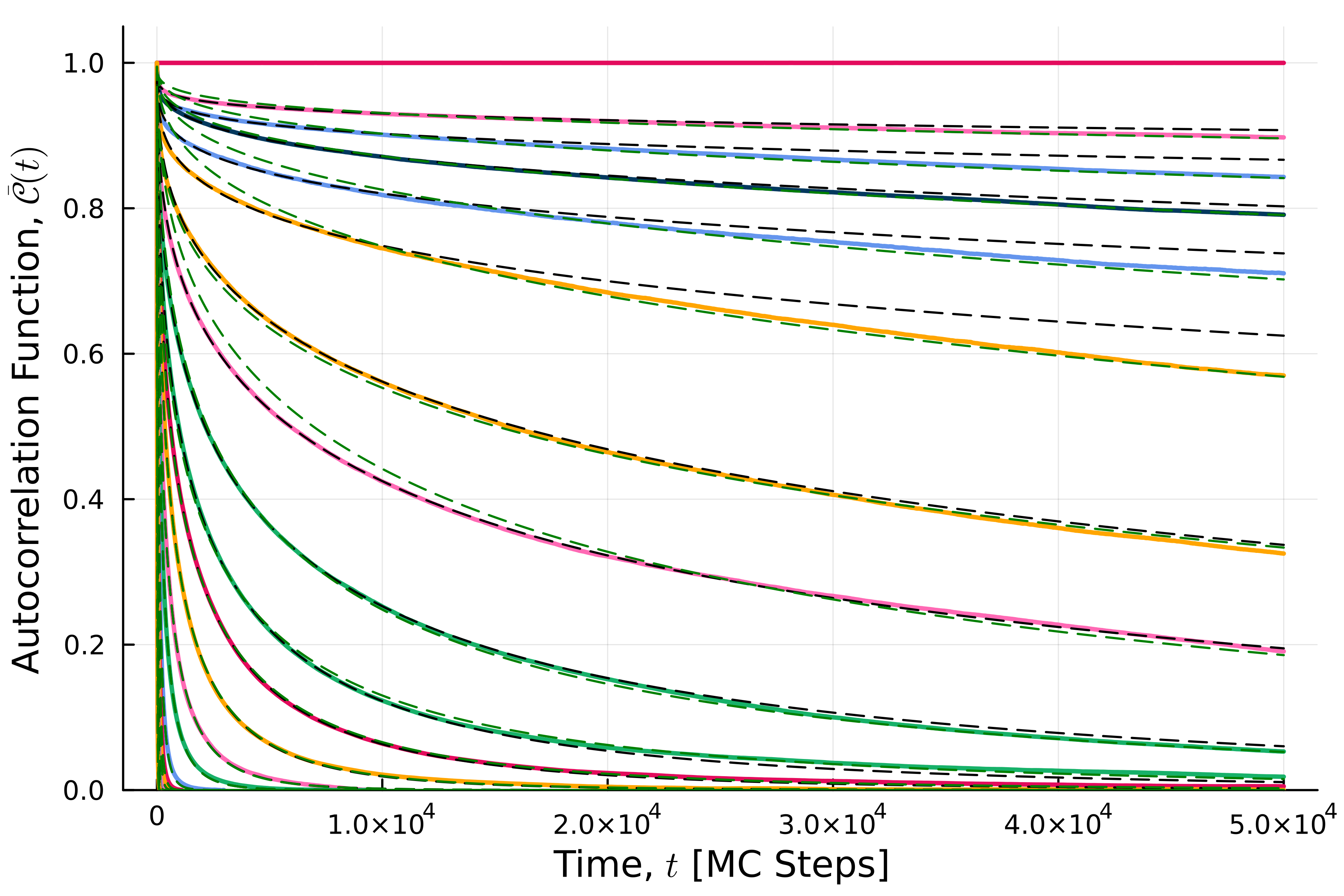}
    \caption{\label{fig:all_autocorr_fits}
    Top panel: Autocorrelation function $\overline{\mathcal{C}}(t)$ plotted for all temperatures used in the main text, for the $L=11$ randomised RC. Black dashed lines display stretched exponential fits up to $t=10^4$, green dashed lines display stretched exponential fits up to $t=10^5$ (as used in each fitting procedure). Bottom panel: Same data on a reduced time range up to $t=5 \times 10^4$ emphasise that $t=10^4$ fits are dominated by shorter time behaviour (and indeed depart when extrapolated to $t=5 \times 10^4$ at low temperatures) and $t=10^5$ fits are dominated by longer time behaviour. Nonetheless, the extracted $\tau(T)$ data is reasonably similar between the two fitting timescales, as seen in the main text and in Fig.~\ref{fig:tau_all_fits}.}
\end{figure}

In Fig.~\ref{fig:all_autocorr_fits}, we show fits to time $t=10^4$ and $t=10^5$ of the autocorrelation function $\overline{\mathcal{C}}(t)$ for all temperatures used to generate Fig.~\ref{fig:relaxation-time-stretching-exponent} in the main text.
These curves were obtained for the $L=11$ randomised RC, where we first anneal and equilibrate the cube at a given temperature using swap-moves and then evolve it for time $t$ using slice-rotations. 
%
%

\section{\texorpdfstring{$\tau(T)$}{tau(T)} fit parameters}\label{sm-sec:tau-fit-parameters}
In this section we present the temperature dependence of the relaxation time $\tau(T)$ obtained from single stretched exponential fits, as in Fig.~\ref{fig:relaxation-time-stretching-exponent} in the main text, but now with stretched exponentials fits up to $t=10^4$ and up to $t=10^5$ (the former being dominated by shorter time behaviour, the latter by longer time behaviour within the accessible time window of our simulations) in separate panels. 
\begin{figure}[ht]
    \centering
    \includegraphics[width=0.45\textwidth]{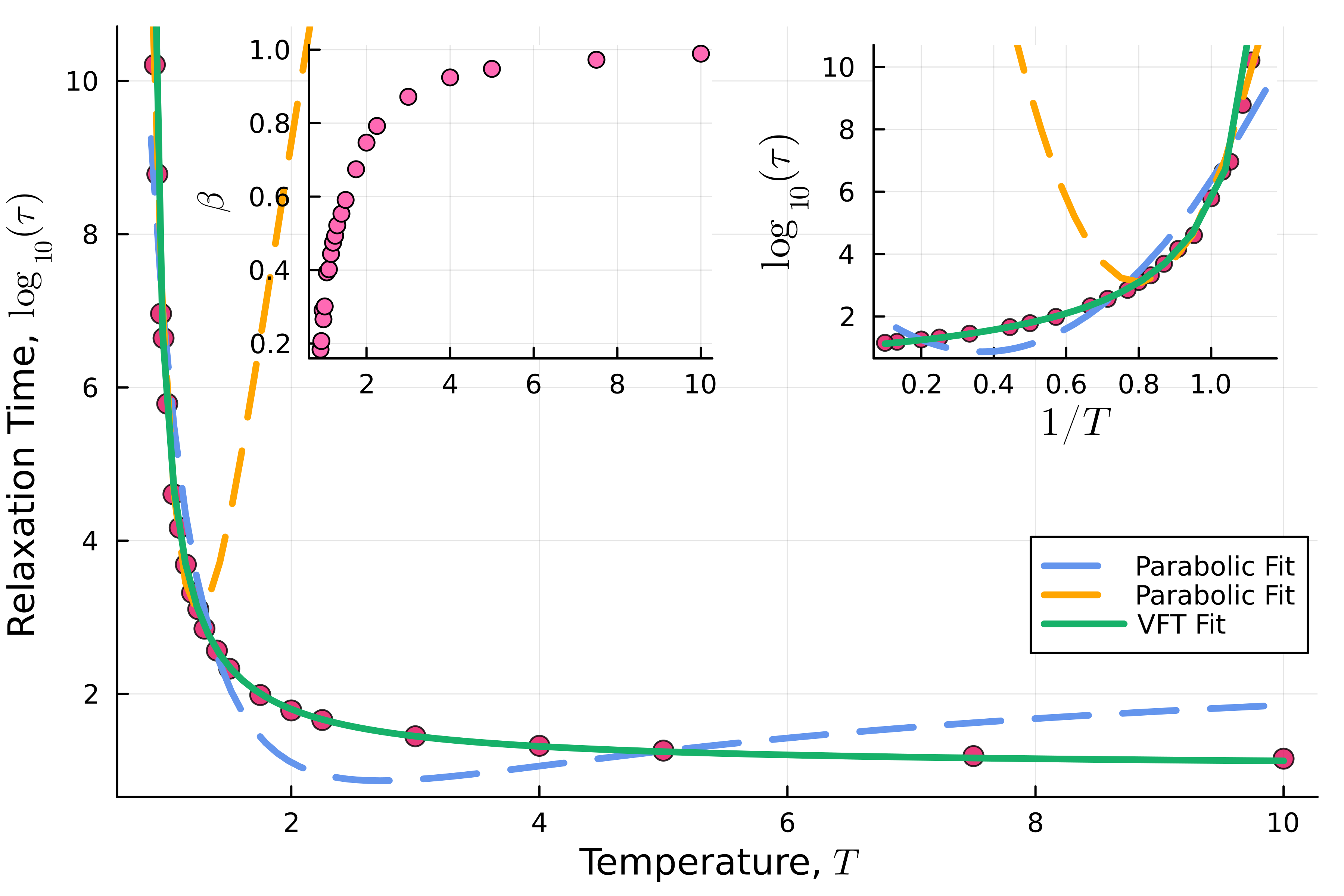}
    \\
    \includegraphics[width=0.45\textwidth]{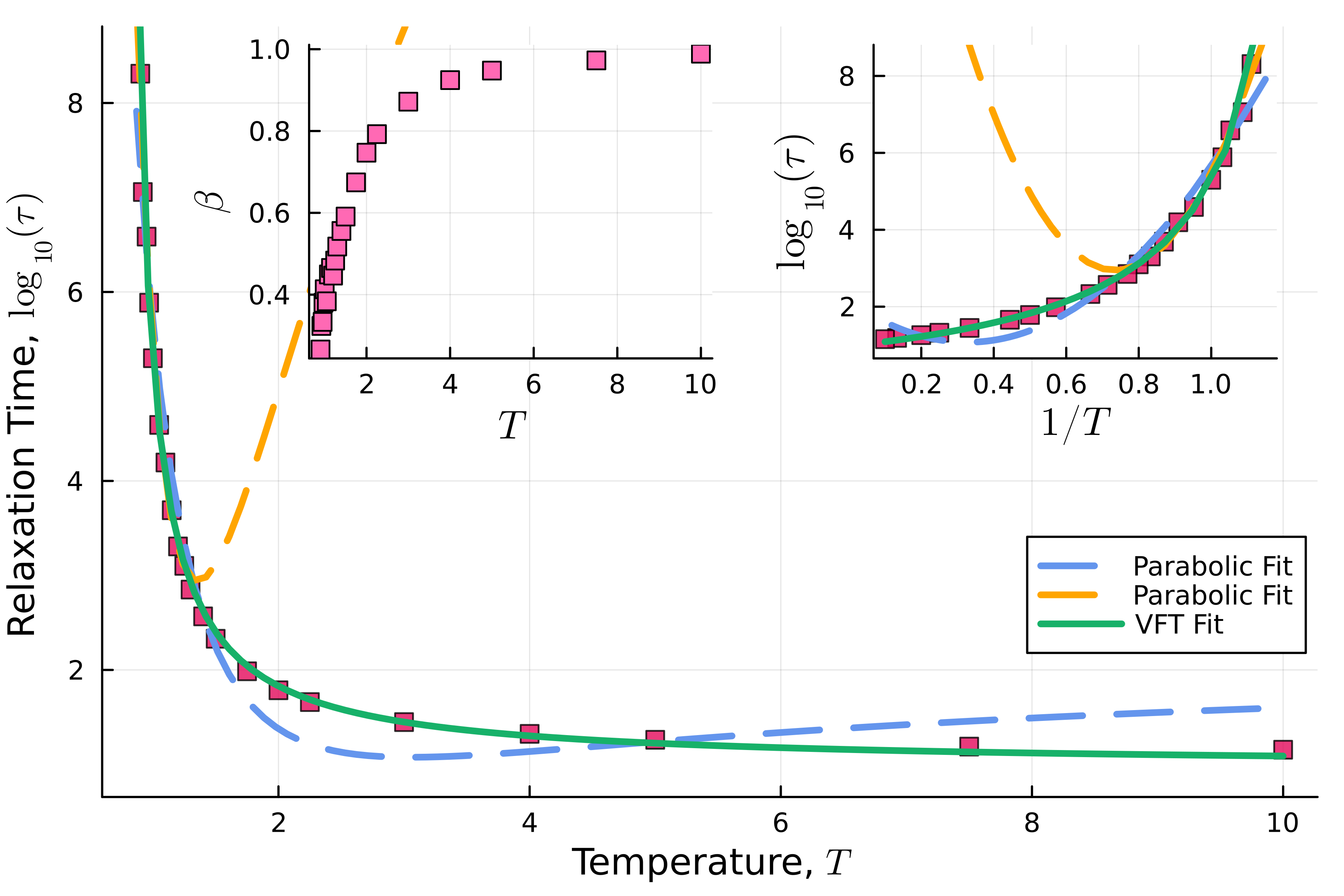}
    \caption{\label{fig:tau_all_fits}
    Temperature dependence of the relaxation time $\tau(T)$ obtained from stretched exponential fits up to $t=10^4$ (top panel, circles) and up to $t=10^5$ (bottom panel, squares) of the autocorrelation function data illustrated in Fig.~\ref{fig:autocorrelation-function} in the main text, for the $L=11$ randomised RC. Fits are presented for the VFT functional form (green solid line) and for the parabolic functional form (blue and orange dashed lines), as discussed in the main text.  Left insets: Temperature dependence of the stretching exponent $\beta(T)$. Right insets: Same $\tau(T)$ data as in the main panel, plotted against $1/T$, to highlight the clear departure from Arrhenius behaviour.}
\end{figure}

In Tables~\ref{table:vft},~\ref{table:pb1}, and~\ref{table:pb2}, we list the VFT and parabolic fit parameters which minimise the least-squares error for the $\tau(T)$ data. In each table, we provide optimal fit parameters when considering $\tau(T)$ data extracted from: $t=10^4$ stretched exponential fits (circles in Fig.~\ref{fig:tau_all_fits}), $t=10^5$ stretched exponential fits (squares in Fig.~\ref{fig:tau_all_fits}), and their combination (as presented in Fig.~\ref{fig:relaxation-time-stretching-exponent} in the main text).

\begin{table}[ht!]
\centering
\begin{tabular}{|c|c|c|c|}
\hline
 & $A$ & $\Delta$ & $T_0$ \\ \hline
$t=10^4$ & 1.02 & 0.93 & 0.80 \\ \hline
$t=10^5$ & 0.97 & 1.07 & 0.75 \\ \hline
Combined & \VFTA & \VFTDelta & \VFTT \\ \hline
\end{tabular}
\caption{\label{table:vft}
Vogel-Fulcher-Tammann (VFT): $A \exp[\Delta / (T-T_0)]$, (green line in Fig.~\ref{fig:tau_all_fits}).}
\end{table}

\begin{table}[ht!]
\centering
\begin{tabular}{|c|c|c|c|}
\hline
 & $A$ & $\Delta$ & $\Delta'$ \\ \hline
$t=10^4$ & 2.73 & -10.10 & 13.73 \\ \hline
$t=10^5$ & 2.16 & -6.63 & 10.12 \\ \hline
Combined & 2.44 & -8.37 & 11.92\\ \hline
\end{tabular}
\caption{\label{table:pb1}
Parabolic fit: $A \exp[ \Delta / T + \Delta' / T^2]$ (blue dashed line in Fig.~\ref{fig:tau_all_fits}).}
\end{table}

\begin{table}[ht!]
\centering
\begin{tabular}{|c|c|c|c|}
\hline
 & $A$ & $\Delta$ & $\Delta'$ \\ \hline
$t=10^4$ & 46.81 & -109.50 & 68.55 \\ \hline
$t=10^5$ & 22.43 & -52.91 & 35.91 \\ \hline
Combined & 34.62 & -81.20 & 52.23 \\ \hline
\end{tabular}
\caption{\label{table:pb2}
Parabolic fit, including only $T \leq 4$: $A \exp[ \Delta / T + \Delta' / T^2]$ (yellow dashed line in Fig.~\ref{fig:tau_all_fits}).}
\end{table}
%
%

\section{\texorpdfstring{$K=1$}{K=1} Configurations}\label{sm-sec:K=1_configurations}
In the main text, we note that $K\geq 2$ and not $K=1$ configurations are identified as generic saddles. 
This classification scheme is consistent with those used for systems with continuous degrees of freedom in Refs.~\onlinecite{broderixEnergyLandscapeLennardJones2000,angelaniSaddlesEnergyLandscape2000,cavagnaFragileVsStrong2001,grigeraGeometricApproachDynamic2002,coslovichLocalizationTransitionUnderlies2019}. There the saddle index is defined as the number of negative eigenvalues of the Hessian at a stationary point; since each negative eigenvalue is associated with an eigenvector corresponding to two energy-decreasing directions in configuration space, stationary points with a single energy-decreasing direction have a minimal eigenvalue of $\lambda =0$ and are therefore classed as minima instead of saddles in this scheme. 
The equivalent of a $K=1$ configuration in systems with continuous degrees of freedom is a stationary point where the energy-decreasing directions coincidentally lead, via different paths, to the same stationary point of lower energy. This is indeed a possibility that was observed in earlier studies (A.Cavagna, private communication), and may be worth exploring further in the future. 
%
%

\section{Growth of number of configurations with energy} 
\label{sm-sec:exponential-growth-of-configurations-with-energy}
\begin{figure}[ht]
    \centering
    \includegraphics[width=\linewidth]{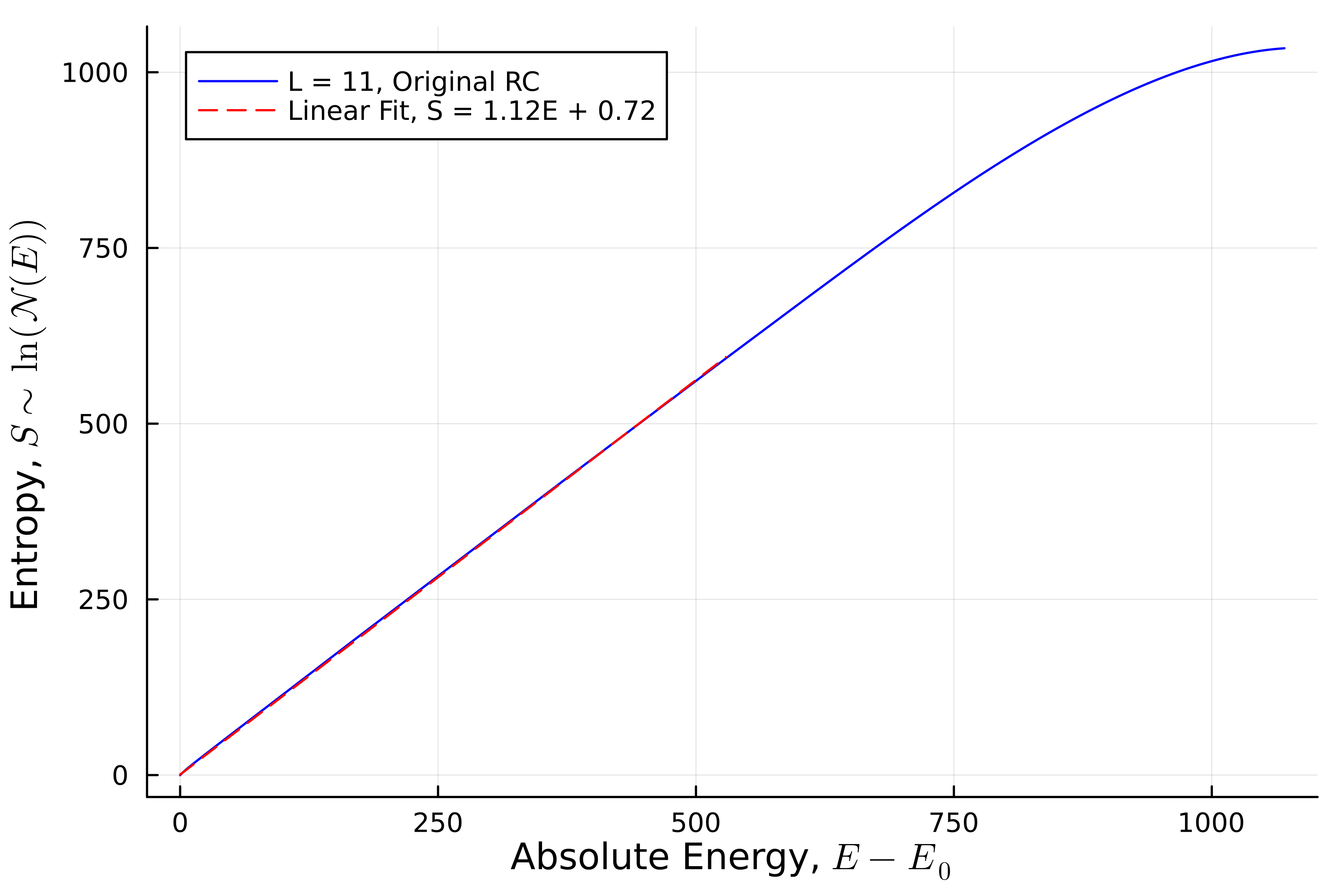}
\includegraphics[width=\linewidth]{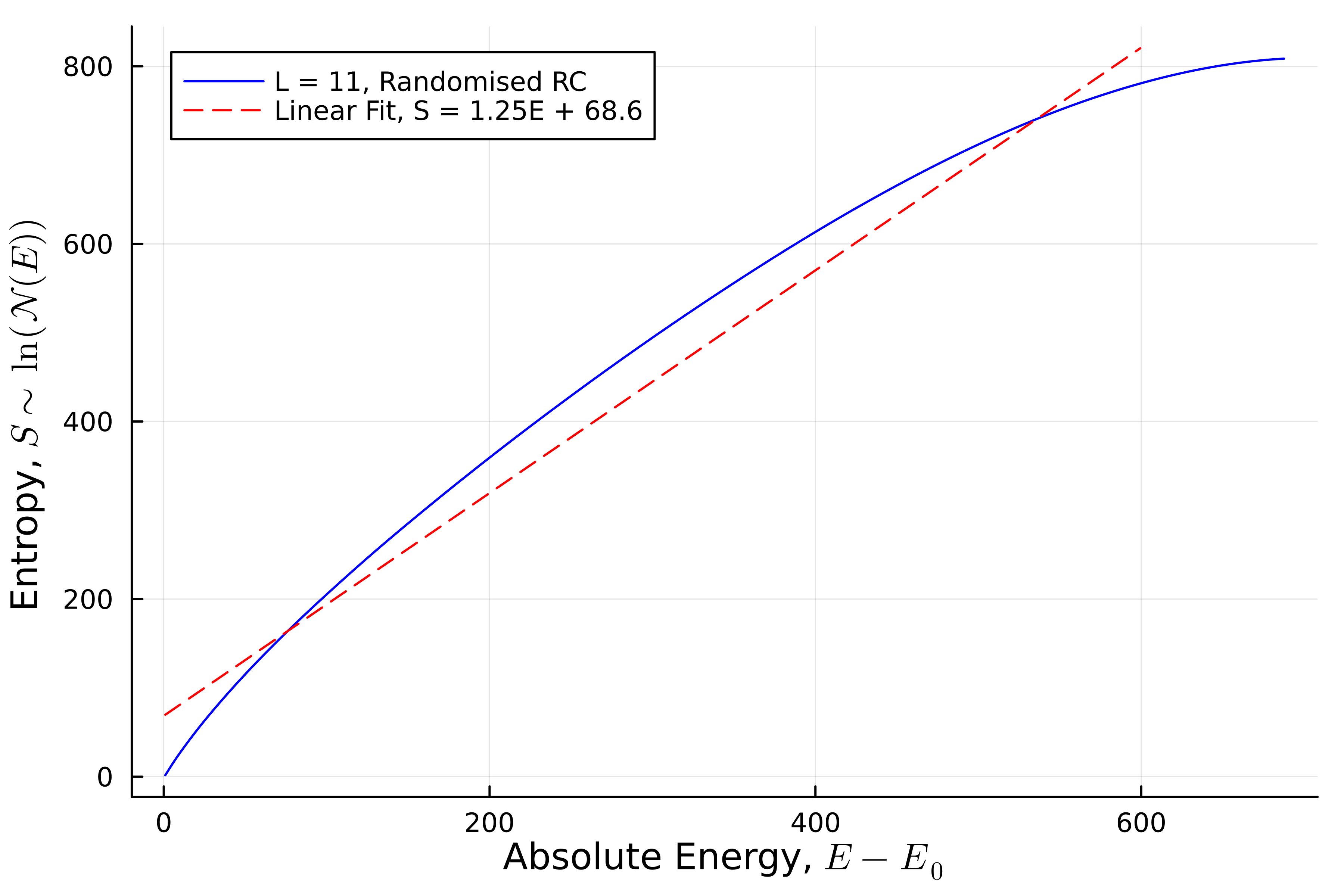}
    \caption{Entropy vs energy curves for the original $L=11$ RC model (top panel) and randomised RC model (bottom panel). Linear fits (dashed red lines) are included as a guide to the eye.}
    \label{fig:entropy-against-energy}
\end{figure}

In this section we calculate the configurational entropy as a function of energy, $S(E)$, in order to estimate the growth of RC configurations with energy $\mathcal{N}(E) \sim e^{S(E)}$. 
We first invert the energy vs temperature curves $E(T)$ shown in Fig.~\ref{fig:relaxed-anneals} to obtain $T(E)$ and then employ the following relation to calculate the entropy:
\[
S(E) = \int_0^E \frac{1}{T(E')} dE' 
\, .
\]
The $S(E)$ curves for both the original RC and randomised RC models are displayed in Fig.~\ref{fig:entropy-against-energy}. Linear fits with their associated parameters are added as a guide to the eye, and their gradients ($\eta \approx 1.12$ for the original RC and $\eta \approx 1.25$ for the randomised RC) can be used as a rough estimate for the  growth of RC configurations with energy  as $\mathcal{N}(E) \sim e^{\eta E}$. 


\section{Growth of number of swap-moves with \texorpdfstring{$L$}{L}}\label{sm-sec:growth-of-swap-moves-with-L}
\begin{figure}[ht]
    \centering
    \includegraphics[width=0.5\textwidth]{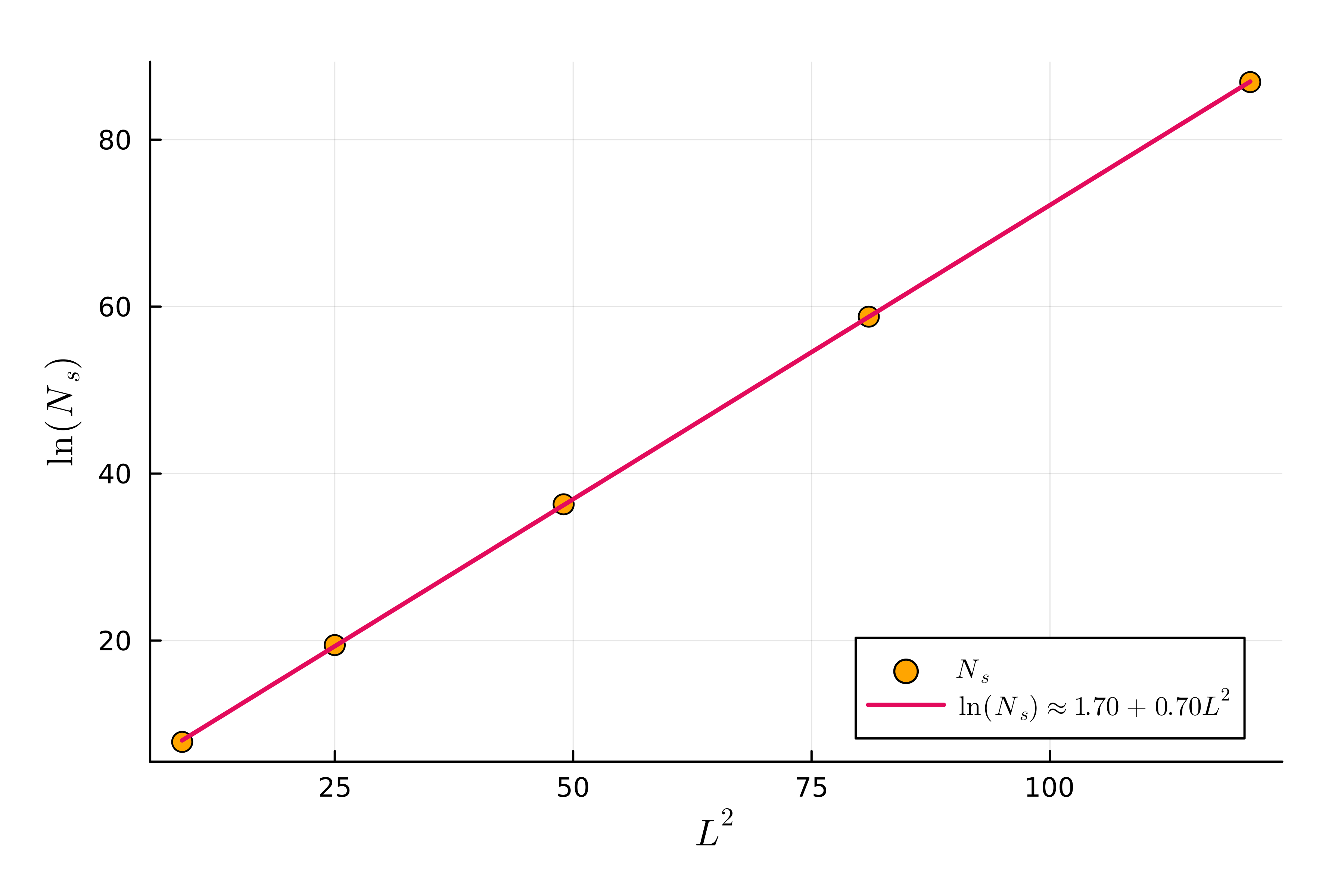}
    \caption{\label{fig:growth-of-swap-moves-with-L}
    Growth of number of swap-moves, $N_s$, against number of facelets, $N=L^2$. The linear growth of $\ln(N_s)$ with $L^2$ demonstrates that $N_s \sim e^{0.70L^2}$}
\end{figure}

\begin{figure}[ht]
    \centering
    \includegraphics[width=0.5\textwidth]{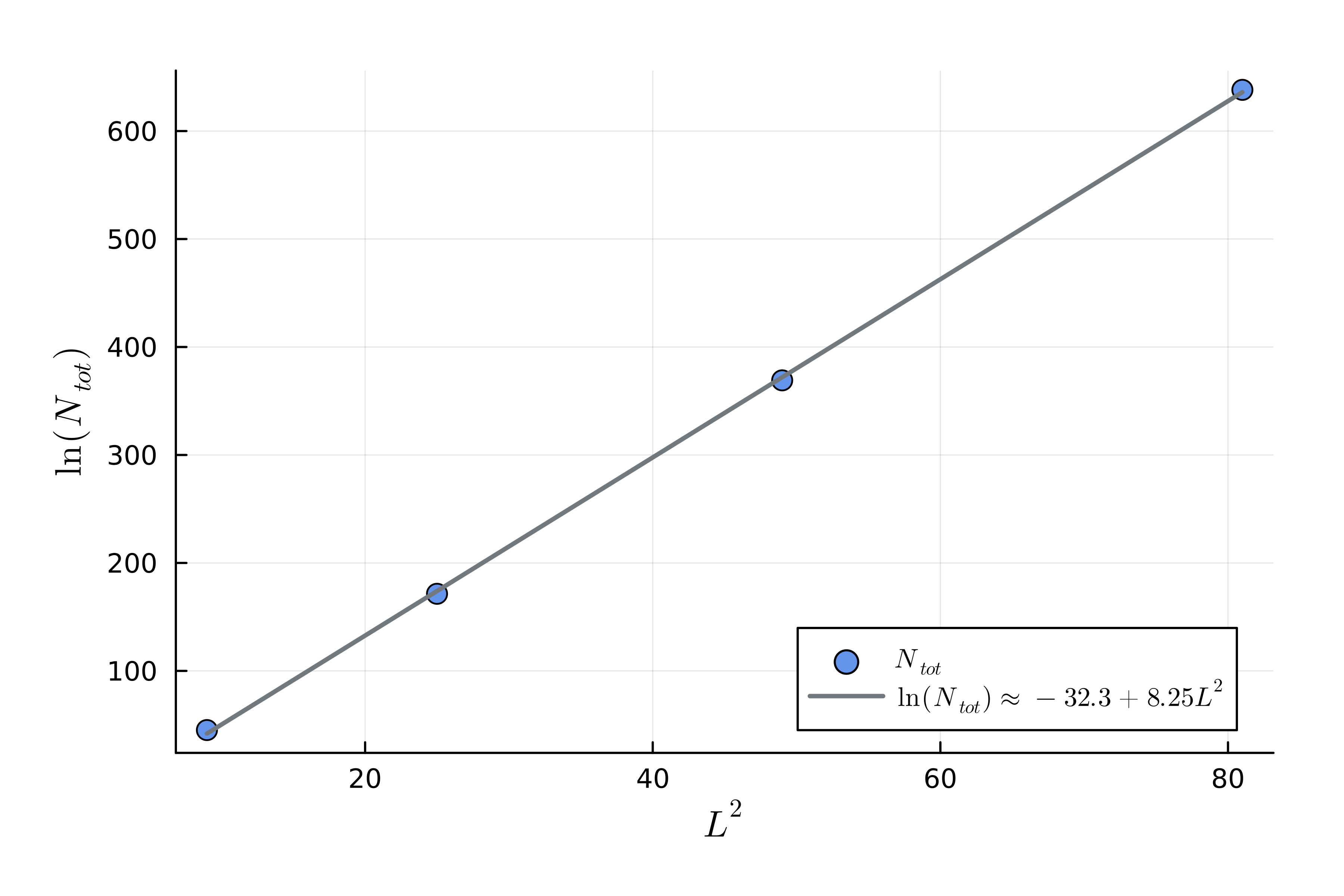}
    \caption{\label{fig:growth-of-total-configurations-with-L}
    Growth of the number of configurations, $N_{\rm tot}$, against the number of facelets, $N=L^2$. The linear growth of $\ln(N_{\rm tot})$ with $L^2$ demonstrates that $N_{\rm tot} \sim e^{8.25L^2}$.}
\end{figure}

In Fig.~\ref{fig:growth-of-swap-moves-with-L} we show that the number of swap-moves per configuration grows exponentially with the number of facelets $N=L^2$, i.e., $N_s \sim e^{\zeta L^2}$. We also show in Fig.~\ref{fig:growth-of-total-configurations-with-L} that the total number of configurations in the configuration network also grows as $N_{\rm tot} \sim e^{\kappa L^2}$, but with $\kappa \gg \zeta$ such that $N_s/N_{\rm tot} \rightarrow 0$ as $L \rightarrow \infty$. Therefore, even though the number of swap-moves per configuration is many orders of magnitude greater than the number of slice-rotations for a cube of given $L$, it is still negligible compared to the total number of configurations in the network. For convenience, we also display the raw data in Tables.~\ref{tab:number-of-swap-moves},~\ref{tab:number-of-configurations} and \ref{tab:ratio-swap-moves-to-configurations}. As in the rest of the paper, we consider only cubes of odd $L$.

It should be noted that the exponential growth of the swap-move connectivity is purely from the Class 3 swap-moves (Class 1 swap-moves scale as $O(L^2)$ and Class 2 scale as $O(1)$) and, as mentioned in Supp.~Mat.Note~\ref{sm-sec:swap-moves}, this formulation of swap-moves is by no means unique; in this work we adopted the simplest option of including all swap-moves defined previously, but we speculate that a smaller subset of the swap-moves with a different connectivity scaling may be equally efficient at thermalising the RC within a reasonable computation cost. An investigation of such sufficient subsets is left for future work. 

\begin{table}[ht]
\centering
\begin{tabular}{|c|c|}
\hline
$L$ & Number of Swap-Moves, $N_S$ \\
\hline
1 &  0 \\
3 &  2,588 \\
5 &  281,635,556 \\
7 &  5,940,848,633,430,244 \\
9 &  34,350,321,361,181,406,176,576,276 \\
11 & 55,012,222,938,456,767,359,870,599,949,634,752,100 \\
\hline
\end{tabular}
\caption{Scaling of the number of swap-moves, $N_S$, with $L$.}
\label{tab:number-of-swap-moves}
\end{table}

\begin{table}[ht]
\centering
\begin{tabular}{|c|c|}
\hline
$L$ & Number of Configurations $N_\text{tot}$\cite{total-number-of-ruibks-cube-configuraitons} \\
\hline
1 & 1 \\
3 & 43,252,003,274,489,856,000 \\
5 & $2.8 \times 10^{74}$ \\
7 & $2.0 \times 10^{160}$ \\
9 & $1.4 \times 10^{277}$\\
11 & $1.1 \times 10^{425}$ \\
\hline
\end{tabular}
\caption{Scaling of the total number of configurations, $N_\text{tot}$, with $L$.}
\label{tab:number-of-configurations}
\end{table}

\begin{table}[ht]
\centering
\begin{tabular}{|c|c|}
\hline
$L$ & $N_S/N_\text{tot}$ \\
\hline
1 & 0 \\
3 & $6.0 \times 10^{-17}$ \\
5 & $1.0 \times 10^{-66}$ \\
7 & $3.0 \times 10^{-145}$ \\
9 & $2.4 \times 10^{-252}$ \\
11 & $5.1 \times 10^{-388}$ \\
\hline
\end{tabular}
\caption{Ratio of the number of swap-moves to the total number of configurations, $N_S/N_\text{tot}$, with $L$.}
\label{tab:ratio-swap-moves-to-configurations}
\end{table}
%
%

\section{Swap-move effective saddle index proportions}
\label{sm-sec:swap-move-saddle-index-proportions}
\begin{figure}[ht]
    \centering
    \includegraphics[width=\linewidth]{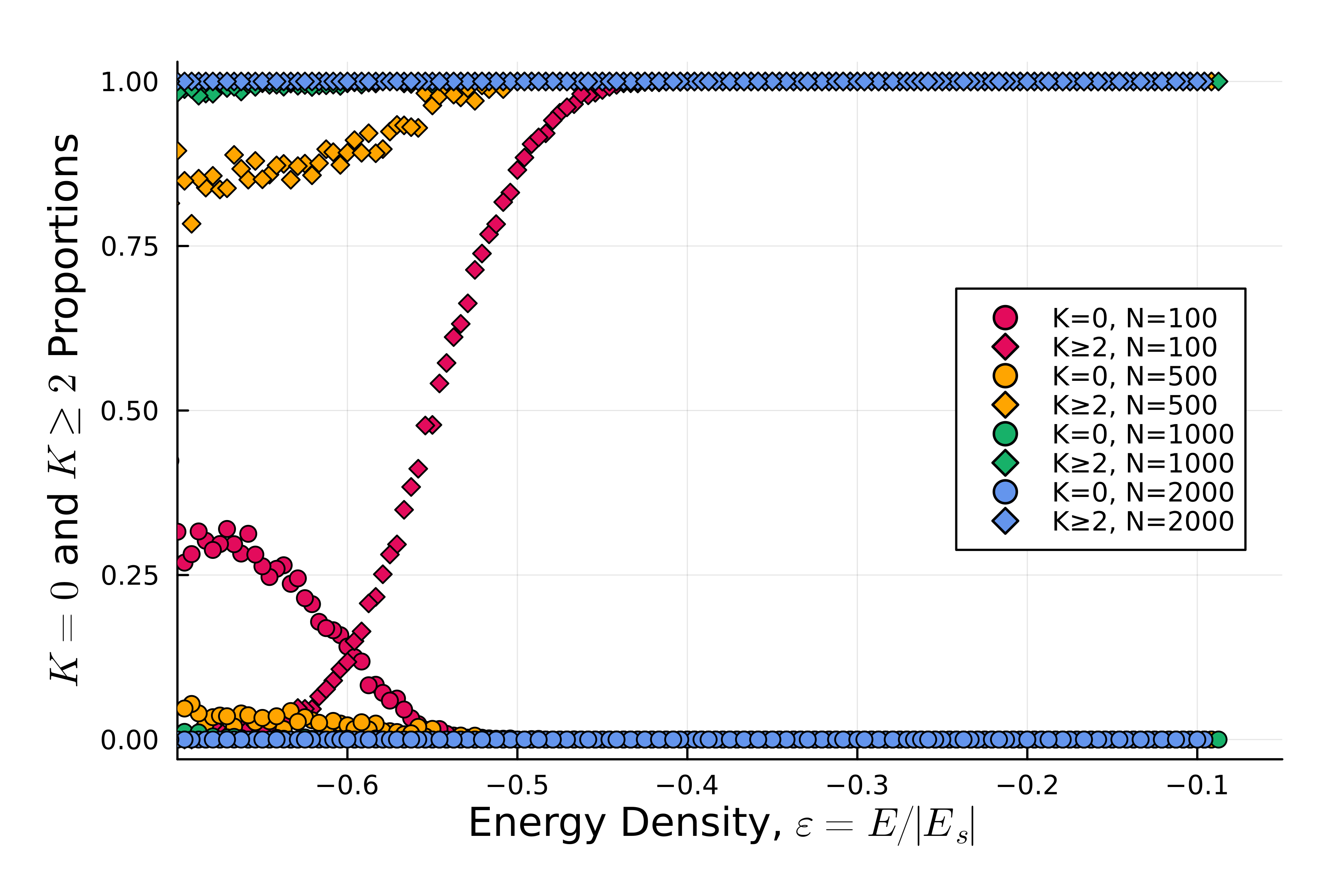}
    \caption{Effective $K$ when sampling $n$ random swap-move neighbours per stochastically sampled configuration of the $L=5$ randomised cube, plotted as a proportion $K/n$.}
    \label{fig:saddle-index-proportions-swap}
\end{figure}

As for slice-rotations in Fig.~\ref{fig:saddle-index-proportions} in the main text, we plot in Fig.~\ref{fig:saddle-index-proportions-swap} the `effective' saddle index proportions for the swap-move configuration network of the $L=5$ randomised cube~\footnote{We used $L=5$ instead of $L=11$ to illustrate this point for ease of computational resources, but the same behaviour is expected to hold for the $L=11$ system.}. 
By `effective' we mean that here $K$ represents the number of energy-decreasing connections in a sample of $n$ random swap-moves applied to each equilibrium $\epsilon^{(0)}$ configuration. Since we are not able to exhaustively sample all swap-move connections for each configuration ($z \sim 10^8 \,\, [\sim 10^{37}]$ for the swap-move configuration network of the $L=5 \,\, [L=11]$ cube), this $K$ does not reflect the true saddle indices of the configurations: the value of $K$ for each sampled configuration increases monotonically with an increased sample size $n$ of swap-move connections per configuration. For sufficiently large $n$, we seem to find that all configurations have $K \geq 2$, and therefore $\overline\epsilon_\times = 0$ and $\overline\epsilon_{\otimes} = 0$ for swap-moves. A study of the scaling behaviour with $n$ (and $L$) to confirm this observation is left for future work. 
%
%

\section{Typical single-move energy density changes from equilibrium configurations}
\label{sm-sec:boltzmann-factor-reweighting}
By re-weighting the histogram probabilities $p(\epsilon^{(0)} \rightarrow \epsilon^{(1)})$, which are displayed in Fig.~\ref{fig:E0-E1-connectivities} of the main text, by the relevant Boltzmann factor 
\begin{equation*}
    \mathcal{P}(\epsilon^{(0)} \rightarrow \epsilon^{(1)}) = \frac{1}{Z} p(\epsilon^{(0)} \rightarrow \epsilon^{(1)}) \min\left\{e^{\frac{\epsilon^{(1)}-\epsilon^{(0)}}{T(\epsilon^{(0)})}}, 1 \right\}
    \, ,
\end{equation*}
where $Z$ is chosen such that $\int \mathcal{P}(\epsilon^{(0)} \rightarrow \epsilon^{(1)}) d\epsilon^{(1)} = 1$, we can calculate the typical energy density change of a single move for the system initialised at an equilibrium configuration of energy density $\epsilon^{(0)}$ (corresponding to temperature $T(\epsilon^{(0)})$), 
\begin{equation*}
    \llangle\Delta\epsilon\rrangle (\epsilon^{(0)}) = \int (\epsilon^{(1)}-\epsilon^{(0)})P(\epsilon^{(0)} \to \epsilon^{(1)}) d\epsilon^{(1)} \, .
\end{equation*}

\begin{figure}[ht]
    \centering
    \includegraphics[width=\linewidth]{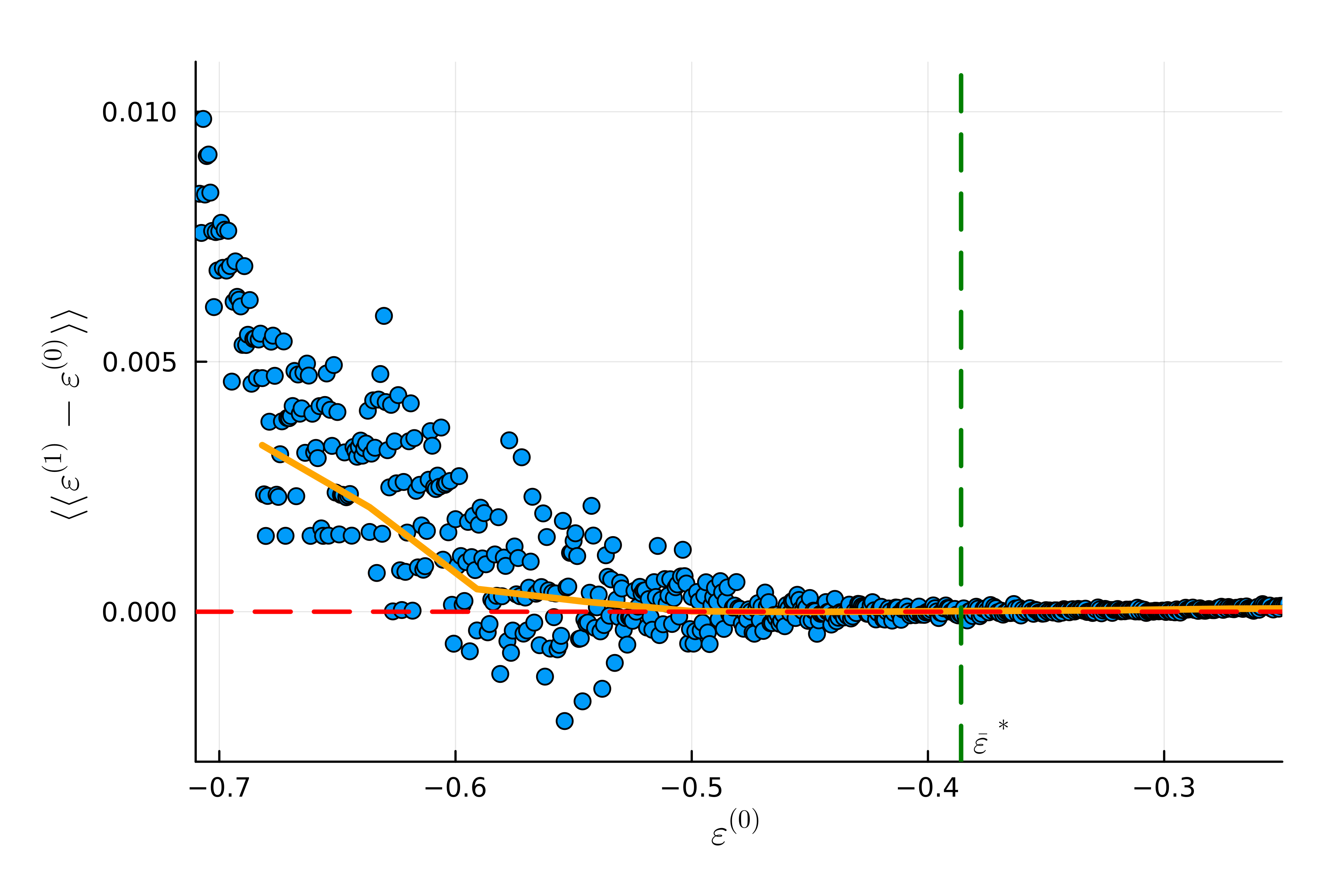}
    \includegraphics[width=\linewidth]{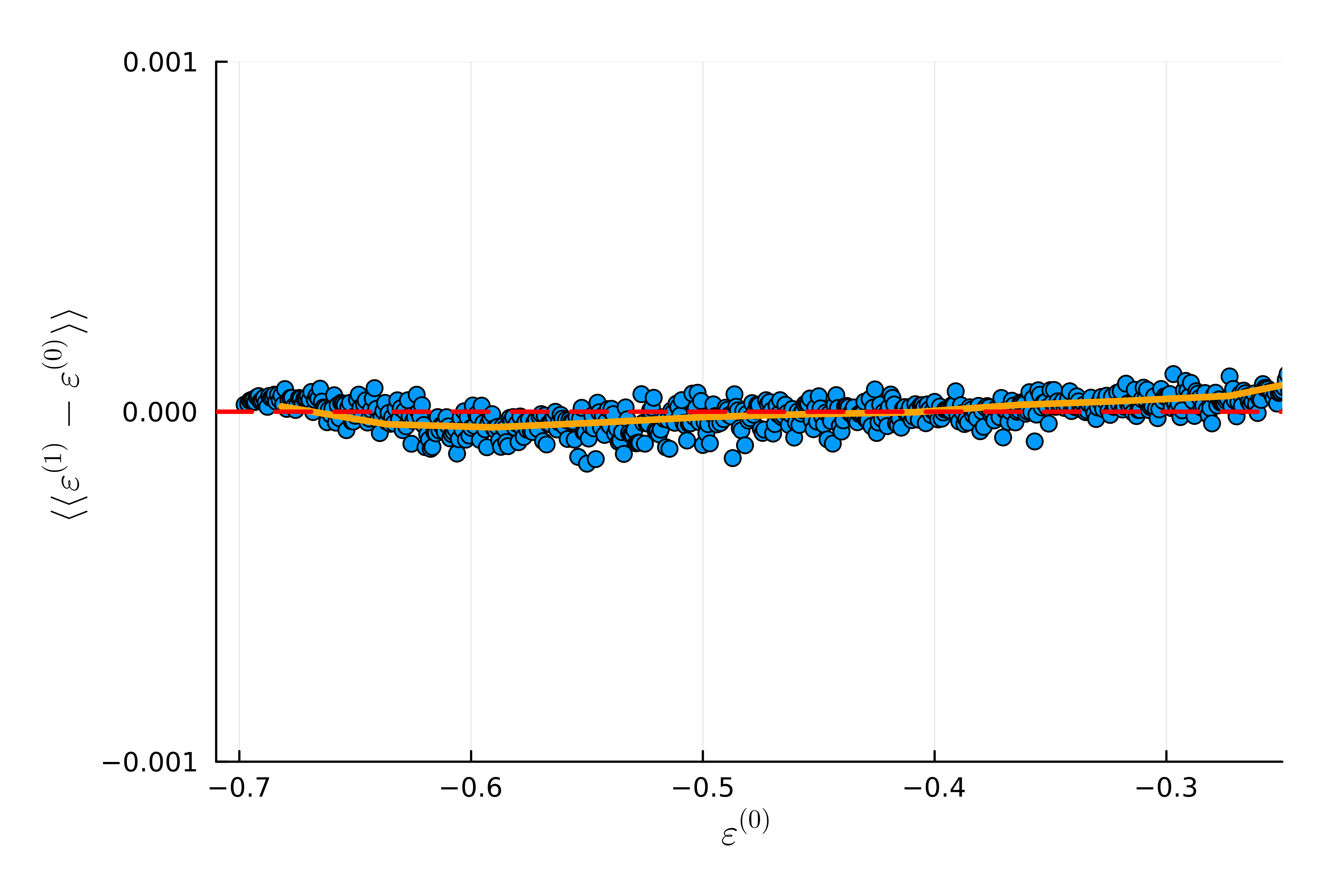}
    \caption{Typical energy density changes $\llangle \Delta \epsilon \rrangle$ of a single move from an equilibrium configuration for both slice-rotations (top panel) and swap-moves (bottom panel). A smoothed average over a $\Delta \epsilon = 0.05$ window is overlaid in orange, as a guide to the eye. The slice-rotation behaviour is dominated by an entropic drive towards energy-increasing connections, which is absent for swap-moves.}
    \label{fig:boltzmann-factor-reweighting}
\end{figure}

As displayed in the insets of Fig.~\ref{fig:boltzmann-factor-reweighting}, we observe that $ \llangle \Delta \epsilon \rrangle \sim 0$ for swap-moves for all energy densities and for slice-rotations at high energy densities, but $\llangle \Delta \epsilon \rrangle > 0$ for slice-rotations at low energy densities, thereby demonstrating that the entropic drive towards energy-increasing connections dominates over the exponential energetic bias towards energy-decreasing connections in this regime. 
%
%
%

\section{Predicted modal energy density costs, \texorpdfstring{$\epsilon^{(1)} - \epsilon^{(0)}$}{epsilon(1) - epsilon(0)}, for slice-rotations}
\label{sm-sec:predicted-energy-density-costs}
The modal neighbouring energy density differences of the slice-rotation cube, $\epsilon^{(1)} - \epsilon^{(0)}$, displayed in Fig.~\ref{fig:E0-E1-connectivities} of the main text, can be well fit by the following analytic arguments.

\begin{figure}[ht]
    \centering
    \includegraphics[width=\linewidth]{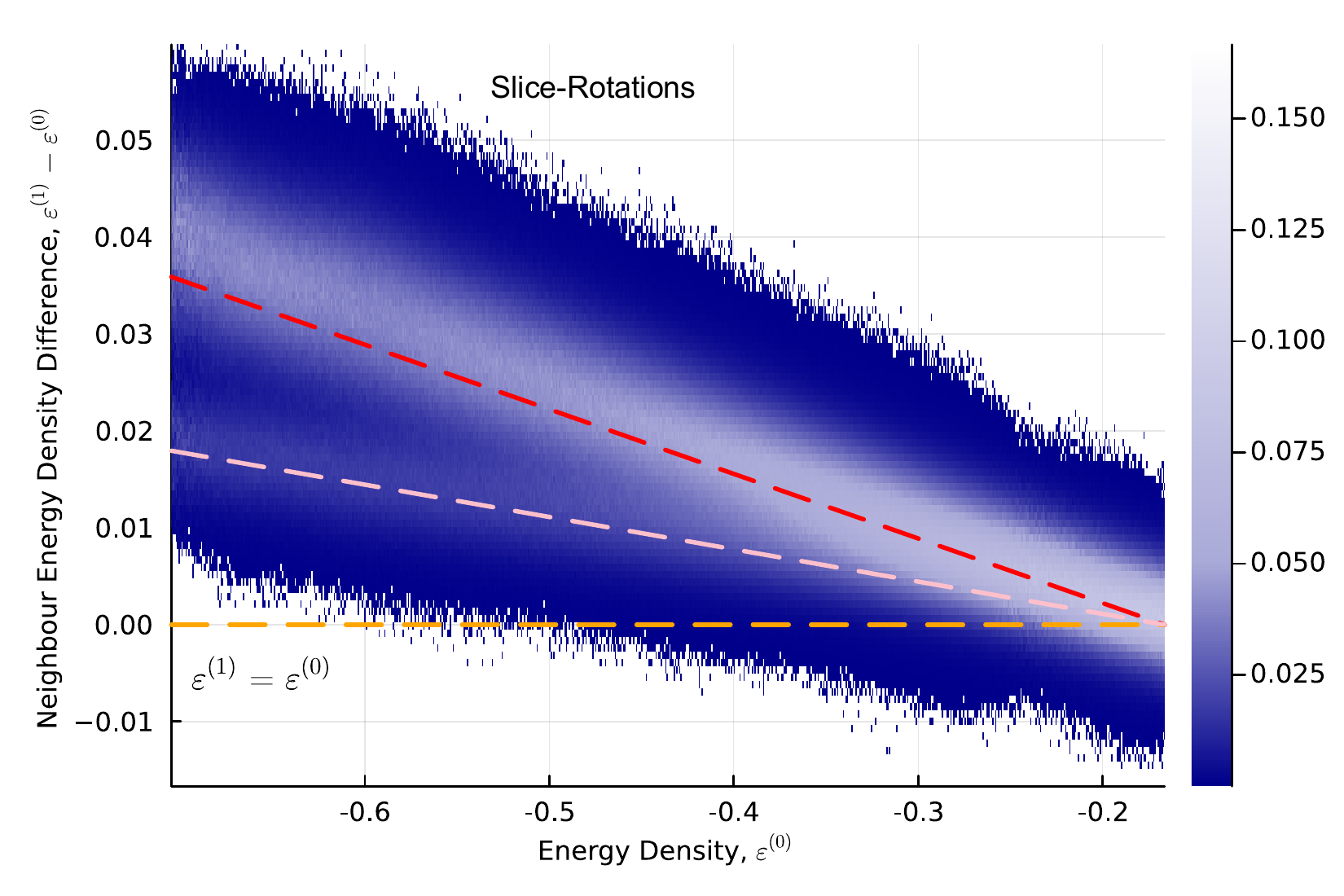}
    \caption{Histograms of the difference $\epsilon^{(1)} - \epsilon^{(0)}$ between energy densities, where 
    $\epsilon^{(0)}$ corresponds to stochastically sampled $L=11$ randomised RC configurations in equilibrium (using swap-move MC) and $\epsilon^{(1)}$ corresponds to the energy density obtained from each such configuration by applying single slice-rotations, as in Fig.~\ref{fig:E0-E1-connectivities} in the main text. The superimposed lines correspond to the predicted energy density costs of an internal layer slice-rotation, $8L(|\epsilon| - 1/6)/|E_s|$ (red), and an external layer slice-rotation, $4L(|\epsilon| - 1/6)/|E_s|$ (pink).}
    \label{fig:prediction-E0-E1-connectivities}
\end{figure}

A single internal layer slice-rotation disrupts $2 \times 4 \times L = 8L$ bonds. If a configuration has energy density $\epsilon = E/|E_s|$ then, by definition, a proportion $|\epsilon|$ of the bonds are satisfied and typically $8L |\epsilon|$ bonds will be broken by the slice-rotation. Neglecting correlations, we can naively assume that $\sim 1/6$ of these bonds will be satisfied in the new configuration by chance. Therefore the net energy cost of such slice-rotation is $+8L|\epsilon| - 8L/6 = 8L (|\epsilon| - 1/6)$, or equivalently the net energy density cost is $8L(|\epsilon| - 1/6)/|E_s|$. This line is plotted in red in Fig.~\ref{fig:prediction-E0-E1-connectivities}. 

By analogous arguments, since external layer slice-rotations disrupt $4L$ bonds instead of $8L$, their predicted energy density cost is $4L(|\epsilon| - 1/6)/|E_s|$. This line is plotted in pink in Fig.~\ref{fig:prediction-E0-E1-connectivities}. 
Both lines are in good agreement with the features observed in the histograms. 
%
%

\section{Histograms of energy density differences \texorpdfstring{$\epsilon^{(2)} - \epsilon^{(1)}$}{epsilon(2) - epsilon(1)} from two slice-rotations}
\label{sm-sec:E2-E1-connectivities}
\begin{figure}[ht!]
    \centering
    \includegraphics[width=0.5\textwidth]{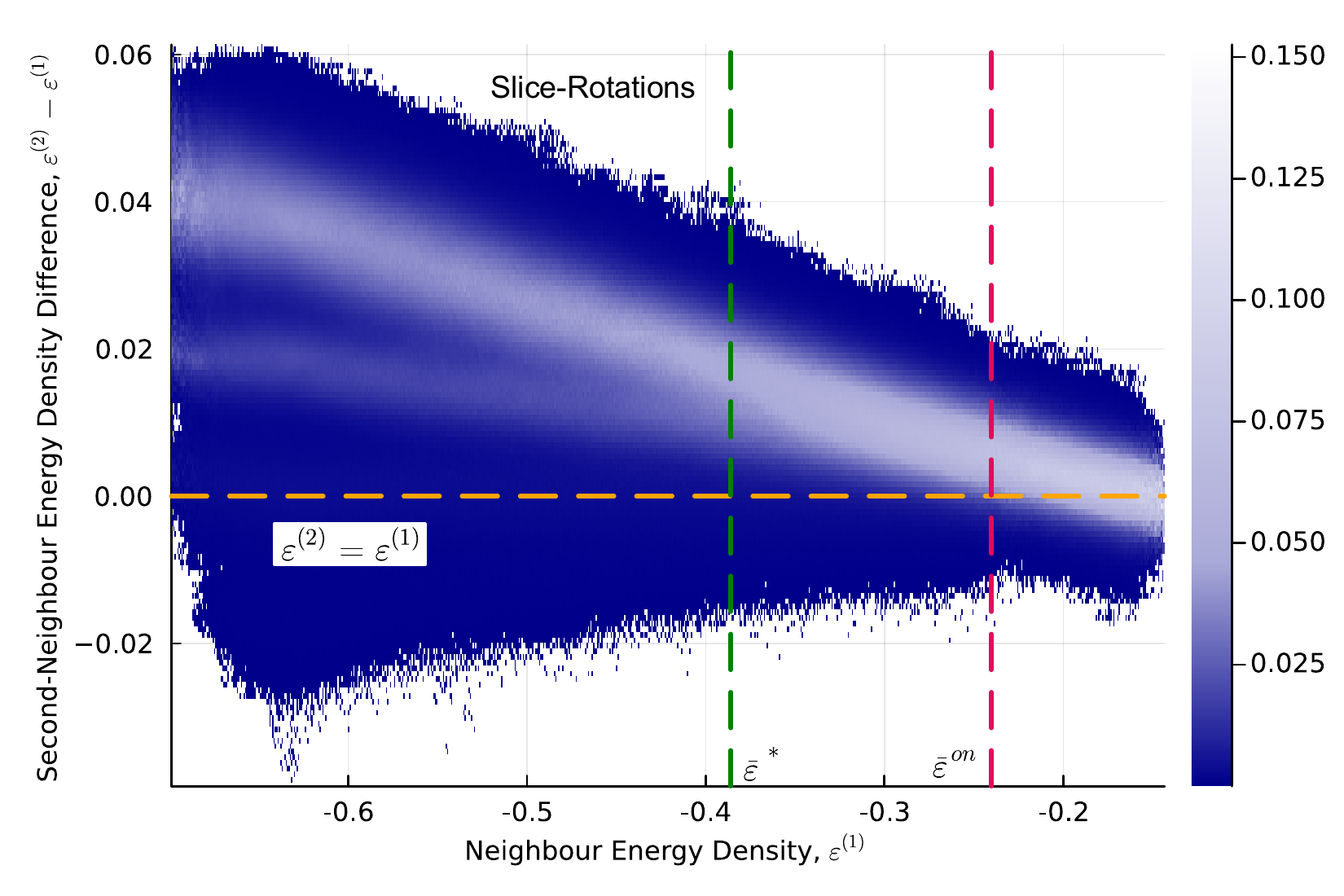}
    \caption{Histograms of the difference $\epsilon^{(2)} - \epsilon^{(1)}$ between energy densities where $\epsilon^{(2)}$ is obtained by applying single slice-rotations to the $L=11$ RC configurations of energy density $\epsilon^{(1)}$ discussed in Fig.~\ref{fig:E0-E1-connectivities} of the main text (top panel). The histograms are normalised independently for each $\epsilon^{(1)}$. By construction, these configurations have one previously-known neighbouring configuration of energy density $\epsilon^{(0)}$, which at low energies is typically $< \epsilon^{(1)}$ and we explicitly exclude it from the histograms. We clearly see that the histograms span across the $\epsilon^{(2)} = \epsilon^{(1)}$ line at all energy densities, indicating that we find a finite density of saddles with $K\geq2$ even for $\epsilon^{(1)} \lesssim \overline\epsilon^*$ (green dashed vertical line). For comparison, we also indicate the value of $\overline\epsilon^{\rm on}$ (red dashed vertical line).}
    \label{fig:E1-E2-connectivities}
\end{figure}

It is informative to look at configurations one further step away from those with energy $\epsilon^{(0)}$, i.e., building histograms of the energy densities $\epsilon^{(2)}$ of configurations obtained by applying single slice-rotations to the configurations with energy density $\epsilon^{(1)}$ in the study discussed above. The resulting histograms are shown in Fig.~\ref{fig:E1-E2-connectivities}.
Contrary to the top panel of Fig.~\ref{fig:E0-E1-connectivities} of the main text, we do not observe any departure of the histograms from the $\epsilon^{(2)} = \epsilon^{(1)}$ line. Given that we explicitly exclude the trivial $\epsilon^{(2)} = \epsilon^{(0)}$ configuration (obtained upon applying one slice-rotation and its inverse), 
this behaviour implies the existence of a finite density of saddles with $K\geq2$ even for $\epsilon^{(1)} < \overline\epsilon^*$, as in Refs.~\onlinecite{doyeSaddlePointsDynamics2002, fabriciusDistanceInherentStructures2002}, thereby indicating that the sampled configurations with energy density $\epsilon^{(1)}$ are atypical compared to the thermodynamic equilibrium ensemble. 
%
%

\section{Finite size scaling collapse of the proportion of saddles with \texorpdfstring{$K \geq 2$}{K >= 2}.}
The dashed lines in Fig.~\ref{fig:combined-saddle-index-proportions-finite-size-scaling} of the main text demonstrate that the proportion of stochastically sampled configurations with saddle index $K \geq 2$ can be well fit by generalised logistic functions of the form $p_{K\geq 2}(\epsilon) = 1 - (1 + e^{\alpha(\epsilon - \overline\epsilon_{\rm log})})^{-\gamma}$, where $\alpha$ scales linearly with $L$.

\begin{figure}[ht]
    \centering
    \includegraphics[width=\linewidth]{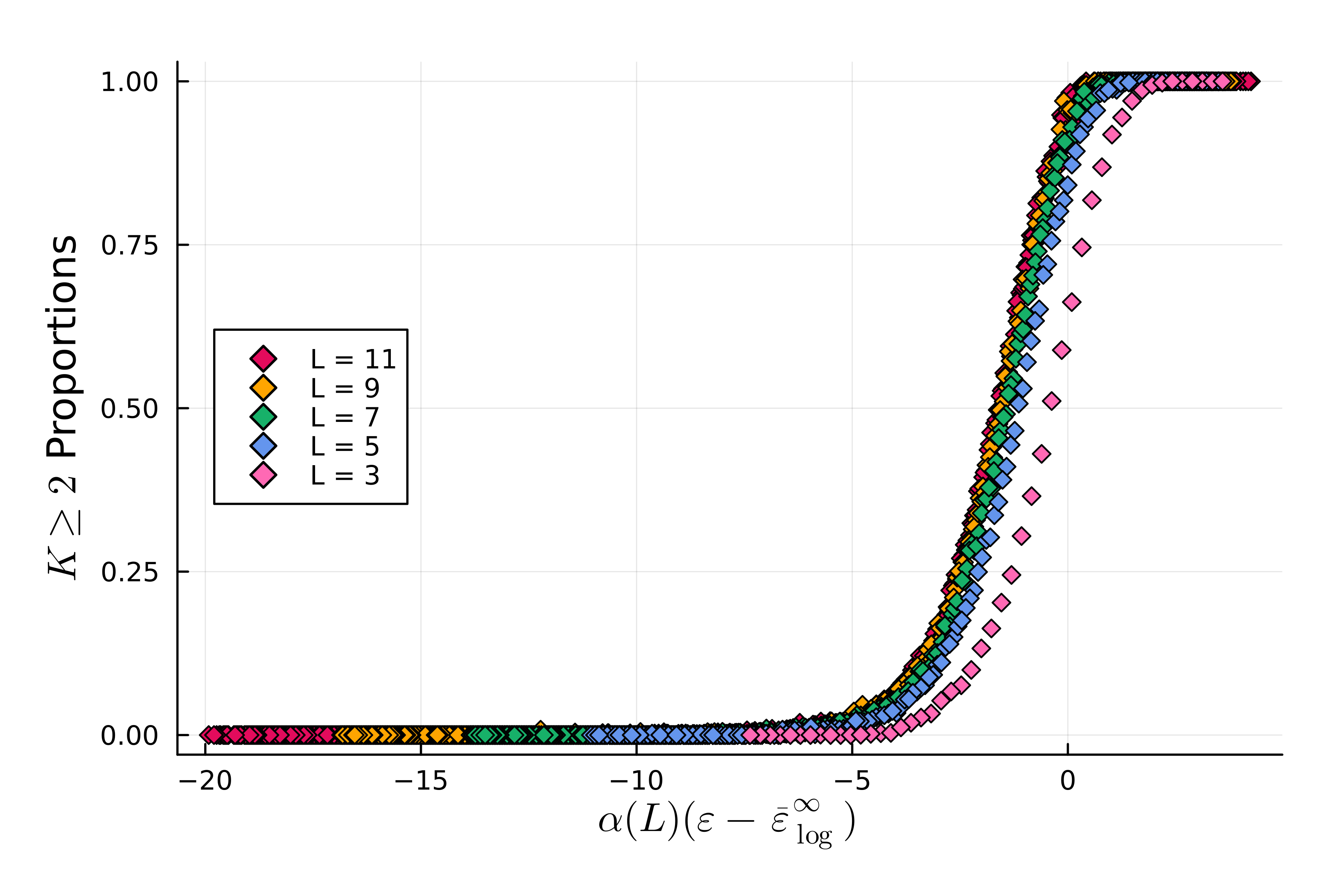}
    \caption{Proportion of stochastically sampled configurations with saddle index $K \geq 2$ for varying system size $L$,  as in Fig.~\ref{fig:combined-saddle-index-proportions-finite-size-scaling} of the main text, but plotted as a function of $\alpha(L)(\epsilon - \overline\epsilon_{\rm log}^{\infty})$ instead of energy density $\epsilon$. The rescaling results in a collapse onto a universal curve.}
    \label{fig:collapsed-saddle-index-proportions}
\end{figure}

In Fig.~\ref{fig:collapsed-saddle-index-proportions} we plot the same data as a function of $\alpha(L)(\epsilon - \overline\epsilon_{\rm log}^{\infty})$ to illustrate their scaling collapse. 
We use the function $\alpha(L) \simeq 7.93 + 2.95L$ (taken from the linear fit presented in Fig.~\ref{fig:combined-saddle-index-proportions-finite-size-scaling}) and $\overline\epsilon_{\rm log}^{\infty} \simeq -0.24$ (taken from the $\frac{1}{L} \rightarrow 0$ limit of the linear fit also presented in Fig.~\ref{fig:combined-saddle-index-proportions-finite-size-scaling}), and observe a good collapse onto a universal curve.
%
%
%

\section{Histograms of energy density differences \texorpdfstring{$\epsilon^{(2)} - \epsilon^{(0)}$}{epsilon(2) - epsilon(0)} from two slice-rotations}
\label{sm-sec:E2-E0-connectivities}
The histograms in Fig.~\ref{fig:E0-E1-connectivities} of the main text depict an exponential suppression of energy-decreasing slice-rotation connections for configurations below the topological crossover (which occurs at an energy density near $\overline\epsilon^*$). However, Fig.~\ref{fig:E1-E2-connectivities} in the main text shows that configurations which are connected by a single slice-rotation to equilibrium configurations lying below the crossover energy are highly atypical, with a surplus of energy-decreasing connections. 

\begin{figure}[ht]
    \centering
    \includegraphics[width=\linewidth]{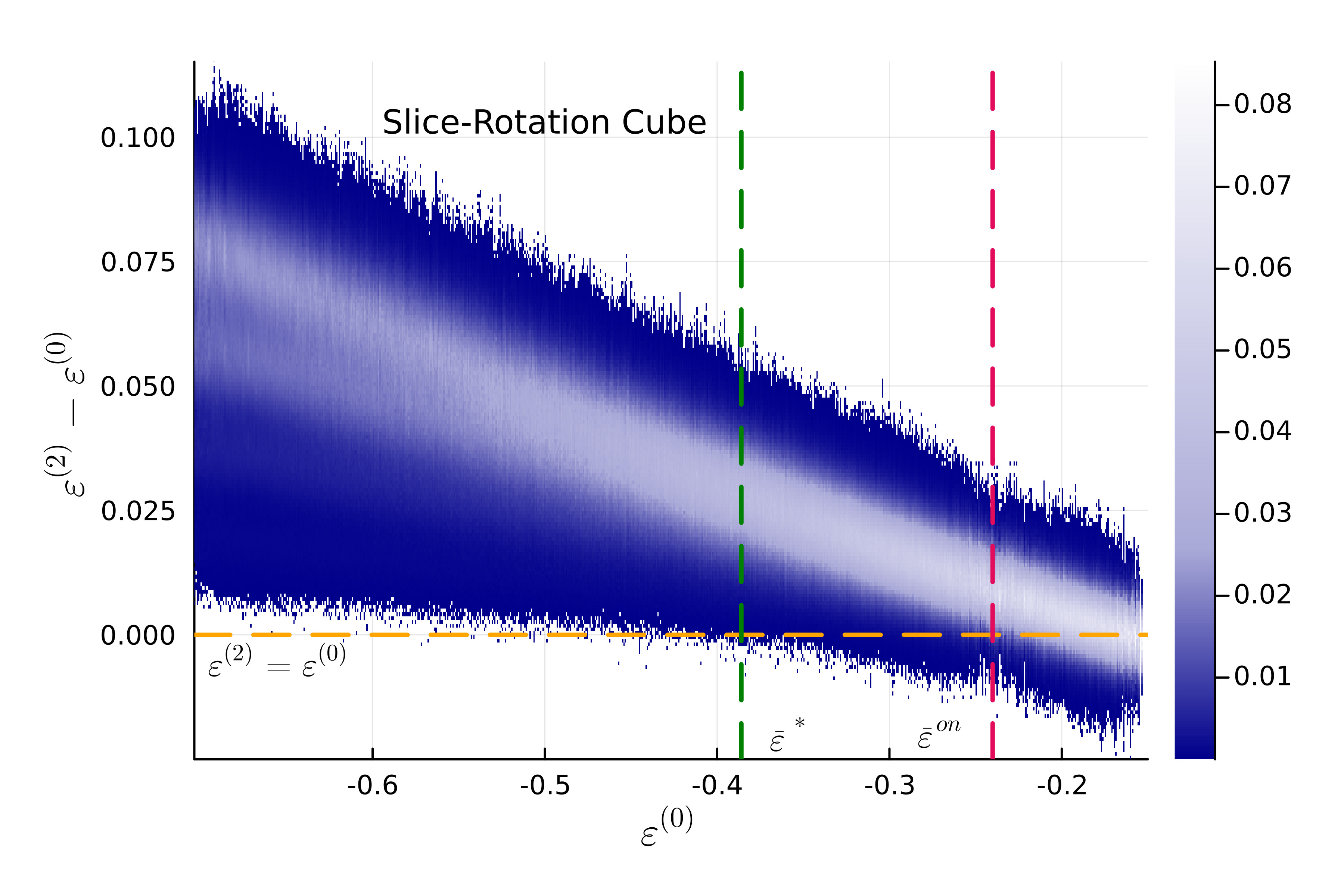}
    \caption{Histograms of the difference $\epsilon^{(2)} - \epsilon^{(0)}$ between energy densities, where 
    $\epsilon^{(0)}$ corresponds to stochastically sampled $L=11$ randomised RC configurations in equilibrium (using swap-move MC) and $\epsilon^{(2)}$ corresponds to the energy density obtained from each such configuration by applying two slice-rotations.}
    \label{fig:E0-E2-connectivities}
\end{figure}

In Fig.~\ref{fig:E0-E2-connectivities} we plot, for completeness, the histograms corresponding to the net energy density differences $\epsilon^{(2)} - \epsilon^{(0)}$ between equilibrium configurations of energy density $\epsilon^{(0)}$ (sampled stochastically using swap-moves in equilibrium), and configurations which are connected to them by two slice-rotations. 
(Namely, this is a combination of the data in Fig.~\ref{fig:E0-E1-connectivities} and Fig.~\ref{fig:E1-E2-connectivities} in the main text.) 
The overall behaviour is similar to that observed in Fig.~\ref{fig:E0-E1-connectivities}, thus demonstrating that a study of the one-slice-rotation connectivity of the system is in fact informative of the behaviour after multiple slice-rotations. 
Once again, we find a crossover at $\overline\epsilon^*$ below which energy-decreasing routes produced by two slice-rotations become rare. We also see that the modal energy-density differences from two slice-rotations are larger by a factor of $2$ compared to those from single slice-rotations.
%
%

\section{Modal energy differences and equilibrium temperature for swap-moves and slice-rotations}
\label{sm-sec:modal-energy-differences}
\begin{figure}[ht]
    \centering
    \includegraphics[width=\linewidth]{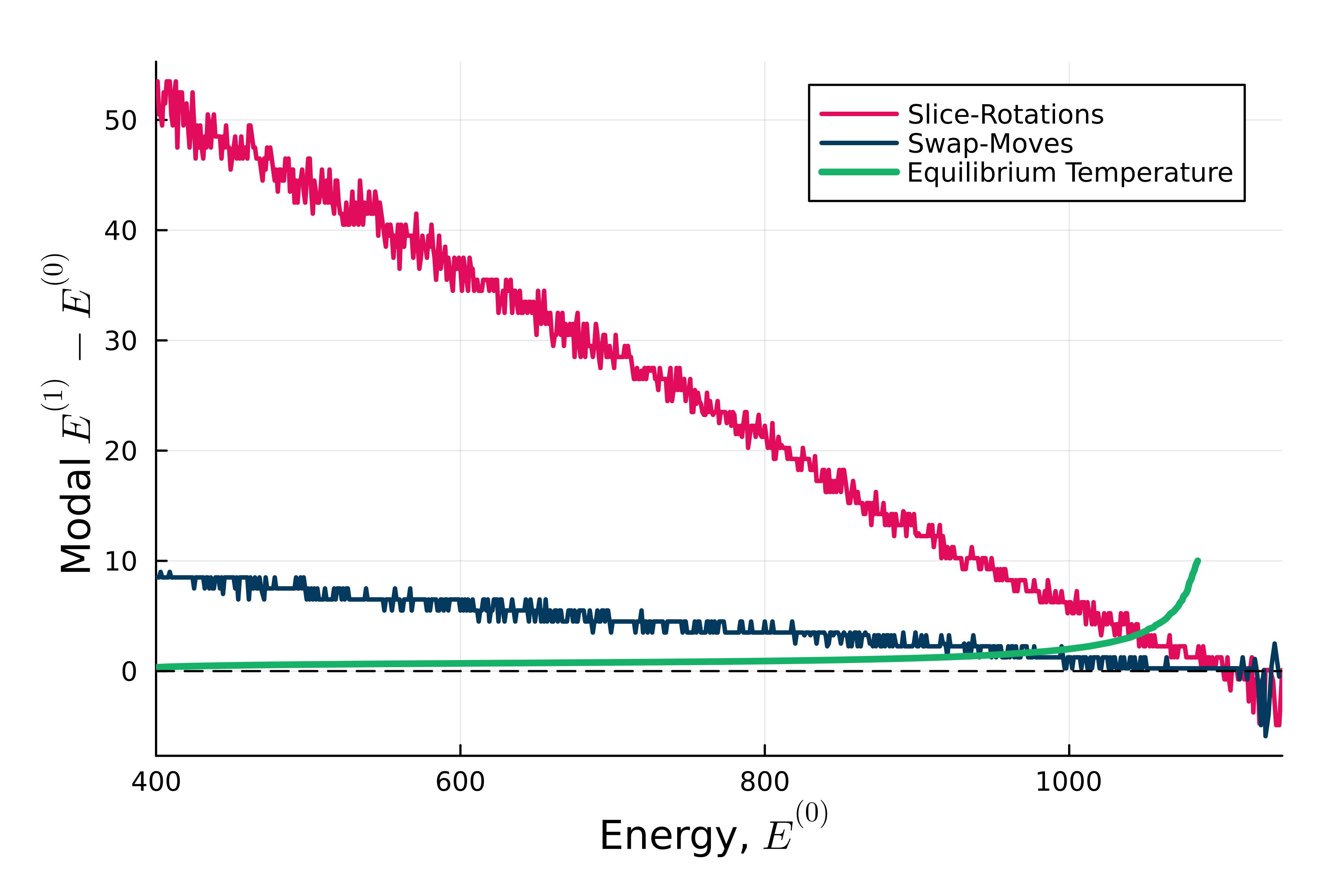}
    \caption{Modal energy differences and equilibrium temperature against energy for both swap-moves and slice-rotations for the $L=11$ randomised cube.}
    \label{fig:modal-energy-difference-and-equilibrium-temperature-against-energy}
\end{figure}

In Fig.~\ref{fig:modal-energy-difference-and-equilibrium-temperature-against-energy} we present the modal energy differences, $E^{(1)} - E^{(0)}$, against energy, $E^{(0)}$ for both slice-rotations and swap-moves for the $L=11$ randomised cube (data from Fig.~\ref{fig:E0-E1-connectivities} in the main text). We observe that the modal energy differences for both swap-moves and slice-rotations grow linearly with decreasing energy, but the slice-rotation energy differences grow at a significantly faster rate.

We also plot the equilibrium temperature against energy (data from Fig.~\ref{fig:relaxed-anneals} in the main text), showing that it intersects the energy difference plots for both systems at relatively high energy. Activated processes (biased by an $\exp[-(E^{(1)} - E^{(0)})/T]$ Boltzmann factor) are therefore generally expected for both systems at low energies, and there is no qualitatively distinct behaviour between the two that we could identify. 
This reaffirms the importance of the saddles-to-minima crossover, present in one system and not in the other, in determining the markedly different dynamical behaviour. 
%
%

\putbib
\end{bibunit}

\end{document}